\documentstyle[epsfig]{mn}
\newcommand{\bn}{{\mathbf{\nabla}}}
\newcommand{\aap}{A\&A}

\newcommand{\apj}{ApJ}

\newcommand{\mnras}{\mbox{MNRAS}}

\newcommand{\nat}{Nature}

\begin{document}

\title[Bright and Dark Matter in Ellipticals] {Bright and Dark Matter in Elliptical Galaxies: Mass and Velocity Distributions from Self-consistent Hydrodynamical Simulations}
\author[O\~norbe et al.]
{J. O\~norbe$^1$, R. Dom\'{\i}nguez-Tenreiro$^1$, A. S\'aiz$^1$\thanks{Current address:
Dept.\ of Physics, Mahidol University, Bangkok 10400, Thailand},  
   and A. Serna$^2$\\
  $^1$Departamento de F\'{\i}sica Te\'orica, C-XI. Universidad
  Aut\'onoma de Madrid, Madrid, E-28049, Spain\\
$^2$Departamento de F\'{\i}sica y A.C., Universidad Miguel Hern\'andez, Elche,
Spain
 }
\maketitle

\begin{abstract}
We have analysed  the mass and velocity
distributions of two samples of relaxed elliptical-like-objects (ELOs)
identified, at $z=0$, in a set of self-consistent hydrodynamical simulations
operating in the context of a concordance cosmological model.
ELOs have been identified as those  virtual galaxies having
a prominent, dynamically relaxed
stellar spheroidal component, with no extended discs and very low
gas content.  
 Our analysis shows that they
 are embedded in extended, massive dark matter haloes,
 and they also have an extended corona of hot diffuse gas.
Dark matter haloes have experienced adiabatic contraction along their assembly
process. The relative ELO dark- to bright-mass content and
space distributions show broken homology, and they are consistent with
observational results on the dark matter fraction at the central
regions, as well as on the gradients of the mass-to-light
ratio profiles for boxy ellipticals,
as a function of their stellar masses.
These results indicate that massive ellipticals miss stars
(i.e., baryons) at their
central regions, as compared with less massive ones. Our simulations
indicate that these missing baryons could be found beyond the virial
radii as a hot, diffuse plasma.
This mass homology breaking could have important implications to
explain the physical origin of the Fundamental Plane relation.
The projected stellar mass profiles of our  virtual
ellipticals can be well fit by the S\'ersic
formula, with shape parameters $n$ that agree,
once a stellar mass-to-light ratio independent of position is assumed,
with those obtained
from surface brightness profiles of ellipticals.
The agreement includes the empirical correlations of
$n$ with size, luminosity and velocity dispersion.
The total mass density profiles show a power-law behaviour
over a large $r/r_{\rm vir}$ interval, consistent with data
on massive lens ellipticals at shorter radii.
The velocity dispersion profiles show kinematical segregation,
with no systematic mass dependence (i.e., no dynamical homology
breaking) and a positive anisotropy (i.e., radial orbits),
roughly independent
of the radial distance outside the central regions.
The LOS velocity dispersion profiles
are declining.
These results give, for the first time from cosmological 
simulations, a rather detailed insight into the intrinsic mass and
velocity distributions of the dark, stellar and gaseous components  of
virtual  ellipticals. The consistency with observations
strongly suggests that they could also describe important 
intrinsic characteristics
of real ellipticals, as well as some of their properties
recently inferred from observational data (for example,
downsizing, the appearance
of blue cores, the increase of the stellar mass contributed by the elliptical
population as $z$ decreases).

\end{abstract}


\section{Introduction}
\label{Intro}
Among all galaxy families, elliptical galaxies (EGs) are the simplest ones
and those that  show the most precise empirical regularities,
such as:
the very similar shapes  their surface brightness profiles
show (i.e., the S\'ersic law, S\'ersic 1968), 
the  power-law form of their  three-dimensional mass density
profiles emerging from some strong lensing analyses (Koopmans et al. 2006),
 and the relations
 among some of their observable parameters.
The importance of these regularities lies in
that they very likely encode a lot of 
relevant informations on 
the physical processes underlying the histories of the 
mass assembly, the rates of dissipation  and
the rates of star formation (SF) of EGs.

Despite of their interest, very few is known, both from the theoretical or
observational points of view, about the mass or velocity distributions
of the different ellipticals mass components (stars, hot and cold gas and dark matter).
There has been, nevertheless, an important recent progress on the photometric
characterisation  of EGs, and, in fact,
authors now agree that the S\'ersic law
adequately describes the optical surface brightness profiles
of most of them (Caon, Capaccioli, \& D'Onofrio 1993;
 Trujillo, Graham, \& Caon 2001; Bertin, Ciotti, \& Del Principe 2002).
The S\'ersic law can be written

\begin{equation}
I^{\rm light}(R) = I_0^{\rm light} exp [-b_{n}(R/R_e^{\rm light})^{1/n}],
\label{sersic}
\end{equation}

where $I^{\rm light}(R)$ is the surface brightness at projected distance $R$
from the ellipticals centre, $R_e^{\rm light}$ is the effective half-light radius, 
encompassing half the total galaxy luminosity, 
$b_n \simeq 2n -1/3 + 0.009876/n$,
and $n$ is the S\'ersic shape parameter. Putting $n=4$ the largely
used de Vaucouleurs $R^{1/4}$ law (de Vaucouleurs 1948) is recovered.

It is generally assumed that galaxies of any type are embedded
in massive haloes of dark matter.
However, from the observational point of view,  the
importance and the distribution of dark matter in EGs
is still a matter of a living debate. Data on stellar kinematics from
integrated-light spectra are very scarce beyond 2$R_e^{\rm light}$,
making it difficult even to establish the presence of a dark matter
halo (Kronawitter et al. 2000;
Magorrian \& Ballantyne 2001) through this method.
Otherwise,  the lack of mass tracers at larger distances that can
be interpreted without any ambiguity, has historically hampered
the proper mapping of  the mass distribution at the outer regions of EGs.
The situation is changing and a dramatic improvement is expected
in the near future. In fact, several ongoing projects have already
produced high quality data on samples of ellipticals  through
different methods, 
for example: stellar kinematics from integral-field
spectroscopic measurements
(SAURON; de Zeeuw et al. 2002; Cappellari et al.
2006); strong gravitational lensing (CLASS; Myers et al. 1995;
LSD; Koopmans \& Treu 2003, Treu \& Koopmans 2004; SLACS; Koopmans et al. 2006);
stellar kinematics from planetary nebulae 
(PNs; Douglas et al. 2002), or globular cluster 
(Bergond et al. 2006) observations;  and X-rays (O'Sullivan \& Ponman 2004a,b).
In particular, the combination of high-quality stellar spectroscopy
and strong lensing analyses breaks the so-called
mass-anisotropy degeneracy, giving strong indications
that constant mass-to-light ratios can be ruled out at $> 99\%$ CL,
consistent with the presence of massive and extended dark matter haloes
around, at least, the massive lens ellipticals analysed
so far (Treu \& Koopmans 2004; Koopmans et al. 2006).
Moreover, these authors have also found that the
dark matter and the baryons mass density profiles
combine in such a way that the total mass density profiles can be
fit by power-law expressions within their Einstein radii,
whose slopes are consistent with isothermality.
Similar conclusions on the important amounts of dark matter 
inside the virial radii of ellipticals have been reached from
weak lensing of $L_*$ galaxies 
(Guzik \& Seljak 2002; Hoekstra, Yee \& Gladders 2004),
dynamical satellite studies (van den Bosch et al. 2004)
and X-ray analyses (Humphrey et al. 2006).
Other observational results or some of their interpretations, however,
could suggest that the amounts of  dark matter in the haloes
of some ellipticals are not that important.
For example,  Napolitano et al. (2005)
have analysed the mass-to-light gradients
of a sample of elliptical + SO galaxies, and found that these are positive and
important in massive, boxy EGs, 
but no very important for faint, discy EGs. This has been confirmed by
 Ferreras, Saha \& Williams 2005 using lensing analyses.
This result is similar to  what  Romanowsky et al. 
(2003; see also Romanowsky 2006)  have propounded
from the study of random velocities at
the outskirts of EGs through PN, found to be low, and first interpreted by these
authors as proving a dearth of dark matter in EGs,
while Dekel et al. (2005)
explain these large-radii low velocity dispersions as an effect of anisotropy
and triaxiality of the halo stellar populations of these galaxies.

Assuming that ellipticals are embedded in massive haloes of dark matter,
a second important concern is the possibility that their
profiles have near-universal shapes. Here most inputs come
from  numerical simulations because observational inputs are scarce.
When no dissipative processes are taken into
account, spherically averaged dark matter density  profiles of relaxed haloes
produced in N-body simulations
have been found to be well fitted by analytical expressions
  such that, once rescaled, give essentially a unique
   mass density profile, determined by two parameters. These two parameters are
    usually taken to be the total mass,
     $M_{\rm vir}$, and the concentration, $c$ or the energy content, $E$.
      These  two parameters are, on their turn, correlated
       (i.e., the mass-concentration relation, see, for example,
        Bullock et al. 2001; Wechsler et al. 2002;
	 Manrique et al. 2003).
When hydrodynamical forces and cooling processes enter the assembly of these
haloes and the baryonic objects they host, the
 dark matter profiles could be modified in the regions where
baryons are dynamically dominant, due the so-called adiabatic contraction
(see, for example, Blumenthal et al. 1986;
   Dalcanton, Spergel, \& Summers 1997; Tissera \& Dom\'{\i}nguez-Tenreiro 1998;
    Gnedin et al. 2004; Gustafsson, Fairbain \& Sommer-Larsen 2006).
So, the shapes of dark matter haloes in ellipticals could
deviate from the near-universal behaviour of dark matter haloes produced
in purely N-body simulations. 
    
Another important issue concerns the three dimensional cold baryon
mass (i.e., stellar mass) distribution, and, more particularly, 
its distribution relative to the dark matter haloes: are ellipticals
homologous systems or is the homology broken in their stellar mass
distribution or in their {\it relative} dark- versus bright-mass distribution?

Concerning the three dimensional velocity distributions of the different 
elliptical components,  very few is known either.
In particular, the anisotropy of the 
stellar three-dimensional velocity dispersion tensor
is hard to be observationally characterised.
This is an important issue, however, not only because anisotropy  plays
an important role in the analyses of the elliptical dark matter content
at several effective radii, but also because it 
could keep fossil informations about the physical processes
involved in mass assembly and stellar formation in EGs.
The relative behaviour of the three-dimensional velocity dispersion tensors
for the stellar and the dark mass components (i.e., the
so-called kinematical segregation) is still more uncertain.
There is not an unambiguous observational input about its 
presence in ellipticals, or about its possible systematic dependence
with the elliptical mass scale. This is an important point 
because of its possible connection with the physical origin
of the so-called Fundamental Plane (FP) relation,
as different authors have suggested (Graham \& Colless 1997;
Busarello et al. 1997; Pahre, de Carvalho \& Djorgovski 1998). 

Understanding the FP relation (Djorgovski \& Davis 1987; Dressler et al. 1987;
        Faber et al. 1987; Bernardi et al. 2003 and references therein)
is in fact a milestone in the physics of elliptical formation.
The FP is defined by the three observational  parameters characterising 
 the mass and velocity distributions of the stellar component:
 luminosity $L$,
  radius at half projected light $R_e^{\rm light}$, and
        central line-of-sight  velocity dispersion $\sigma_{\rm los, 0}$.
	  The FP is tilted relative to the virial plane.
	   Different authors interpret this tilt
	   as caused by different misassumptions relative
to the constancy of the dynamical mass-to-light ratios, $M_{\rm vir}/L$,
or the mass structure coefficients,
$c_{\rm M}^{\rm vir} = {G M_{\rm vir} \over 3 \sigma_{\rm los, 0}^{2} R_e^{\rm light}}$, with the mass scale
(see discussion in O\~norbe et al. 2005, 2006, and references cited therein).  
Recently,  O\~norbe et al. (2005, 2006)  have found that the
	   samples of elliptical-like objects identified at $z=0$
	   in their fully-consistent cosmological hydrodynamical simulations
show systematic trends with the mass scale in both, the relative
global dark-to-bright mass
content,  as well as in the relative sizes of the
stellar and the dark mass components, giving rise to {\it dynamical} FPs.

  We see that the mass or velocity  distributions of the different 
  elliptical mass components encode a lot of informations about
the physical origin of the FP, and, consequently, on the physics of elliptical formation.
We see also that, unfortunately,
observational methods, by themselves, suffer from some drawbacks
to deepen into these issues. 
 A major drawback
is that the informations on the intrinsical
 mass distribution are not directly available:
we see the projections 
of luminosity (not three-dimensional mass) distributions.
Another major drawback  is that  the intrinsic 3D velocity distribution 
of galaxies is  severely limited by projection,
only 
 the line-of-sight velocity distributions 
can be inferred from galaxy spectra.
And, so, the interpretation  of observational data
is not always straightforward.
To complement the informations provided by data and circumvent these drawbacks,
analytical modelling is largely used in literature 
(Kronawitter et al. 2000; Gerhard et al. 2001;
Romanowsky \& Kochanek 2001;
Borriello, Salucci \& Danese 2003; Padmanabhan et al. 2004;
Mamon \& Lockas 2005a, 2005b).
They give very interesting insights into
mass and velocity distributions, as well as the physical
processes causing them, but are somewhat
limited by symmetry  considerations and other necessary  simplifying
hypotheses.
These difficulties and limitations could be circumvented should we have
at our disposal complete  informations on the phase-space
of the galaxy constituents.
This is not possible through observations,
but can be attained, at least in a virtual sense, through
numerical simulations.

The first authors who studied  the
formation and properties of EGs by means of numerical
methods used purely gravitatory pre-prepared simulations. The origin of the FP
was addressed, among others, by  Capelato, de Carvalho \& Carlberg (1995);
Gonz\'alez-Garc{\'\i}a \& van Albada (2003);
Dantas et al. (2003); Nipoti, Londrillo \& Ciotti (2003, 2006) and 
Boylan-Kolchin, Ma \& Quataert (2005).
Some of these authors (Dantas et al. 2003, Nipoti et al. 2003) conclude that
dissipation must be a basic ingredient in elliptical formation.
Bekki (1998) first considered the role of dissipation in elliptical
formation through pre-prepared simulations. He adopts the merger hypothesis
(i.e., ellipticals form by the mergers of two equal-mass gas-rich
spirals) and he focuses on the role of the timescale for SF
in determining the structural and kinematical properties of the
merger remnants.
He concludes that the slope of the FP reflects the
difference in the amount of dissipation the merger end products
have experienced according with their luminosity (or mass).
Recently, Robertson et al. (2006) have confirmed
this conclusion on the role of dissipative dynamics
to shape the FP, again  through  pre-prepared
mergers of disc galaxies.

Apart from the FP relation origin, other aspects of the formation and
evolution of EGs have been analysed through pre-prepared
simulations. For example, the different isophotal shapes of ellipticals
 and their kinematical
support (either intrinsic rotation or anisotropic dispersion) have been
addressed by A. Burkert and coworkers, who analysed in detail binary
mergers of virtual galaxies with different morphologies and initial
conditions (Naab \& Burkert 2003; Naab \& Trujillo 2006;
Naab, Khochfar \& Burkert 2006).  
Otherwise, Gonz\'alez-Garc\'{i}a, Balcells \& Olshevsky (2006) analysed the
velocity distribution of dissipationless
binary  merger remnants involving  galaxies with different morphologies.

We see that pre-prepared simulations of merger events provide
a very useful tool  to work out
the mass and velocity distributions of EGs.
They allow also to find out their links with the processes involved
in galaxy assembly, but they are somewhat limited,
for example 
by the fact that
the probability of a particular initial setup at a given $z$ is not
known a priori, and that mergers involving more than two
objects also occur and are frequent at high $z$s,
so that some complementary informations must be
provided, for example through semi-analytical models (Khochfar \&
Burkert 2005; Naab, Khochfar \& Burkert 2006).

To overcome these limitations, a convenient method is to study
the processes involved
in galaxy formation  in a {\it cosmological context}
through {\it self-consistent} gravodynamical simulations.
The method works as follows:
initial conditions are set at high
$z$, in an homogeneously sampled box, as a Montecarlo
realization of the field of primordial fluctuations
to a given cosmological model;
 then the evolution of these fluctuations is numerically followed
 up to $z =0$ by means of a computer code that solves,
 in a periodic box, the
 gravitational and  hydrodynamical
 evolution equations.
 This method allows us to follow the evolution
 of the dynamical and thermohydrodynamical
 properties of matter in the universe;
 individual galaxy-like objects  naturally appear as a consequence 
 of this evolution.
  No  prescriptions  are needed
  as far as their mass  assembly processes
  are concerned, only SF processes need further modelling.
  The important point here is that self-consistent simulations   
 {\it directly}  
provide  with complete 6-dimensional
phase-space informations on each constituent particle sampling a
given galaxy-like object formed in the simulation,
that is, they give directly the mass and velocity  distributions
of dark matter, gas and stars of each object.
 
Kobayashi (2005)  has simulated the chemodynamical evolution of 74
fields with different cosmological cold dark matter  initial spectra set  in
slowly rotating spheres, each of them  with a 1.5  Mpc comoving radius
and vacuum boundaries. So, these simulations are not yet fully self-consistent.
She succeeded in reproducing the observed
global scaling relations shown by EGs, and, in particular, the FP relation,
and the surface-brightness profiles, as well as the colour-magnitude and
the mass-metallicity relations. She  also analyses the role of major
merger events and the timescales for SF
in shaping the mass and sizes of remnants.

Concerning self-consistent hydrodynamical simulations,
Sommer-Larsen, Gotz \& Portinari (2002) present first results on early-type
galaxy formation in a cosmological context.
Meza et al. (2003) present results of the dissipative formation of a 
compact elliptical galaxy in the $\Lambda$CDM scenario.
Kawata \& Gibson (2003, 2005) and Gibson et al. (2006) have studied the
X-ray and optical properties of virtual ellipticals formed in different 
simulations run with their chemodynamical Tree/SPH code. 
Romeo, Portinari \& Sommer-Larsen (2005) analyse the galaxy stellar  populations
formed in their simulations of galaxy clusters.
Naab et al. (2005) got, from cosmological initial conditions,
a spheroidal system whose photometric
and kinematical properties agree with observations of ellipticals,
in a scenario  not including feedback from supernovae or AGN
and not requiring recent major mergers.
Interesting results on elliptical formation have also been obtained
by de Lucia et al. (2006),  from a semi-analytic model
of galaxy formation grafted to the {\it Millennium Simulation}.

However, detailed analyses of the mass and velocity distributions
of samples of virtual ellipticals formed in fully self-consistent
hydrodynamical simulations,
and, in particular, of the amount and distribution of dark matter
relative to the bright matter distribution,
as well as of the kinematics of the dark and bright components,
and their successful comparison with observational data,
were still missing.
This is the work we present in this paper.
To be specific, we have analysed the samples of virtual ellipticals  formed in
ten self-consistent hydrodynamical   
simulations, run in the framework of
a flat $\Lambda$CDM cosmological model characterised
by cosmological parameters consistent with their
last determinations (Spergel et al. 2006). 
Galaxy-like objects of different morphologies appear in these
 simulations at $z=0$: disc-like objects, SO-like objects,
 elliptical-like objects (hereafter,
ELOs) and peculiars.

Our previous work has shown that these ELO samples have properties
that agree with observational data,
so that they have counterparts in the real world.
In fact, an analysis of the structural and dynamical ELO parameters that
can be constrained from observations 
(i.e., stellar masses,  projected half-mass  radii,
central line-of-sight velocity dispersions), has shown that they
are consistent with those measured in the SDSS DR1 elliptical sample 
(S\'aiz, Dom\'{\i}nguez-Tenreiro 
\& Serna 2004), including the FP relation
(O\~norbe et al. 2005), and their lack of evolution at
low and intermediate redshifts 
(Dom\'{\i}nguez-Tenreiro et al. 2006, hereafter DTal06).
Also, their stellar populations have age distributions
showing similar trends as those inferred from observations, i.e.,
most stars have formed at high $z$ on short timescales, and, moreover
more massive objects have older  means  and narrower spreads
in their stellar age distributions than less massive ones \footnote[1]{
this is equivalent to the {\it downsizing} concept introduced by
Cowie et al. (1996) to mean that SF is stronger at
low redshifts in smaller galaxies than in larger ones}
(Dom\'{\i}nguez-Tenreiro, S\'aiz \& Serna, 2004, hereafter DSS04).

The paper is organised as follows:
in $\S$2 we briefly describe the simulations and the SF 
modelling.
In $\S$3, the ELO sample and the generic structure of ELOs
are described.
A brief account on ELO formation is given in $\S$4.
The three dimensional orbital structure of the
baryonic component is briefly described in $\S$5.
$\S$6  is devoted to report on the dark matter and
baryonic mass density profiles of  the ELOs produced in the simulations.
In $\S$7 we report on the kinematics of the different ELO components.  
Finally, in $\S$8 we summarise our results and discuss them 
in the context of theoretical results on halo structure
and dissipation of the gaseous component and their connections with
observational data.

\section{The Simulations}
\label{simula}

We have analysed ELOs identified in ten
self-consistent cosmological simulations run  
in the framework of the same
global  flat $\Lambda$CDM cosmological model,
with $h=0.65$,
$\Omega_{\rm m} = 0.35$, $\Omega_{\rm b} = 0.06$.
The normalisation parameter has been taken slightly high,
$\sigma_8 = 1.18$, as compared with the average fluctuations
of 2dFGRS or SDSS  galaxies
(Spergel et al. 2006)
to mimic an active region of the Universe
(Evrard, Silk \& Szalay 1990).

We have used a lagrangian code
(\textsc{DEVA}; Serna, Dom\'inguez-Tenreiro \& S\'aiz 2003),
particularly designed to study
 galaxy assembly in a cosmological context.
  Gravity is computed through an AP3M-like  method, based
  on Couchman (1991).
  Hydrodynamics is  computed through a SPH technique where 
  special attention has been paid that
  the implementation of conservation laws (energy, entropy
  and angular momentum) is as accurate as possible
   (see Serna et al. 2003 for details,
   in particular for a discussion on the implications of
   violating some conservation laws).
 Entropy
 conservation  is assured by taking into consideration the space variation
 of the smoothing length (i.e., the so-called $\bn h$ terms).
   Time steps are individual for particles
  (to save CPU time, allowing a good time resolution),
 as well as masses. Time
		    integration uses a PEC scheme.
In any run, an homogeneously sampled
periodic box of 10~Mpc side  has been employed
and  64$^3$
dark matter and 64$^3$ baryonic particles,
with a mass of $m^{\rm dark} = 1.29 \times
10^8$ and $m^{\rm b} = 2.67 \times 10^7 $M$_{\odot}$, respectively, have
been used.
The gravitational softening used was $\epsilon = 2.3$ kpc.
The cooling function is that from Tucker (1975)  and Bond et al. (1984)
for an optically thin primordial mixture of H and He ($X=0.76$,
$Y=0.24$) in collisional equilibrium and in absence of any
significant background radiation field
with a primordial gas composition.
Each  of the ten simulations started at a redshift $z_{\rm in} = 20$.

SF processes have been included through a simple 
phenomenological parametrization, as that first used by Katz (1992,
see also Tissera, Lambas \& Abadi 1997
and Serna et al. 2003 for
details)
  that transforms   cold locally-collapsing gas
  at the scales the code resolves,
  denser than a threshold density,
  $\rho_{\rm thres} $,
  into stars at a rate
  $d\rho_{\rm star}/dt = c_{\ast}  \rho_{\rm gas}/ t_g$,
  where  $t_g$ is a characteristic time-scale chosen
  to be equal to the maximum of the local gas-dynamical time,
  $t_{dyn} = (4\pi G\rho_{\rm gas})^{- 1/2}$,
  and the local cooling time;   $c_{\ast}$ is the average
  SF efficiency at resolution  $\epsilon $ scales, i.e.,
  the empirical Kennicutt-Schmidt law (Kennicutt 1998).
It is worth noting that, in the
context of the new sequential multi-scale SF scenarios
(V\'azquez-Semadeni 2004a and 2004b; Ballesteros-Paredes et al. 2006
and references therein),
 it has been argued that
 this law, and particularly so the low $c_{\ast}$ values
 inferred from observations, can be explained as a result of
 SF processes acting on dense molecular cloud core scales
 when conveniently averaged on disc scales
 (Elmegreen 2002, Sarson et al.\ 2004, see below).
Supernova feedback effects or  energy inputs other than gravitational
have not been  {\it explicitly} included in these simulations.
We note that the  role of discrete stellar energy sources at the
 scales resolved in this work is not yet clear,
  as some authors argue that
   stellar energy releases drive gas density structuration
    locally at sub-kpc scales (Elmegreen 2002).
      In fact, recent MHD simulations of self-regulating Type II supernova
        heating in the interstellar medium at scales $<$ 250 pc
	  (Sarson et al.\ 2004), indicate that
	    this process produces a Kennicutt-Schmidt-like law on
	      average. If this were the case, the Kennicutt-Schmidt law
	        implemented in our code would already  {\it implicitly}
		  account for the effects
		    stellar self-regulation has on the scales our code resolves,
		      and our ignorance on sub-kpc scale processes would be
contained in the particular values of $\rho_{\rm thres} $
and $c_{\ast}$.

Five out of the ten simulations
(the SF-A type simulations)
share the SF parameters ($\rho_{\rm thres} = 6 \times 10^{-25} $ gr cm$^{-3}$,
$c_*$ = 0.3) and differ in the seed  used to build up the initial conditions.
To test the role of SF parameterisation,
the same initial conditions
have been run with different SF parameters
($\rho_{\rm thres}$ =  $1.8 \times 10^{-24} $ gr cm$^{-3}$,
$c_*$ = 0.1) making SF  more difficult, contributing another set of five
simulations  (hereafter, the SF-B type simulations).

\section{Simulated Elliptical-like Objects at $z$=0: The ELO Sample}		
\label{ELOSample}

 ELOs have been identified as those galaxy-like-objects
  having a  prominent,  relaxed spheroidal
component, made out of stars, 
with no extended discs and very low cold gas content.
Moreover, their dark-matter haloes must also be relaxed to
allow their meaningful analysis, excluding systems that have just merged
or that are about to merge.
It turns out that, at $z=0$, 
26 (17) objects out of the more massive  formed in
SF-A (SF-B) type simulations fulfil these conditions.
ELOs form two  samples 
(the SF-A and SF-B ELO samples) partially analysed
 in S\'aiz et al. (2004), in DSS04
 and in DTal06.
 In O\~norbe et al. (2005, 2006) it is shown that both samples
 satisfy dynamical FP relations. 

ELOs are embedded in dark matter haloes whose virial radii 
\footnote[2]{The virial radii, $r_{\rm vir}$,  have been calculated
 using the Bryan \& Norman (1998) fitting formula, that yields, at $z=0$,
   a value of $\Delta \simeq 100$ for the mean density within
      $r_{\rm vir}$ in units of the critical density}
are in the range $527 > r_{\rm vir} > 191$  kpc.      
ELO stellar components have ellipsoidal shapes and have a  lower limit
in their stellar mass content of 3.8 $\times$ 10$^{10}$ M$_{\odot}$
(see Kauffmann et al. 2003 for a similar result in SDSS galaxies).
Inner discs, when present, are made out of cold gas. ELOs have
also  hot diffuse gas forming an extended halo (S\'aiz, Dom\'{\i}nguez-Tenreiro 
\& Serna 2003).
The number of dark and baryonic particles within $r_{\rm vir}$ 
are in the ranges $(5.3 \times 10^{4}, 2.4 \times 10^{3})$ and
$(3.1 \times 10^{4}, 2.0 \times 10^{3})$, respectively,
giving a lower limit in the
virial masses of ELOs of $M_{\rm vir} > $ 3.7 $\times$ 10$^{11}$ M$_{\odot}$.
Some ELOs show a clear net rotation,
resulting in an average value of their spin parameter of
$\bar{\lambda} = 0.033$.
ELO  mass function is consistent
with that of a small group, that is a dense, environment
(Cuesta-Bolao \& Serna, private communication).
  ELOs in the SF-B sample tend to be of later type than their
  corresponding SF-A counterparts because
forming stars becomes more
 difficult; this is why many of 
 the SF-B sample counterparts of the less massive ELOs
 in SF-A sample do not satisfy the selection criteria, and the SF-B sample has
 a lower number of ELOs that the SF-A sample.

\section{A brief account on ELO formation}
\label{ELOForm}

The simulations unveil the physical  patterns of
ELO mass assembly, energy dissipation
and SF rate histories
(see Sierra Glez. de Buitrago et al. 2003;
    DSS04, DTal06).
Our simulations indicate that
  ELOs are assembled out of  the mass elements that at high $z$ are
     enclosed by those  overdense regions $R$
         whose local coalescence length $L_c(t, R)$
(Vergassola et al. 1994)
	 grows much faster than average, and whose mass scale
	 (total mass enclosed by $R$, $M_R$)
	 is of the order of an elliptical galaxy virial mass.
Analytical models, as well as N-body simulations indicate that two different
phases operate  along  halo mass assembly: first, a violent fast one,
where the mass aggregation rates are high,
and then, a  slower one, with lower mass aggregation rates
(Wechsler et al. 2002; Zhao et al. 2003; 
Salvador-Sol\'e, Manrique, {\&} Solanes 2005).
  Our hydrodynamical simulations indicate 
that the fast phase occurs through a
multiclump collapse (see Thomas, Greggio \& Bender 1999)
ensuing turnaround of the overdense regions,
and it is characterised by the fast head-on 
(that is,  with very low relative orbital angular momentum)
fusions experienced by
the nodes of the cellular structure these regions enclose,
resulting in strong shocks and  high cooling rates of their gaseous component,
and, at the same time, in strong and very fast star SF bursts
that transform most of the available cold  gas in $R$  into stars.
For the massive ELOs in this work, this happens between $z \sim 6$
and $ \sim 2.5$ and mainly
corresponds to a cold mode of gas aggregation, as in
Keres et al. (2005).
Consequently, most of the dissipation involved in the mass assembly of
a given ELO occurs  in this violent early phase at high $z$;
moreover, its rate history \footnote[3]{That is, the
amount of cooling per time unit
experienced by those gas particles that at $z=0$ form the ELO stellar component}
is reflected by the SF rate history,
as illustrated in figure 1 of DTal06. 
The Fundamental Plane relation
shown by EGs appears in this fast violent phase as
a consequence of dissipation and homology breaking in the mass distribution
(see O\~norbe et al. 2005, 2006 and DTal06).

The slow phase comes after the fast phase.
In this phase, the halo mass aggregation rate is low and the $M_{\rm vir}$
increment results from major mergers, minor mergers or continuous
mass accretion. Our cosmological simulations show that
the  fusion rates are generally low and that
these mergers generally imply only a modest amount of energy dissipation
or SF. 
In fact,  a strong SF burst
and dissipation  follow a major merger
only if enough gas is still available after the early violent phase.
This is  unlikely in any case, and it becomes more and more
unlikely as $M_{\rm vir}$ increases (see DSS04).
A  consequence of the lack of dissipation is that the Fundamental
Plane  is roughly preserved  along  the slow phase
(see DTal06).
We have to point out that mergers play an important role in this slow phase
as far as mass assembly is concerned:
an $\sim $ 50\% of ELOs in the sample have experienced a major merger
event at $ 2 < z < 0$, that result in the increase of
the ELO mass content, size  and 
stellar mean square velocity.
Some of these mergers are multiple, and in some few cases, either binary or
multiple mergers  involve disc galaxies, but none of the ELOs in our
sample has been shaped by a merger of two or more adult disc virtual galaxies,
maybe because our simulated box mimics a dense environment.

So, our simulations suggest  that most of the stars of to-day ellipticals,
or at least of those in dense environments, could have
formed at high redshifts, while they are assembled later on
[see de Lucia et al. (2005), for similar results from a semi-analytic model
of galaxy formation grafted to the {\it Millennium Simulation}].
    This formation scenario
    shares some aspects of both, the hierarchical merging
      and the monolithic
       collapse scenarios,
but it has also significant differences, mainly
that most stars belonging to EGs
form out of cold gas that had never been shock heated
at the halo virial temperature and then formed a disc,
as the conventional recipe for galaxy formation propounds
[see discussion  in Keres et al. (2005) and references therein].
An important point is that  our simulations indicate that
this formation scenario follows from simple physical  principles in the context
of the current $\Lambda$CDM scenario.

\section{Three dimensional  structure of the baryonic constituent}
\label{ThreeDStrucDis}                                                 

A  quantitative description of ELO
mass distributions 
is given
by their 3D density profile and the structure their constituent particles.
We first address the structure of the baryonic particles.

\subsection{Three dimensional  structure for gas particles}
\label{ThreeDStruc}

\begin{figure*}[H]
  \begin{center}
    \includegraphics[width=.7\textwidth]{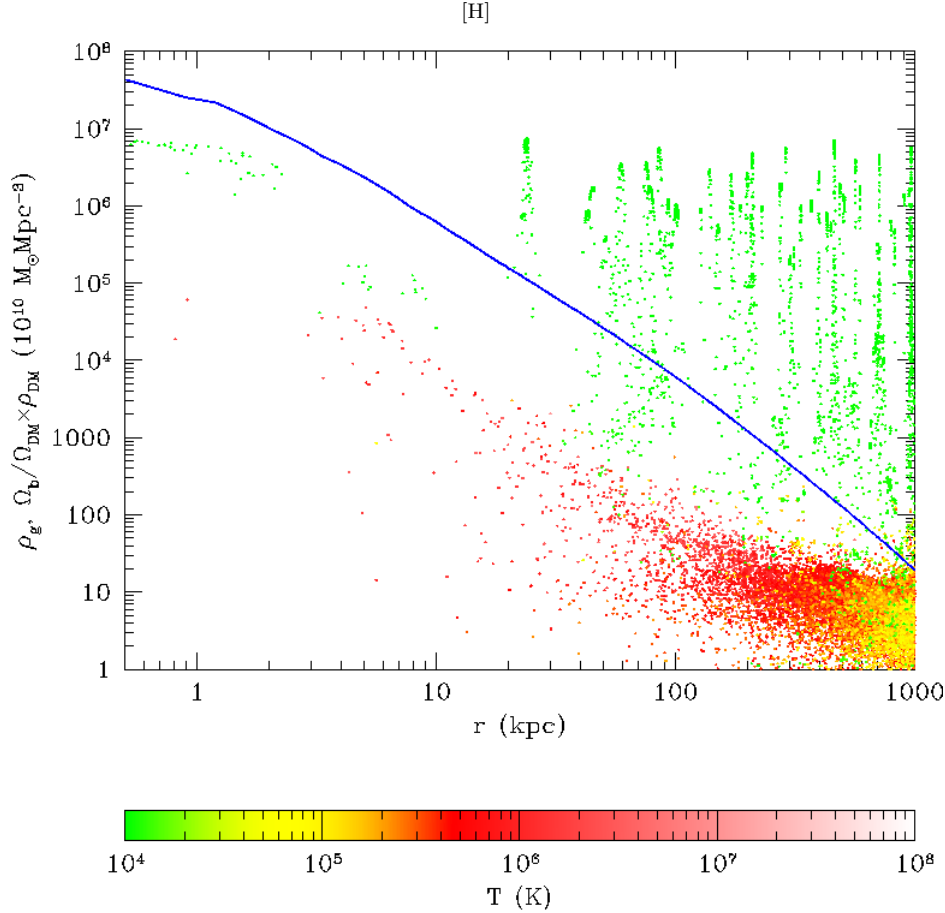}
    \caption[3D gas and dark matter densities for ELO S26/\#353]{
      3D gas (points) and dark matter (blue line)
      density for a typical ELO. Note the dense cold gas clumps embedded in the
      diffuse hot gas component.
      See text for an explanation. 
      }
    \label{Dens3DMonster}
  \end{center}
\end{figure*}

The gas structure  is drawn in Figure~\ref{Dens3DMonster} 
for the second more massive object formed in a SF-B type simulation.
 The 3D density at a given distance,
$r$, from the centre of the object has been calculated
by binning on concentric spherical shells around $r$. 
In this Figure, the line is the density profile
of dark matter around the object, multiplied by
$\Omega_{\rm b} / \Omega_{\rm m}$.
Points represent gas density at the positions of SPH particles,
and colours stand for gas particle temperatures according with
the scale at the bottom of the Figure.

We see in this Figure that very few gas is left at positions
with $r \le 30$ kpc where stars dominate the mass density,
that cold gas at $r \ge 30$ kpc is dense and clumpy,
while hot gas (that is, gaseous particles with $T > 3 \times 10^{4}$K)
is diffuse with an almost isothermal component
at 100 kpc $ \le r \le $ 400 kpc, and a warm component at the outskirts of
the configuration, reaching  outside the virial radius (395.0 kpc).
Two scales stand out in this configuration: the {\it ELO scale} or stellar
component, with a size in this case  of $\sim $ 30  kpc,
and the {\it halo scale}, a halo of dark matter of 395.0 kpc.
Cold dense gas particles  are associated in most cases with small dark matter
haloes (not seen in the Figure);
both gaseous particles in cold clumps and dark matter particles
in their (sub)haloes are {\it shocked} particles, using the
terminology of the adhesion model (see, for example, Vergassola et al. 1994). 
The configuration illustrated by this Figure is generic for ELOs:
we can distinguish an  {\it ELO scale},
with typical sizes of no more than $\sim $ 10 - 40  kpc,
and the {\it halo scale}, a halo of dark matter typically ten  
times larger in size.

\begin{figure*}
  \begin{center}
    \includegraphics[width=.7\textwidth]{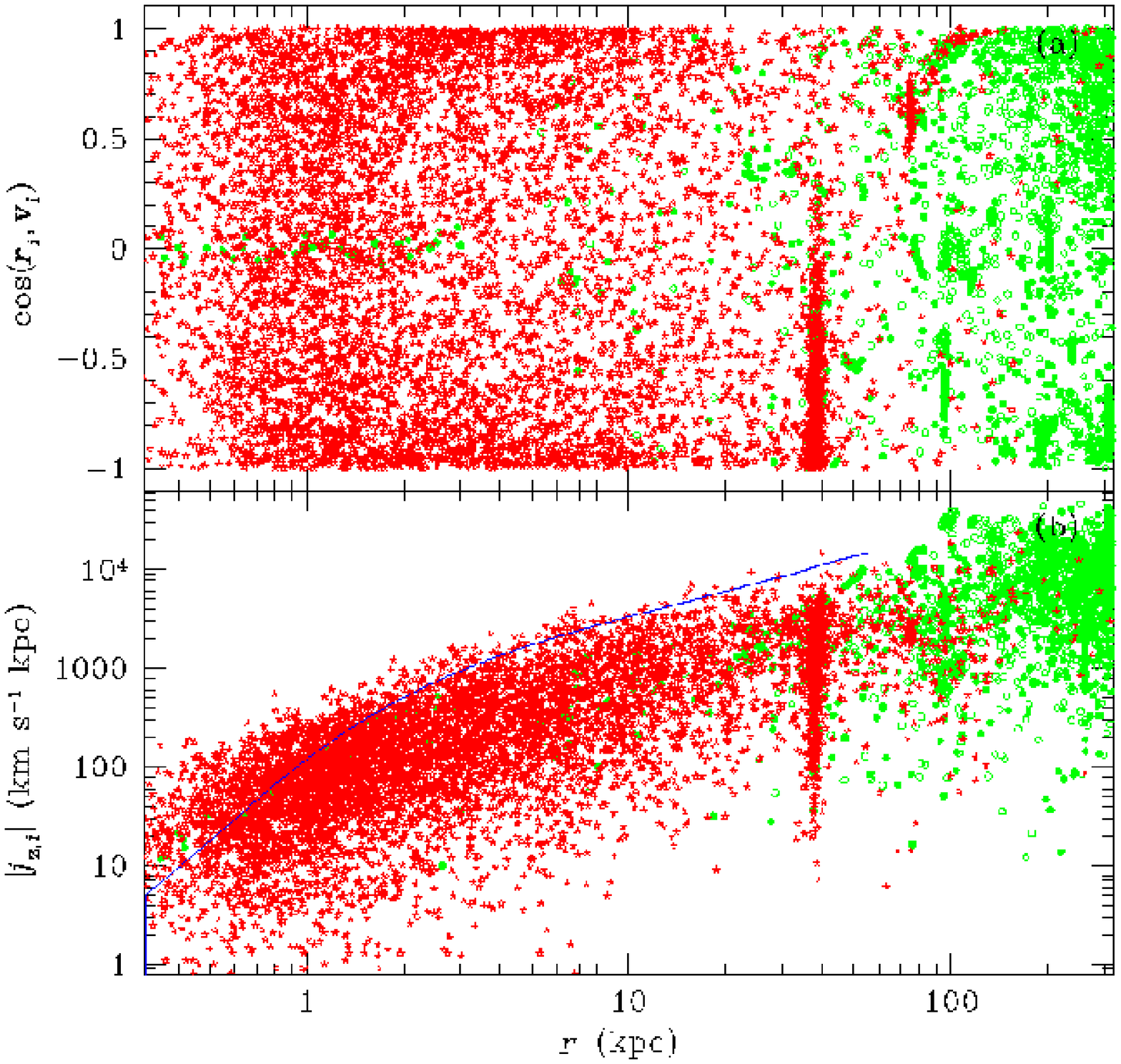}
    \caption[Cosine of position and velocity, and angular momentum,
     of constituent particles of ELO S26/\#353]{ Upper panel:
  the cosine of the angle formed by the position and the velocity
  vectors for each gaseous (green circles) and stellar (starred red symbols)
  particle belonging to a typical ELO.
  Filled (open) symbols stand for particles in (counter)
       corotation with the small inner disc
   }
  \label{CosJzSegundo}
 \end{center}
\end{figure*}

\subsection{Stellar and gaseous particle orbits}
\label{orbits}

ELO constituent particles  of different kinds travel on orbits
that have  different characteristics.
To analyse this point, in the upper panel of
Figure~\ref{CosJzSegundo}
we plot, for each star particle  and
each gaseous particle  of a typical ELO,
the cosine of the angle formed by its position ($\vec{r_i}$)
and its velocity ($\vec{v_i}$) as a function of $r_i$.
Positions and velocities have been taken
with respect to  the centre of mass of the main baryonic
object.  In this plot radial orbits have cosines = $\pm 1$,
while circular orbits have cosines = 0. Starred (circular)
symbols
stand for stellar (gaseous) particles.
We see that cold gas particles at $r \le 4$ kpc form a disc in coherent
circular motion; filled (open) symbols represent particles in
corotation (counterrotation) with respect to this small disc.
We can also see that stellar particle orbits at $\le$ 3 kpc scales
do not show any preference, while those further away,
as well as gaseous particles outside the disc,
show a slight tendency to be on radial orbits providing anisotropy
to the velocity dispersion.
Stellar particles 
constitute  a disordered or dynamically hot component,
showing an important velocity dispersion, and, also, in some cases,
a coherent net rotation.
In \S\ref{Kine} these issues will be addressed in detail.

\section{The dark matter and baryonic  mass distributions}

\subsection{Dark matter profiles}
\label{DMProf}

Spherically averaged dark matter density  profiles of relaxed haloes
formed in N-body simulations
 have been found to be well fitted by analytical expressions
 such that, once rescaled, give essentially a unique
 mass density profile 
 i.e., a two parameter family.
 These two parameters are
 usually taken to be the total mass,
 $M_{\rm vir}$, and the concentration, $c$ or the energy content, $E$.
 These  two parameters are, on their turn, correlated
 (i.e., the mass-concentration relation, see, for example,
 Bullock et al. 2001; Wechsler et al. 2002; 
 Manrique et al. 2003)
 because the assembly process implies a given correlation between
 $M_{\rm vir}$ and $E$.
 Different authors propound slightly different fitting formulae,
 see Einasto (1965, 1968, 1969) or Navarro et al. 2004,
 Hernquist 1990 (Hern90), Navarro, Frenk \& White 1995, 1996 (NFW),
 Tissera \& Dom\'{\i}nguez-Tenreiro 1998 (TD),
 Moore et al. 1999 and Jing \& Suto 2000 (JS),
 that can be written as:
 
\begin{equation}
\rho_{\rm h}^{\rm dark}(r) = \rho_{\Delta}^{\rm aver} \times \frac{c^{3} \rho(r/a_h)}{3 g(c)}
\label{rhoDM}
\end{equation}

where $\rho_{\Delta}^{\rm aver}$ is the average density within the virial
radii, 
$c \equiv r_{\rm vir}/a_h$ is the so-called concentration parameter,
and

\begin{equation}
\rho(y) = y^{- \alpha} (1 + y)^{- \beta},
\label{rhored}
\end{equation}

where  ($\alpha, \beta$) = (1, 3) for
Hern90; ($\alpha, \beta$) = (1, 2) for NFW; 
($\alpha, \beta$) = (2, 2) for TD,
 and $\beta = 3 - \alpha$,
with $\alpha$ left free, for the general formula found by Jing \& Suto (2000)
(note that NFW can be considered as JS with $\alpha = 1$).
In these fitting formulae $\alpha$ is the inner slope ($r << a_h$), 
the outer slope ($r >> a_h$) is $\alpha + \beta$ (3 for JS or NFW), so that
$a_h$ characterises the scale where the slope changes. Other interesting 
scale is $r_{-2}$, the $r$ value where the logarithmic slope,
$d \ln \rho / d \ln r = -2$. 
We have $r_{-2} = a_h  (2 - \alpha)/(\alpha + \beta -2)$  for a profile given
by Eq.~\ref{rhored}, with $r_{-2} = a_h  (2 - \alpha)$ for JS and $r_{-2} = a_h$ for NFW. 
Navarro et al. (2004) propound a different fitting formula of the
form:

\begin{equation}
\rho(y) = \exp (-2 \mu y^{1/\mu}).
\label{NalDM}
\end{equation}

where $d \ln \rho / d \ln r = -2 (r/a_h)^{1/\mu}$ and $r_{-2} = a_h$.
Note that this last fitting formula is similar to the S\'ersic formula (Eq. 1),
as Merritt et al. (2005) first pointed out. It was first used by Einasto 
(1965, 1968, 1969), see also Einasto \& Haud (1989),
so that we will refer to it as {\it Einasto model } 
(Eina), in consistency with the terminology used by other
 authors (Merritt et al. 2006). 

The $g(c)$ functions can be written as:

\begin{equation}
g(y) = y^{2}/2(y+1)^{2}~(Hern90)
\end{equation}

\begin{equation}
g(y) = \ln (y + 1) - y/(y + 1)~(NFW)
\end{equation}

\begin{equation}
g(y) = 9 y/(1+y)~(TD)
\end{equation}

\begin{equation}
g(y) = (3-\alpha)^{-1} y^{3-\alpha} \hspace{0.1cm} _2F_1(3-\alpha,3-\alpha,4-\alpha,-y)~(JS)
\end{equation}

\begin{equation}
g(y) = \frac{1}{2} (2 \mu )^{1-3\mu} \gamma(3\mu,2\mu y^{1/\mu})~(Eina)
\end{equation}

where $_2 F_1$ is the hypergeometric function and $\gamma$ is the lower
 incomplete gamma function.

\begin{figure*}
   \begin{center}
     \includegraphics[width=.45\textwidth]{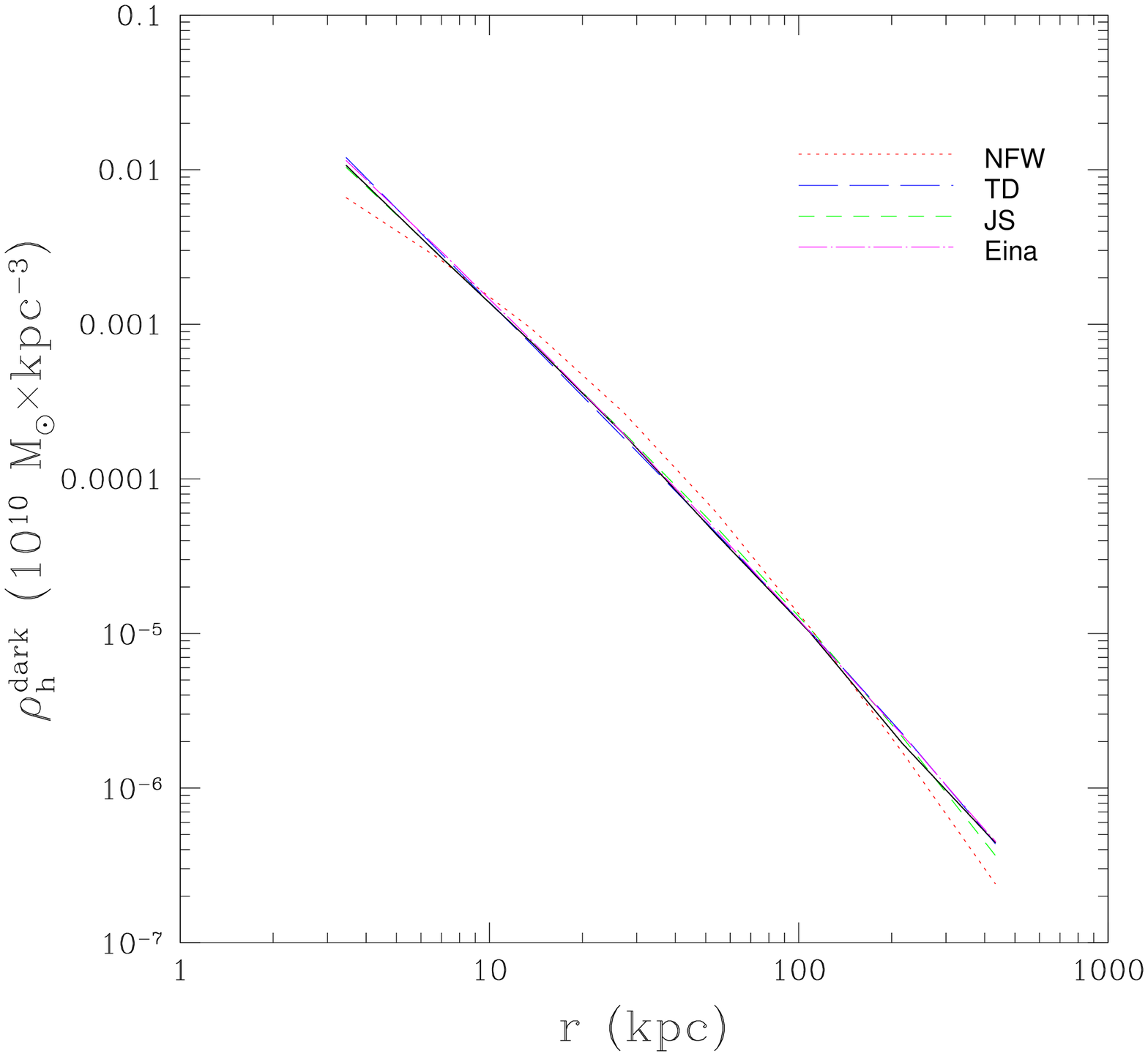}
     \includegraphics[width=.45\textwidth]{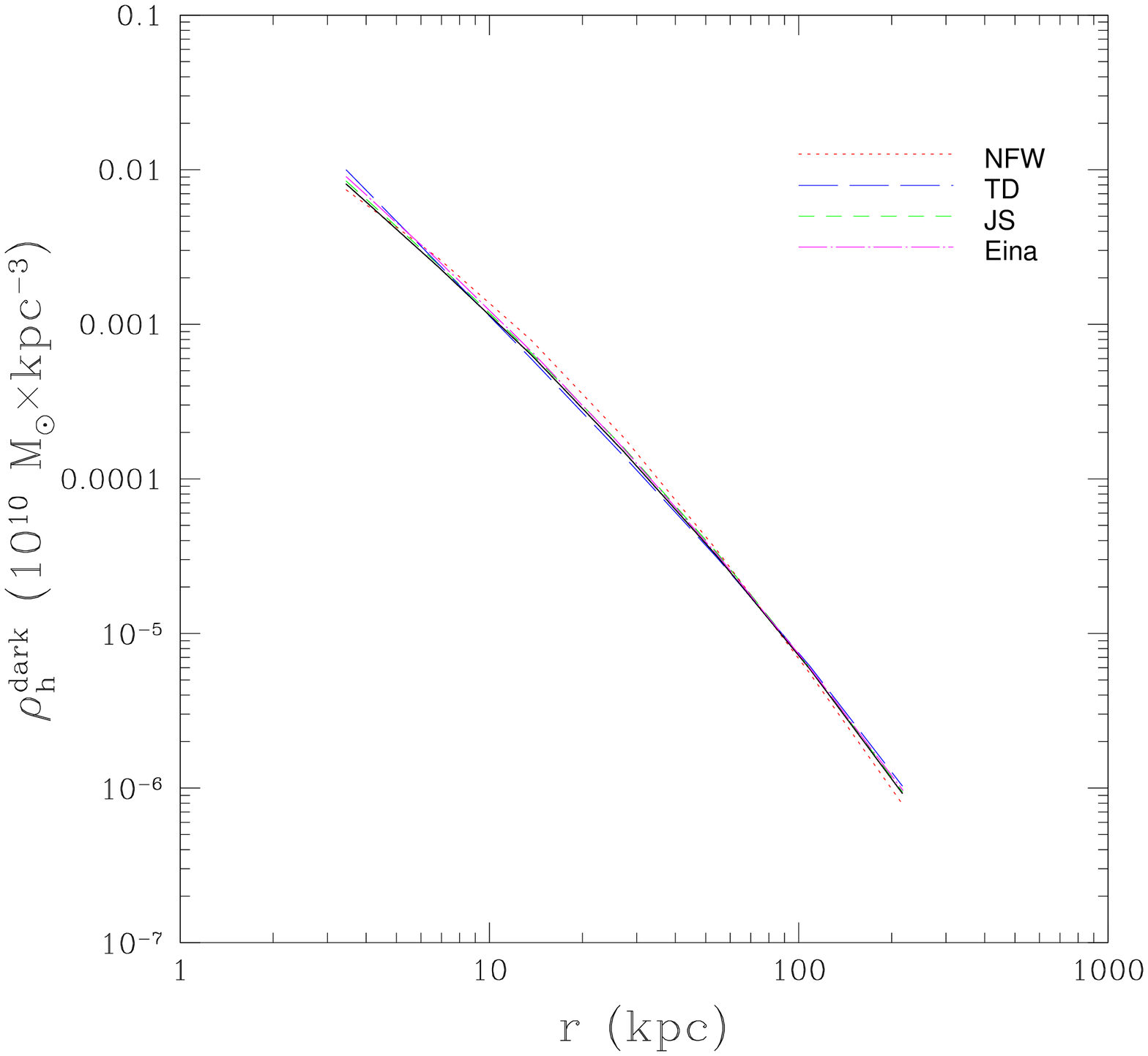}
     \includegraphics[width=.45\textwidth]{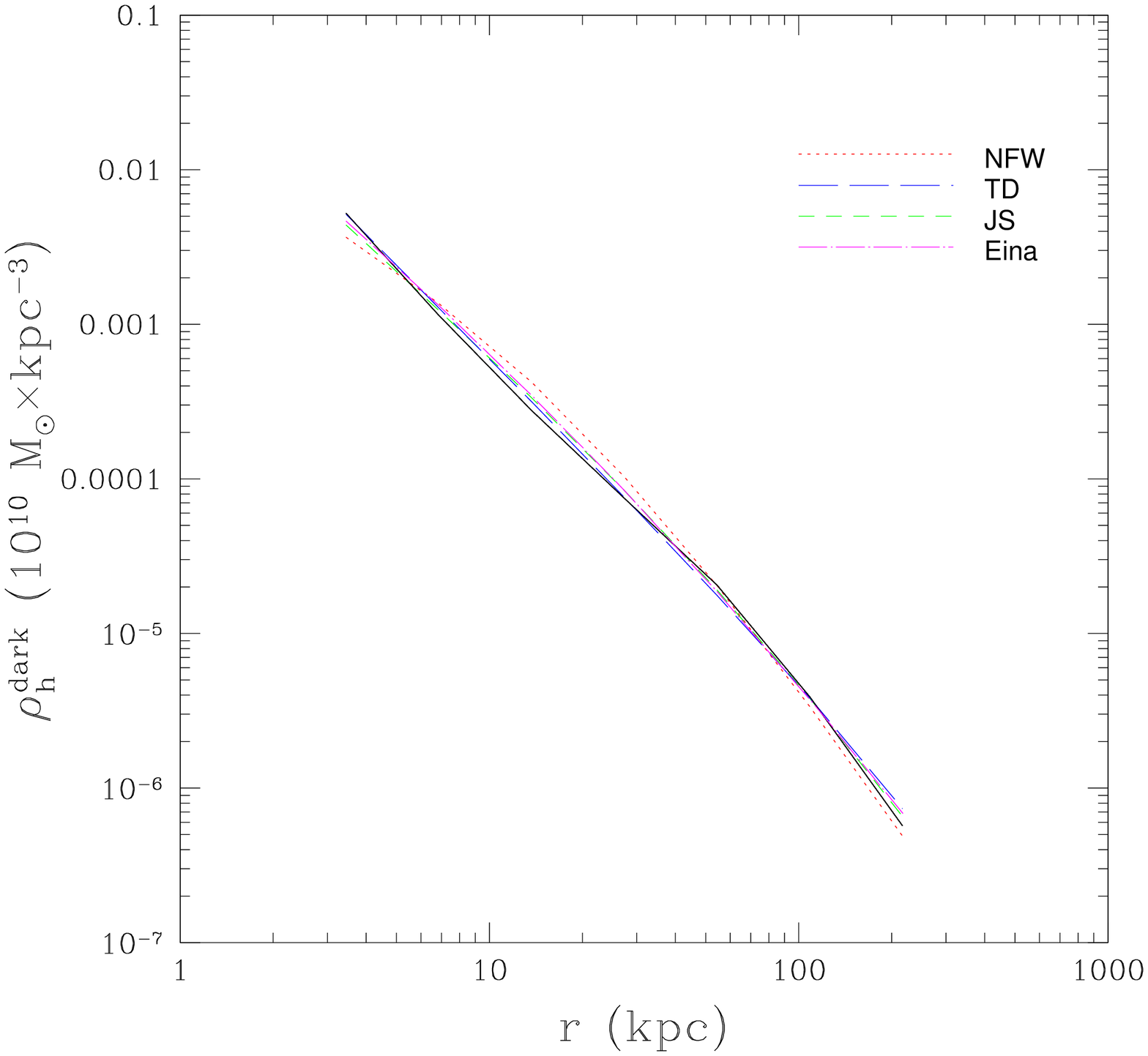}
     \includegraphics[width=.45\textwidth]{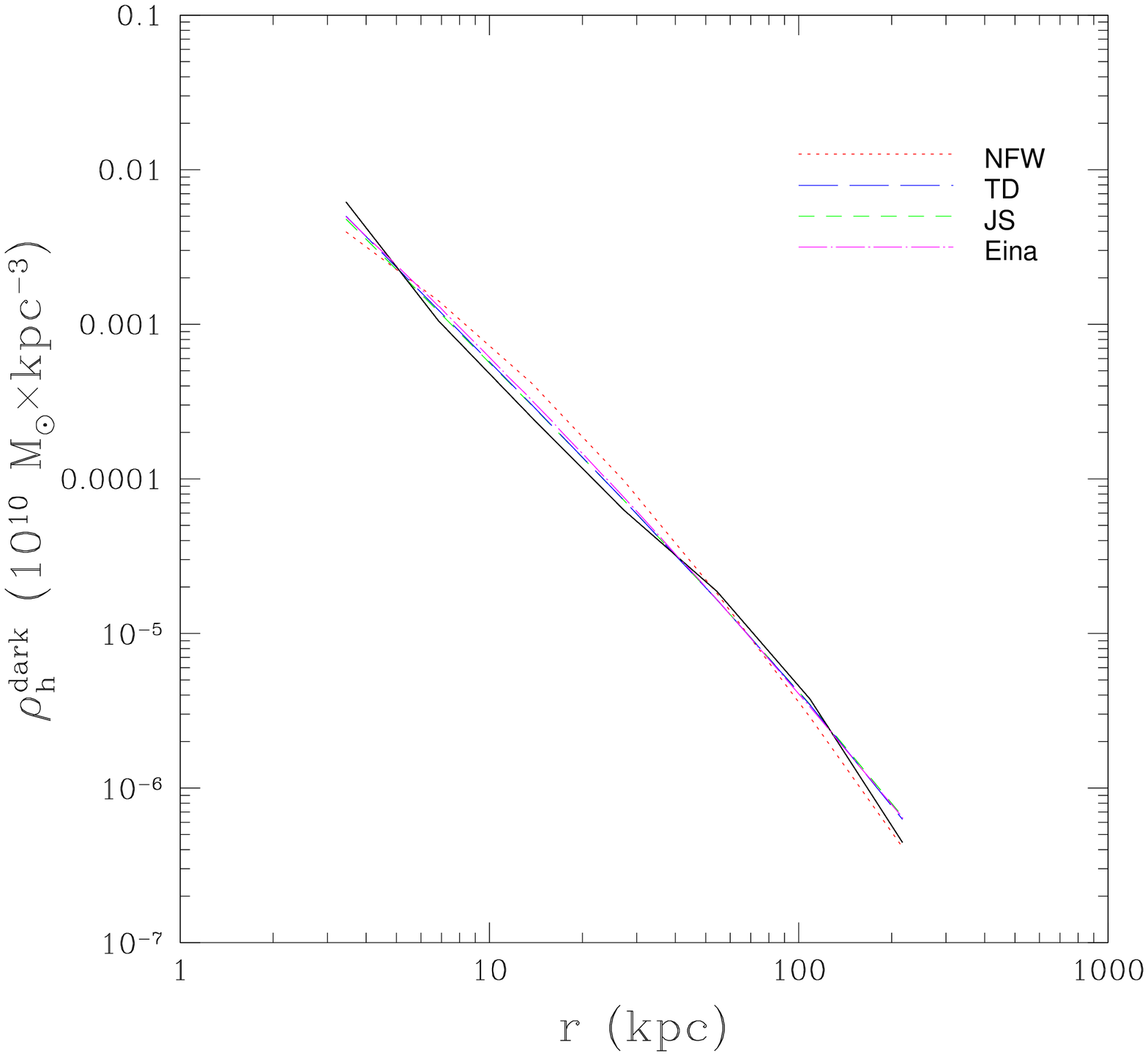}
\caption{Dark matter density  profiles (black full line) 
      for several typical ELOs
      from SF-A and SF-B samples along with their best fits to different
     analytical       profiles:  NFW (red point line), 
     TD (blue long-dashed line),
      JS (green short-dashed line) and Eina (magenta point-dashed line).
	     }   
\label{profDM1}
\end{center}
\end{figure*}

 When processes other than gravitational are involved in mass
 assembly (for example, cooling or heating), the
 dark matter density profiles
 could be modified (see Blumenthal et al. 1986;
 Dalcanton et al. 1997; Tissera \& Dom\'{\i}nguez-Tenreiro 1998;
 Gnedin et al. 2004).
 To analyse this point, in Figure~\ref{profDM1} we plot
 the dark matter density  profiles for several typical ELOs,
 along with their best fit to different analytical profiles.
The optimal fit has been obtained by minimising the statistics:

\begin{equation}
\chi^2 = \Sigma_{i = 1}^{N} [\log M(< r_i) - \log M^{\rm dark}_{\rm ELO}(< r_i)]^{2}/N
\label{chiDef}
\end{equation}

where $M^{\rm dark}_{\rm ELO}(< r_i)$ is the ELO dark matter mass within a 
sphere of radius $r_i$ centred at its centre of mass, $M(< r_i)$ is
the integrated mass density profile corresponding to the
different formulae above,
and the virial radii $r_{\rm vir}$ have been  taken as outer boundaries of the
fitting range. The {\it integrated} dark matter density profiles have been used
as fitting formulae instead of the dark matter density profiles themselves 
because these latter are binning dependent.
An updated version of the MINUIT software from the CERN library has been used
to make these fits as well as the others in this paper.

\begin{figure}
 \begin{center}
     \includegraphics[width=0.45\textwidth]{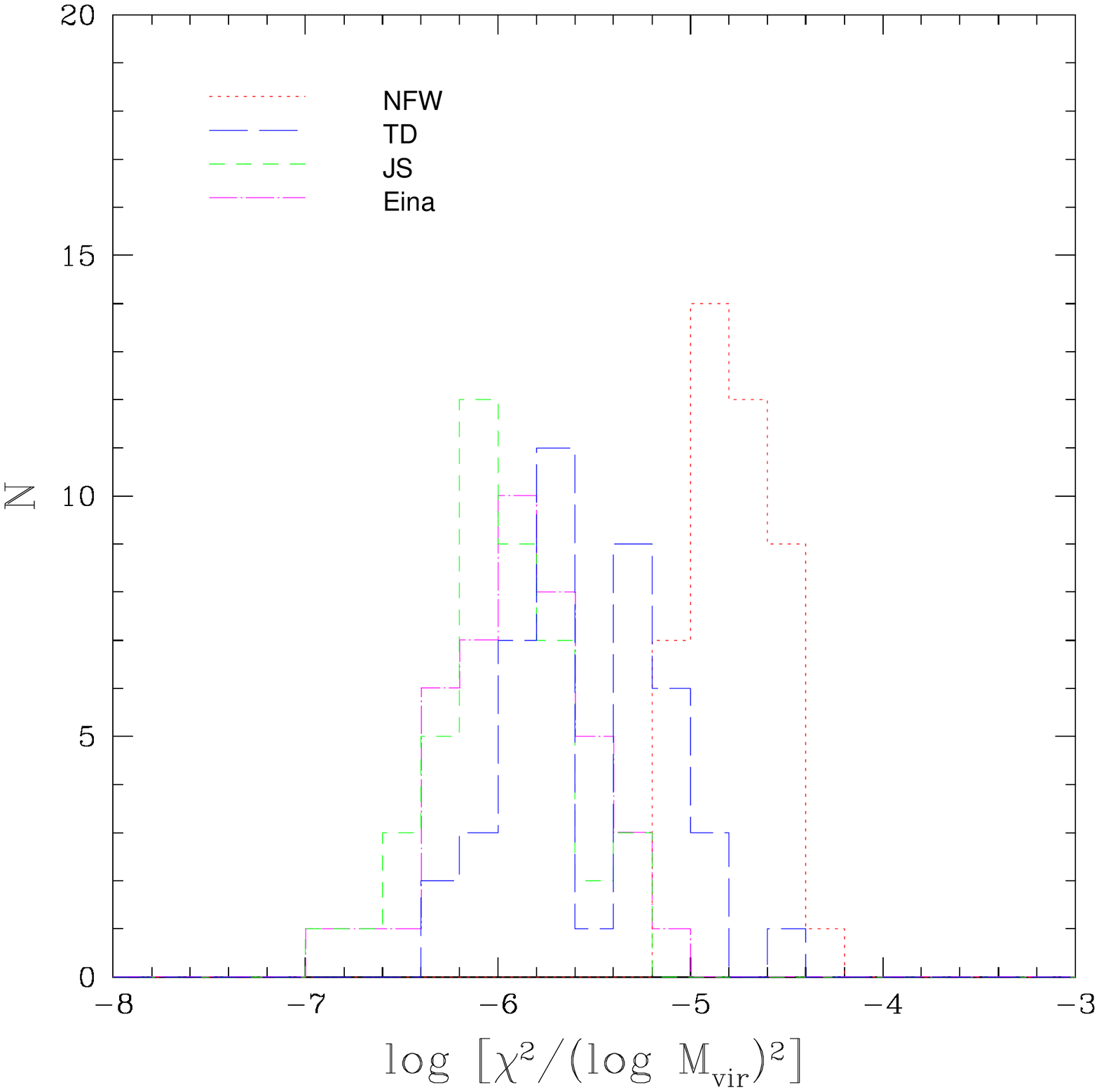}
           \caption{ The distributions of the $\chi^2$ per d.o.f., normalised to the logarithm of their respective mass square,  for the fits of the DM density profiles
	of ELO  haloes 
	(SF-A and SF-B samples) to different analytical profiles
		        }
    \label{Chi2DM}
   \end{center}
\end{figure}

\begin{figure}
   \begin{center}
     \includegraphics[width=.45\textwidth]{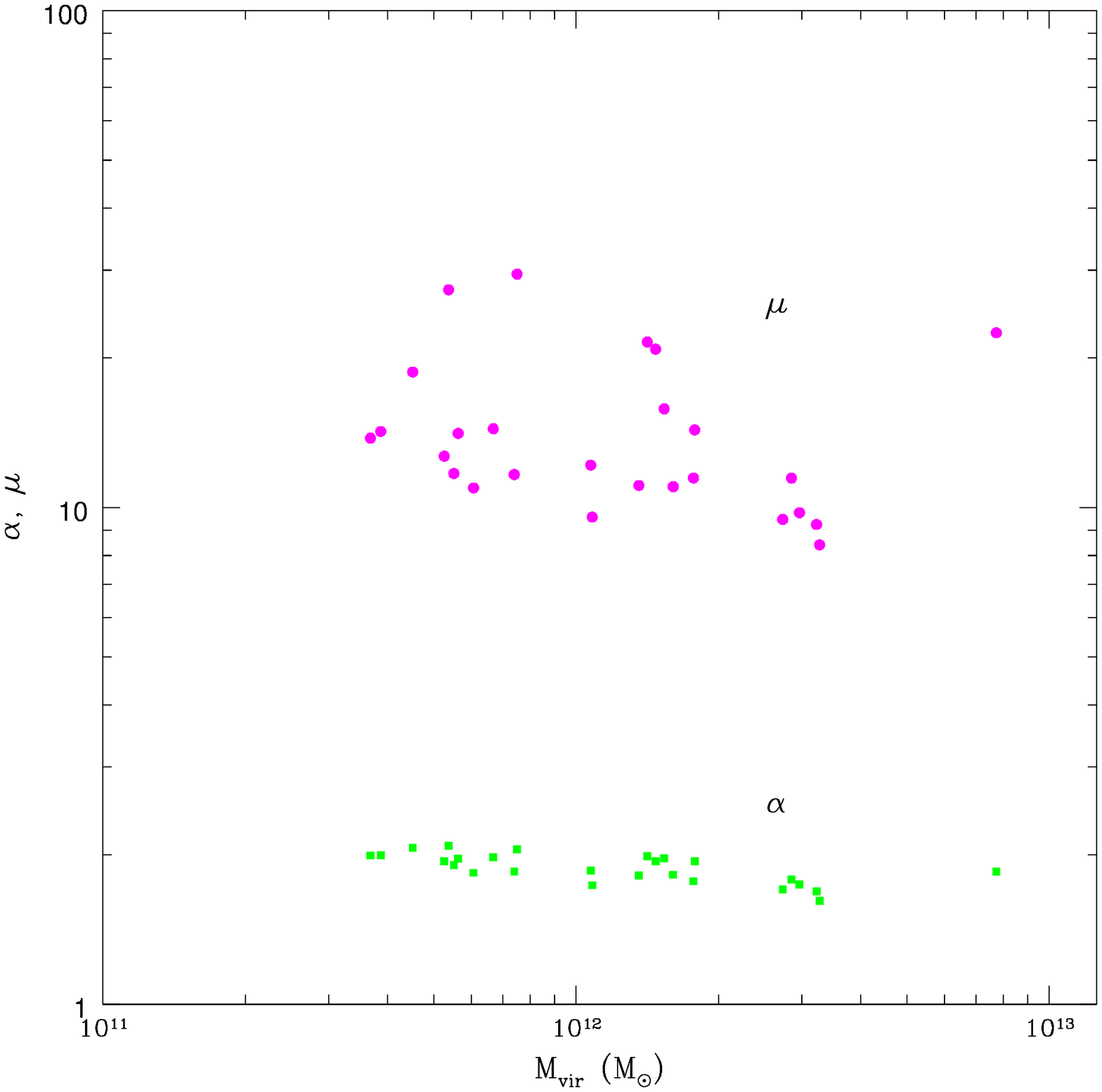}
     \includegraphics[width=.45\textwidth]{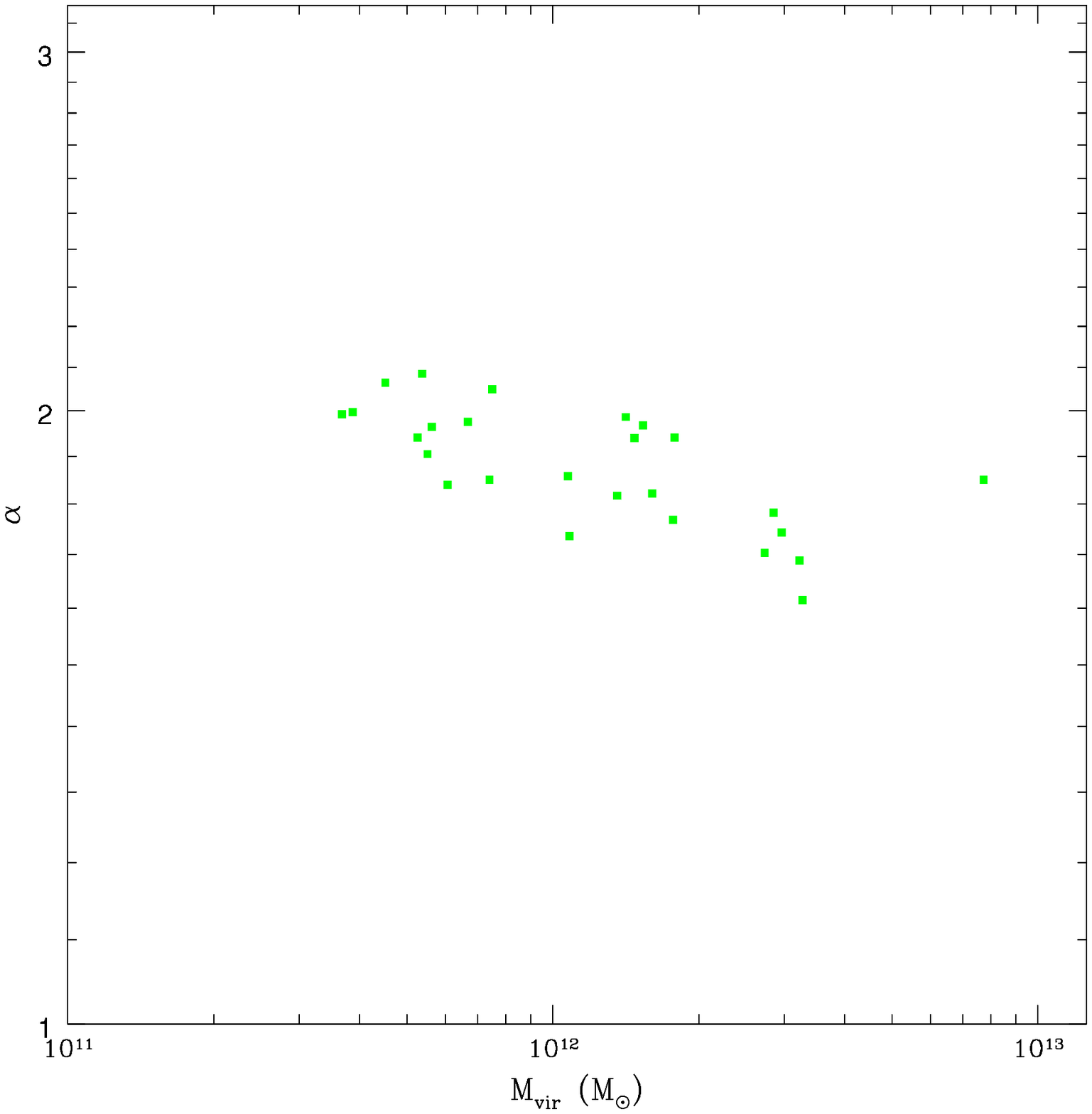}
\caption{Left panel: the optimal inner slope $\alpha$ of the general
Jing \& Suto profile for the DM haloes of ELOs (green filled squares)
and the $\mu$ coefficient of the Einasto analytical profile (magenta filled circles),
versus their virial mass for SF-A sample ELOs.
Right panel: zoom of the $\alpha$ versus virial mass plot to clarify
the mass effect.
}
\label{JSalfaymu}
\end{center}
\end{figure}

Note in Figure~\ref{profDM1} that the quality of the fits
differs from one analytical profile to another.
To quantify this effect, in Figure~\ref{Chi2DM} we plot the  distributions of
the $\chi^2$ per d.o.f.
statistics, normalised to  $(\log M_{\rm vir})^2$, resulting from the fits to
the different profiles above, except for Hern90 one whose results are 
generally poorer.
We see that the lower $\chi^2$ per d.o.f. values generally correspond 
to either the Eina or the JS
 profiles, with the TD profiles in the third position.
 In Figure~\ref{JSalfaymu} we draw the  values of
the $\mu$ (for Eina profiles) and $\alpha$ (for JS profiles) slopes
corresponding to the optimal fits of SF-A sample DM haloes. A
slight mass effect can be appreciated with lower mass ELOs having
steeper DM haloes than more massive ones, presumably due to a more important
pulling in of baryons onto dark matter as they fall to the ELO 
centre with decreasing ELO mass.    
That is, massive haloes are less concentrated than lighter ones,
i.e., the mass-concentration relation.
In any case, the profiles are
always steeper than $\alpha = 1$ (i.e., the NFW profile; see Mamon \& Lockas
2005a).

\begin{figure}
 \begin{center}
   \includegraphics[width=0.45\textwidth]{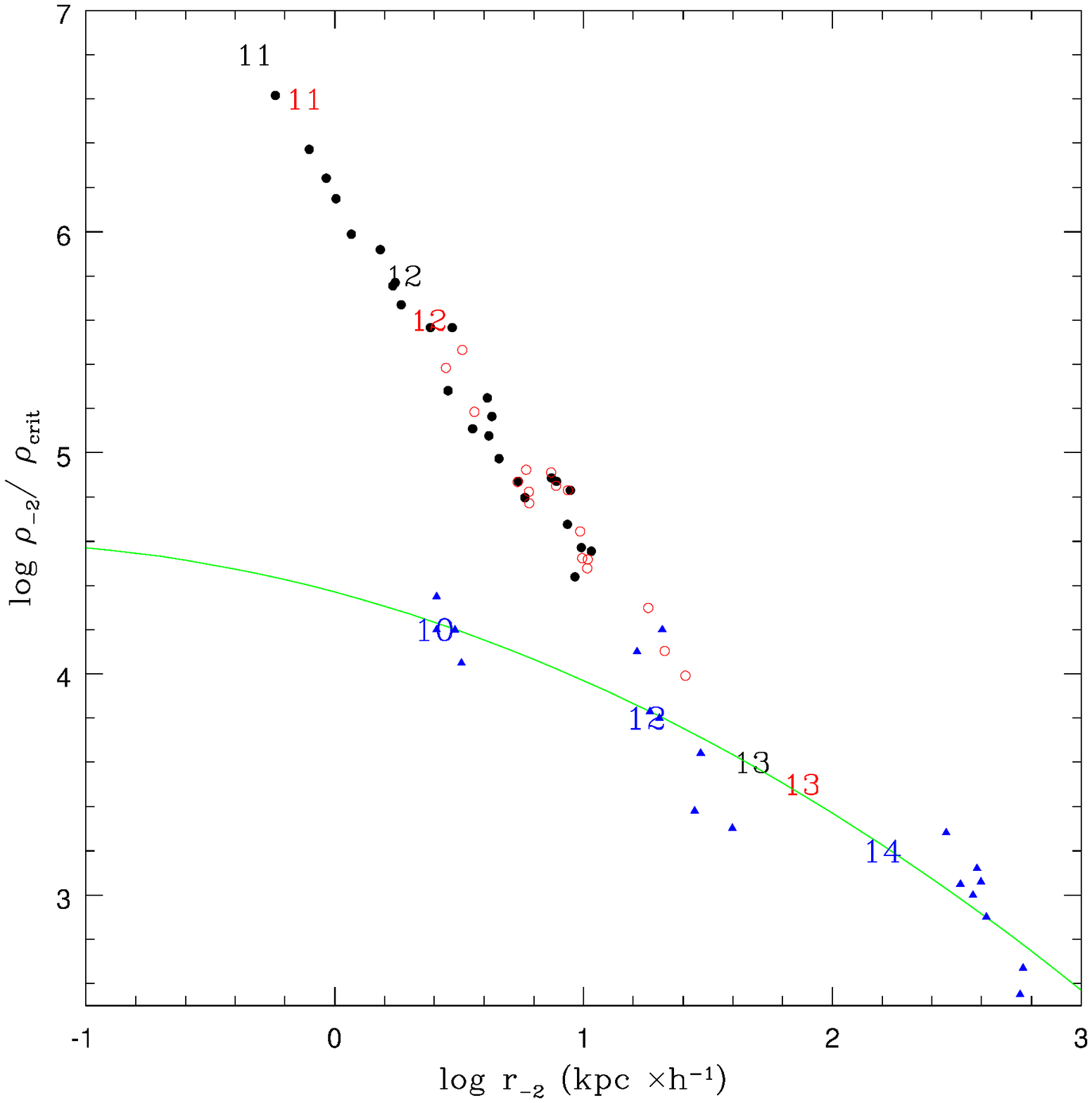}
\caption{  The
$\rho_{-2}$ density parameter versus the $r_{-2}$ scale obtained from
fits to the Einasto model, for ELOs in both the SF-A sample
(filled black circles)  and the SF-B sample (open red circles).
Blue triangles are measurements by Navarro et al.
(2004), onto haloes formed in N-body simulations, with its fit by 
Mamon \& Lockas 2005a (green line). 
Numbers correspond to the logarithms of the virial masses
(in units of  M$_{\odot}$) of haloes
formed in different simulations, according with their respective colours
}
\label{CompNav04}
\end{center}
\end{figure}

To further analyse this effect, we plot in Figure~\ref{CompNav04} the 
$\rho_{-2}$ density parameter versus the $r_{-2}$ scale obtained from
fits to the Einasto model. Blue triangles 
 are measurements by Navarro et al. 
(2004) onto haloes formed in N-body simulations and the green line
is their  best fit. We see that at given $r_{-2}$, 
$\rho_{-2}$ is higher in our hydrodynamical simulations than in those
of Navarro et al. (2004),
presumably due to the pulling in of dark matter by baryon infall.
We also see that at given $M_{\rm vir}$, $r_{-2}$ is shorter in
hydrodynamical simulations than in purely gravitatory ones,
by the same reason.

\subsection{Projected stellar mass density profiles} 
\label{StarProjProf}

Authors now agree that the S\'ersic law given in Eq.~\ref{sersic}
(S\'ersic 1968) is an adequate empirical representation of the
 optical surface brightness profiles
 of most ellipticals (see, for example, Caon et al. 1993;
 Bertin et al. 2002).
 Assuming that the stellar mass-to-light ratio $\gamma^{\rm star}$
 does not appreciably change with ELO projected radius $R$
 \footnote[4]{Hereafter we will use capital $R$ to mean projected radii},
 the projected stellar {\it mass} profile,
 $\Sigma^{\rm star}(R)$ can  be taken as a measure of the
 surface {\it brightness} profile and  
 be written as

\begin{equation}
 \Sigma^{\rm star}(R) = \gamma^{\rm star} I^{\rm light}(R).
\label{SigmaR}
\end{equation}

\begin{figure}
\begin{center}
\includegraphics[width=.45\textwidth]{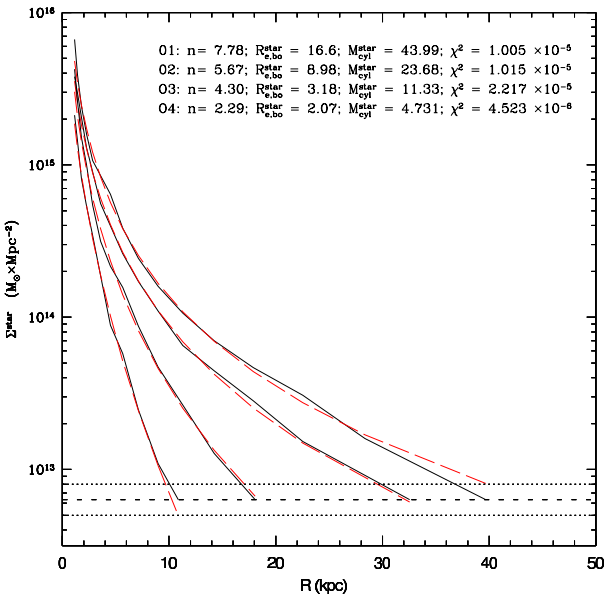}
\caption{Projected stellar mass density profiles for different ELOs
along with their best fit by a S\'ersic law.
The  corresponding shape parameter best values
and minimal $\chi^2$ per-degree-of freedom are also shown 
}
\label{StarProfSersic}
\end{center}
\end{figure}

 One can then  expect that  $\Sigma^{\rm star}(R)$
 can be fitted by a S\'ersic-like law.
 This is in fact the case as shown
in Figure~\ref{StarProfSersic} for several typical ELOs 
drawn from both SF-A and SF-B samples (see Kawata \& Gibson, 2005, for a
similar result concerning one virtual elliptical galaxy).
Some remarks on how our fits have been made are in order.
First, the $\Sigma^{\rm star}(R)$ profiles have been calculated
by averaging on concentric rings centred at the projection
of the centre-of-mass of the corresponding ELO.
Three  projections along orthogonal
directions have been considered for each ELO.
Also, because these projected densities are
binning dependent and somewhat noisy, the integrated projected
mass density in concentric cylinders of radius $R$ and mass

\begin{equation}
M_{\rm cyl}^{\rm star}(R) = 2 \pi \int_0^{R} \Sigma^{\rm star}(R') R' dR'
\end{equation}

has been used as a fitting function, instead of $\Sigma^{\rm star}(R)$
itself. Concerning the fitting range, we have adopted an
outer boundary
$R_{\rm max}$ such  that the corresponding surface brightness
$I^{\rm light}(R_{\rm max})$ (see Eq.~\ref{SigmaR})  gives the 
standard value of $\mu_B = 27$ mag arcsec$^{-2}$. The values for the
stellar mass-to-blue-light $\gamma^{\rm star}_B$ span a range
from $\gamma^{\rm star}_B  = 2$ to 12, depending on the details of its
determination (see discussion in Mamon \& Lockas 2005a), and best values of
$\gamma^{\rm star}_B = 5$ to 8. Their geometric mean
$\gamma^{\rm star}_B = 6.3$ has been used to make the fits
drawn in Figure~\ref{StarProfSersic}, but the results of the fit
do not significantly depend on the particular
$\gamma^{\rm star}_B$ value used within its range of best values.

\begin{figure}
\begin{center}
\includegraphics[width=.45\textwidth]{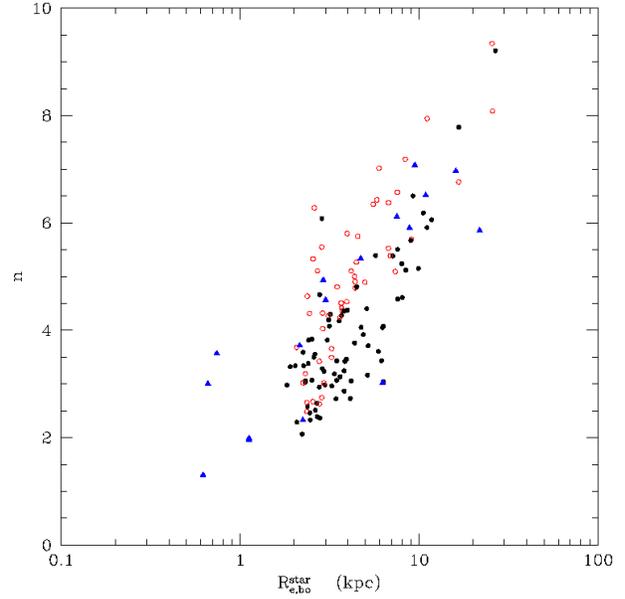}
\caption{The S\'ersic shape parameter $n$ versus the projected stellar half-mass
radii for
SF-A sample  (black filled circles) and SF-B sample (red open circles) ELOs.
For each ELO, the results of projections along three orthogonal directions
are shown.
Blue filled triangles are data on $n$ and $R_e^{\rm light}$ from D'Onofrio (2001)
}
\label{SersicNvsM}
\end{center}
\end{figure}

An interesting result is that
the values of the shape parameter $n$ we have obtained are consistent
with observations, including their correlations with
the effective radii $R_e^{\rm light}$, luminosity $L$
and velocity dispersion
(Caon et al. 1993; Prugniel \& Simien 1997;
Graham 1998;
M\'arquez et al. 2000; D'Onofrio 2001; 
Trujillo et al. 2001; Vazdekis, Trujillo \& Yamada 2004;
Graham et al. 2006),
as illustrated in Figure~\ref{SersicNvsM}.
In this Figure we plot the shape parameter $n$ versus the 
 ELO projected stellar half-mass radii, $R_{\rm e, bo}^{\rm star}$,
 defined by the condition that
 $M_{\rm cyl}^{\rm star}(R_{\rm e, bo}^{\rm star})$
 encloses half the total stellar mass of the system;
 assuming that $\gamma^{\rm star}_B$ does not depend on $R$,
 we will have $R_{\rm e, bo}^{\rm star} \simeq R_e^{\rm light}$. 
 Blue triangles are data on $n$ and $R_e^{\rm light}$
 from D'Onofrio (2001).
Note that  a slight
effect resulting from the different SF parametrization in
SF-A and SF-B sample ELOs is apparent in this plot,
mainly due to the smaller sizes of SF-B sample ELOs as compared with 
their SF-A sample counterparts.

\subsection{Baryonic three-dimensional mass density profiles}
\label{BarProf}

In the last section it has been shown that the projected
stellar mass density profiles are adequately described by 
the standard S\'ersic profiles, that is, that they are
consistent with observational data.
We now analyse the three-dimensional mass density profiles
of baryons. 

\begin{figure}[H]
 \begin{center}
     \includegraphics[width=.45\textwidth]{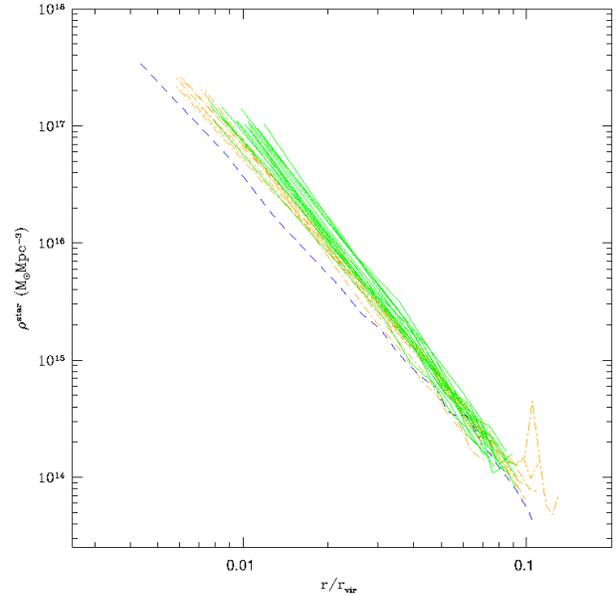}
           \caption{Three-dimensional stellar mass profiles for
ELOs in the SF-A sample: full green lines, ELOs with 
$M_{\rm vir} < 1.5 \times 10^{12}$ M$_{\odot}$; orange point-dashed lines, ELOs with 
$1.5 \times 10^{12}$ M$_{\odot} \leq M_{\rm vir} < 5 \times 10^{12}$ M$_{\odot}$; blue dashed lines:
ELOs with $M_{\rm vir} \geq 5 \times 10^{12}$ M$_{\odot}$.
The stellar mass density
profiles show homology breaking
}
\label{PerSt3D}
\end{center}
\end{figure}

We first analyse the baryon distribution at the ELO scale, where the
main contribution to the mass density comes from stars.
We lack of any observational input on how the three-dimensional stellar-mass
density profiles $\rho^{\rm star}(r)$
can be, except for a deprojection of the S\'ersic profiles
(Prugniel \& Simien 1997; Lima Neto, Gerbal \& M\'arquez 1999).
In Figure~\ref{PerSt3D} we plot 
$\rho^{\rm star}(r)$ for ELOs in the SF-A sample.
Different colours have been used for ELOs in different mass intervals
and a clear mass effect can be appreciated in this Figure, and particularly so
at the inner regions, where at fixed $r/r_{\rm vir}$ the stellar-mass density
of less massive ELOs can be a factor of two or so higher than that of more
massive ones. This means that the mass homology is broken in the 
three-dimensional stellar mass distribution.

\begin{figure*}
 \begin{center}
     \includegraphics[width=0.45\textwidth]{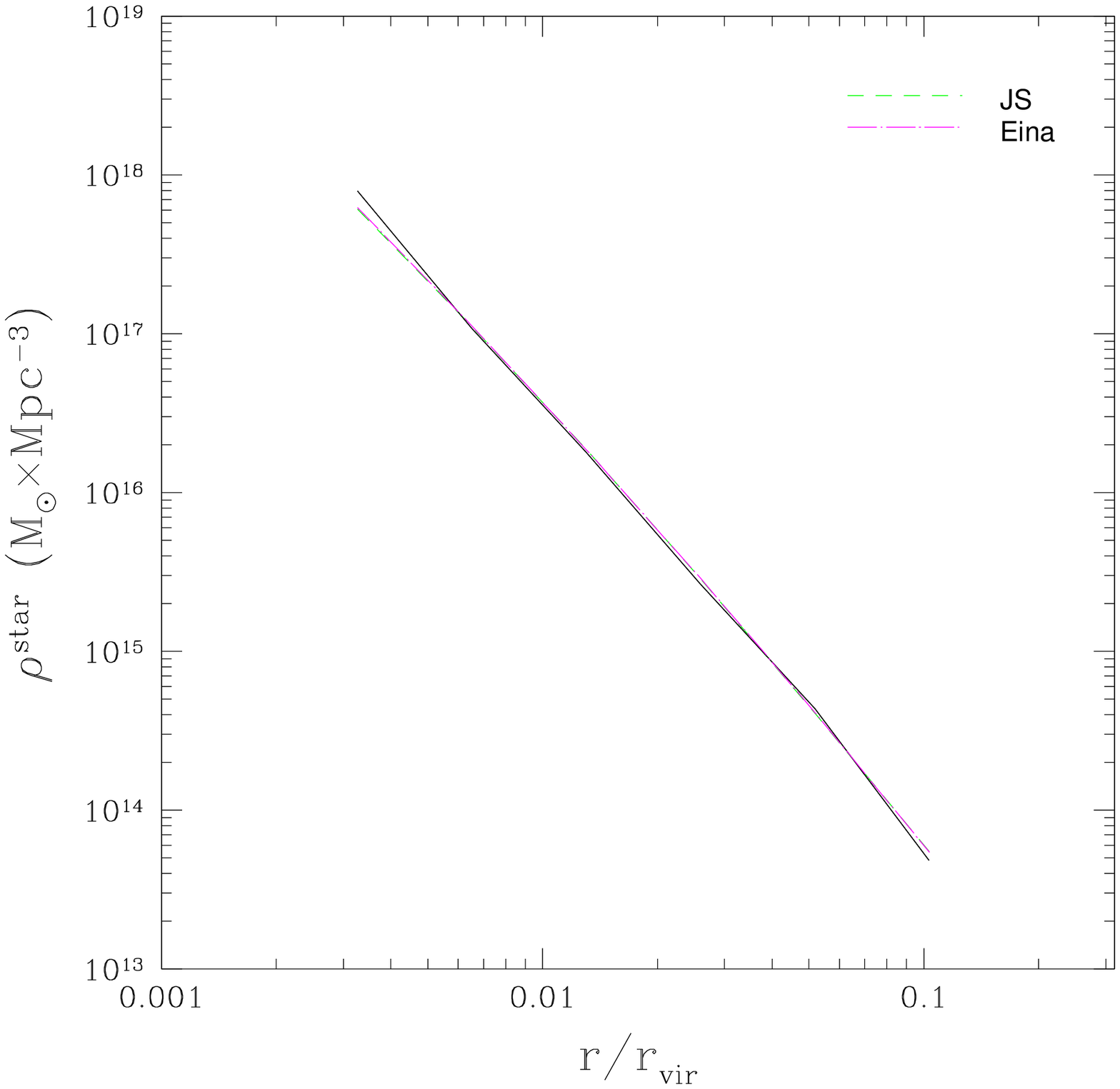}
     \includegraphics[width=0.45\textwidth]{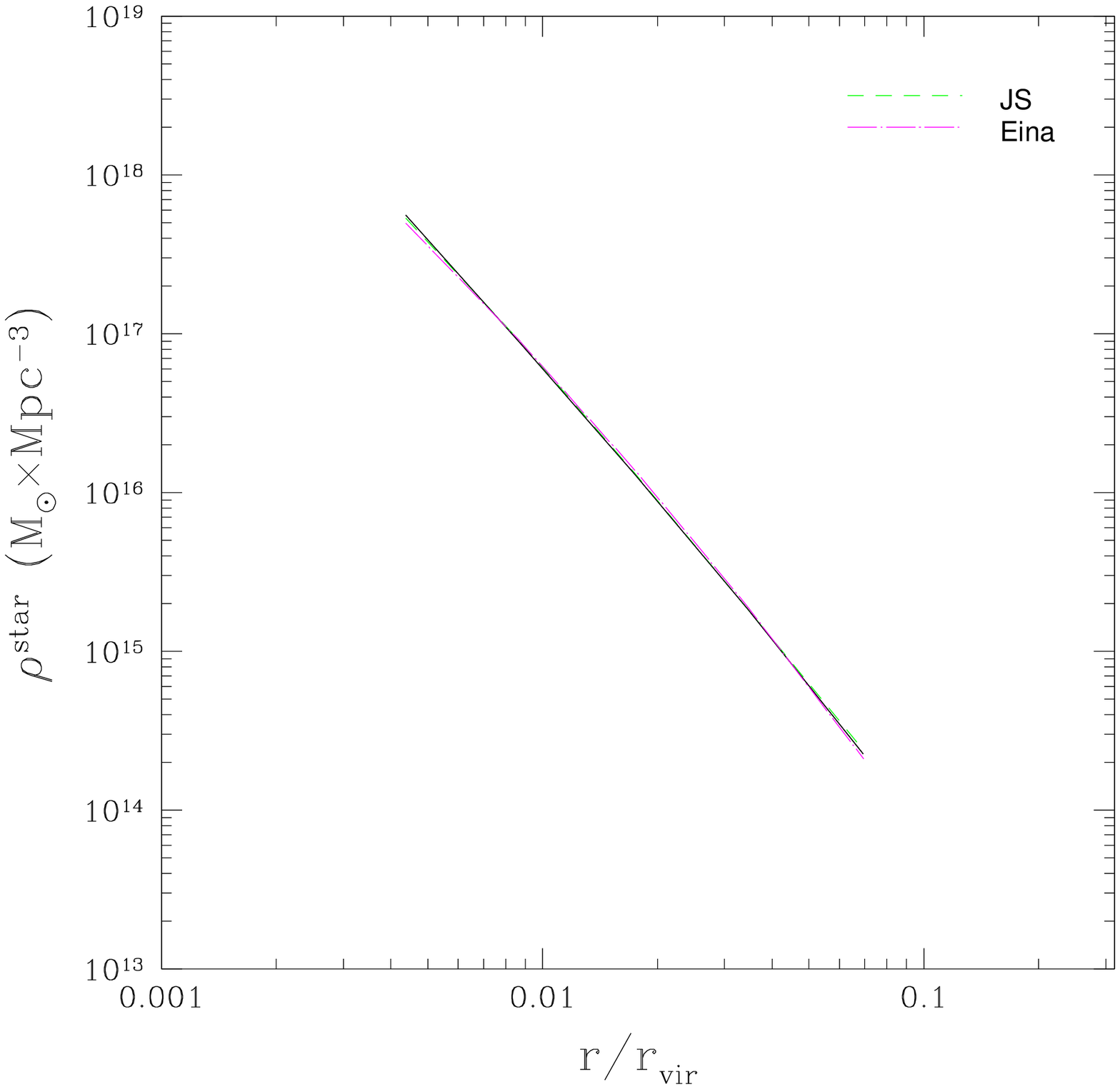}
     \includegraphics[width=0.45\textwidth]{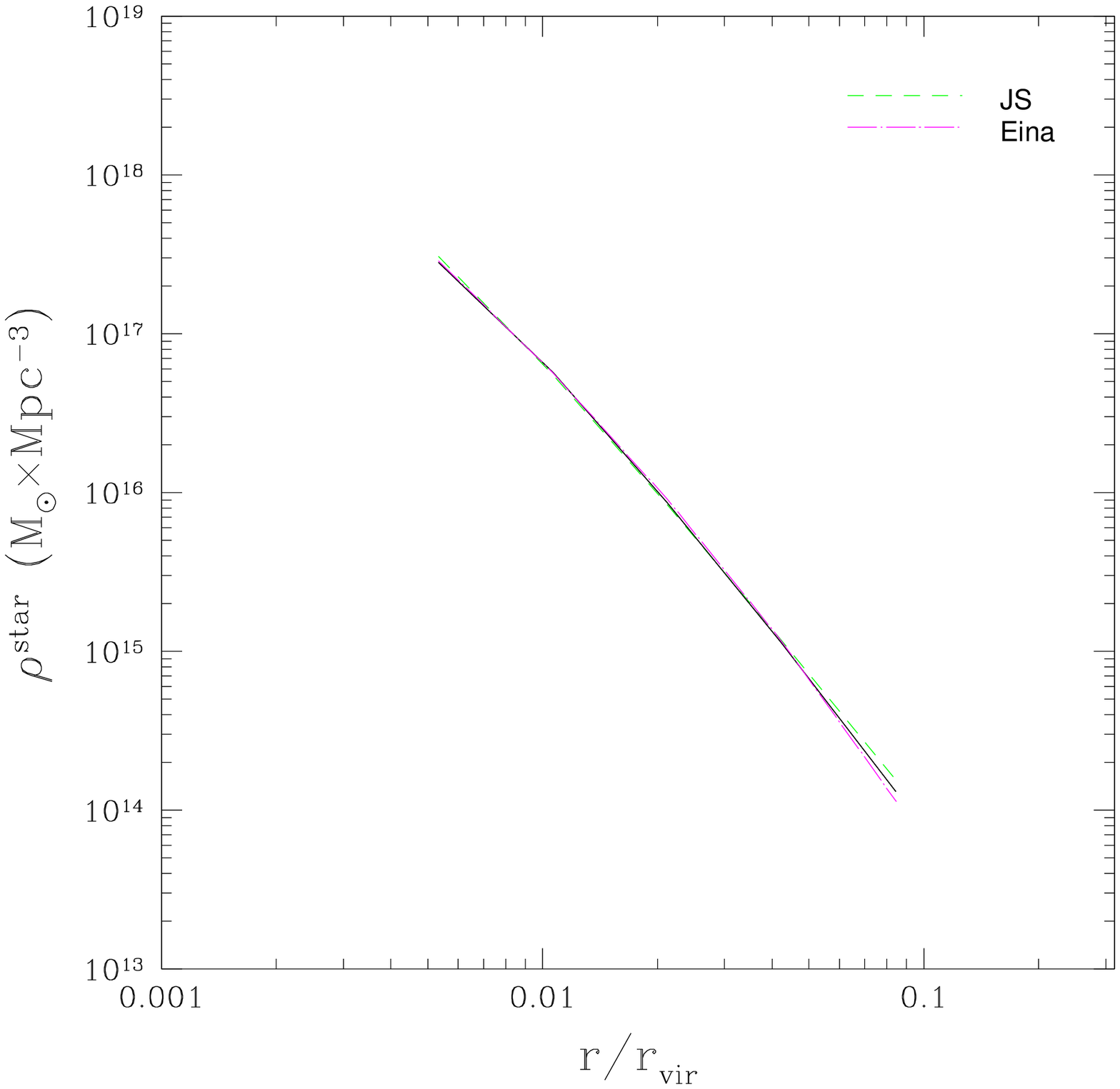}
     \includegraphics[width=0.45\textwidth]{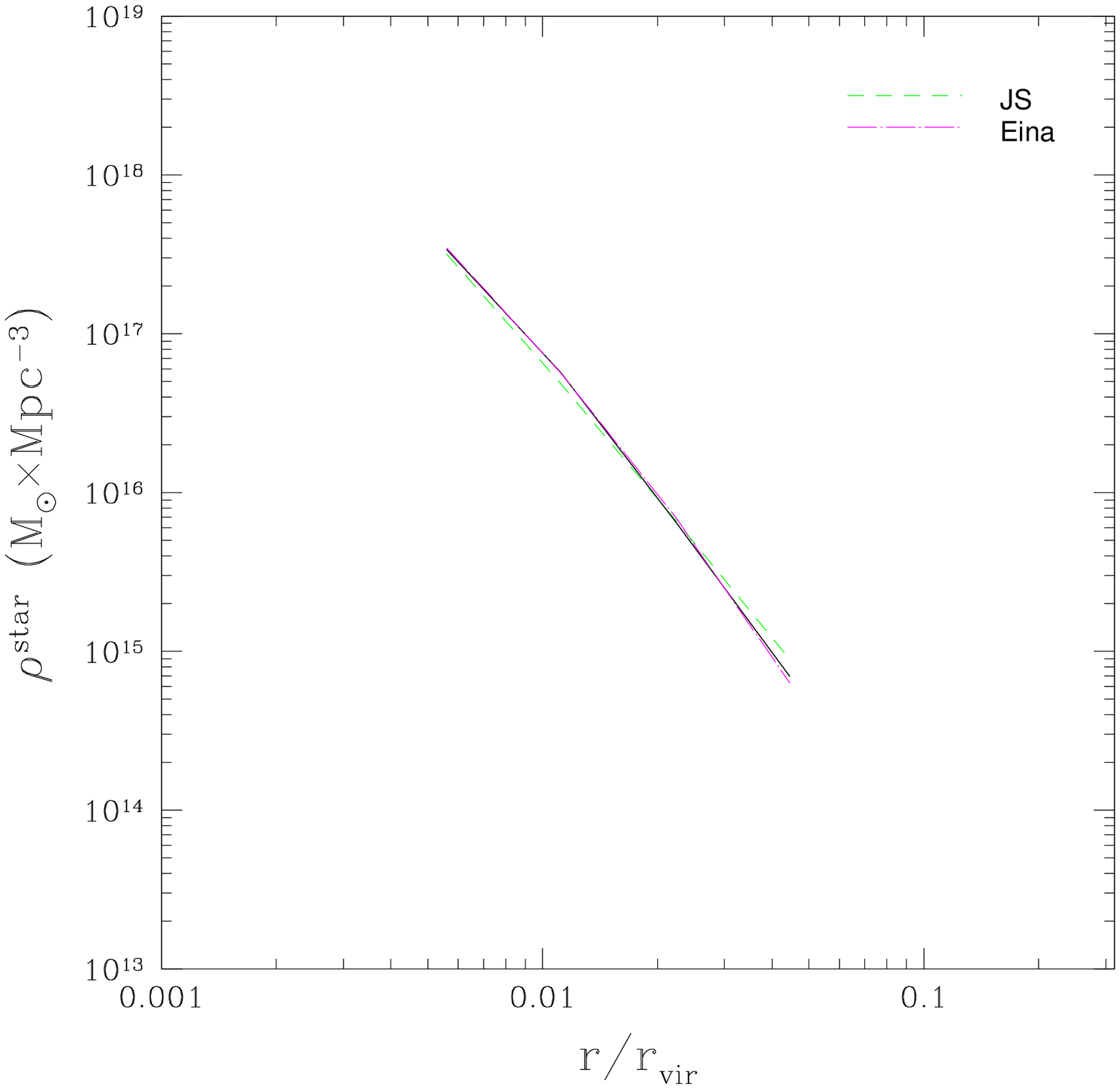}
           \caption{The stellar mass profiles for 4 typical ELOs in
	   SF-A sample (black continuous lines)
	   and their optimal fits to Einasto profiles 
	   (magenta point-dashed lines) and JS profiles (green dashed lines)
}
\label{StarProFit}
\end{center}
\end{figure*}

\begin{figure}
 \begin{center}
      \includegraphics[width=0.45\textwidth]{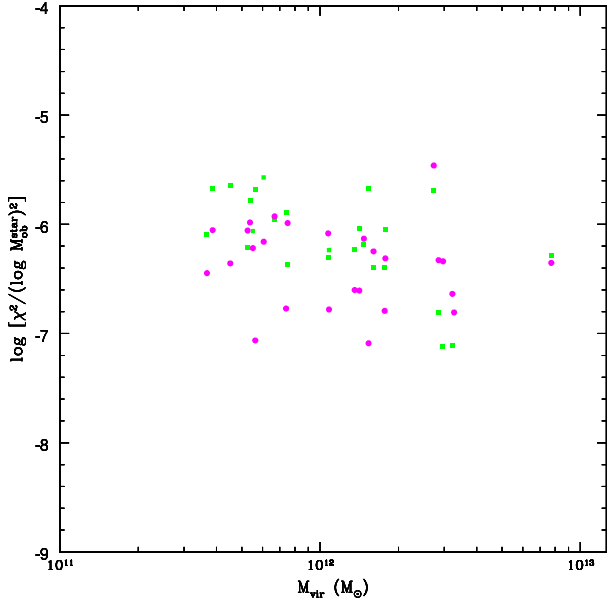}
      \caption{The $\chi^2$ statistics (Eq. \ref{chiDef}) corresponding to the
      fits of the stellar mass profiles for ELOs in the SF-A sample Einasto 
      profiles (magenta filled circles) and JS profiles (green filled squares)
}
\label{StarProChi2}
\end{center}
\end{figure}

To quantify the stellar three-dimensional mass density profiles of ELOs,
they have been fit to JS and Einasto analytical formulae through the statistics
defined in Eq. ~\ref{chiDef} where 
$M^{\rm dark}_{\rm ELO}(< r_i)$ has been replaced by the 
ELO stellar mass within a sphere of radius $r_i$.
The quality of the fits is illustrated in Figure~\ref{StarProFit},
and in Figure~\ref{StarProChi2} the  values of the
$\chi^2$ p.d.o.f. statistics are given, normalised to 
$\log M^{\rm star}_{\rm bo}$.
Both Figures show that these profiles describe adequately well
the spherically-averaged stellar mass distribution
in three dimensions, even if with very small $r_{-2}$ values.

\begin{figure}
 \begin{center}
     \includegraphics[width=0.45\textwidth]{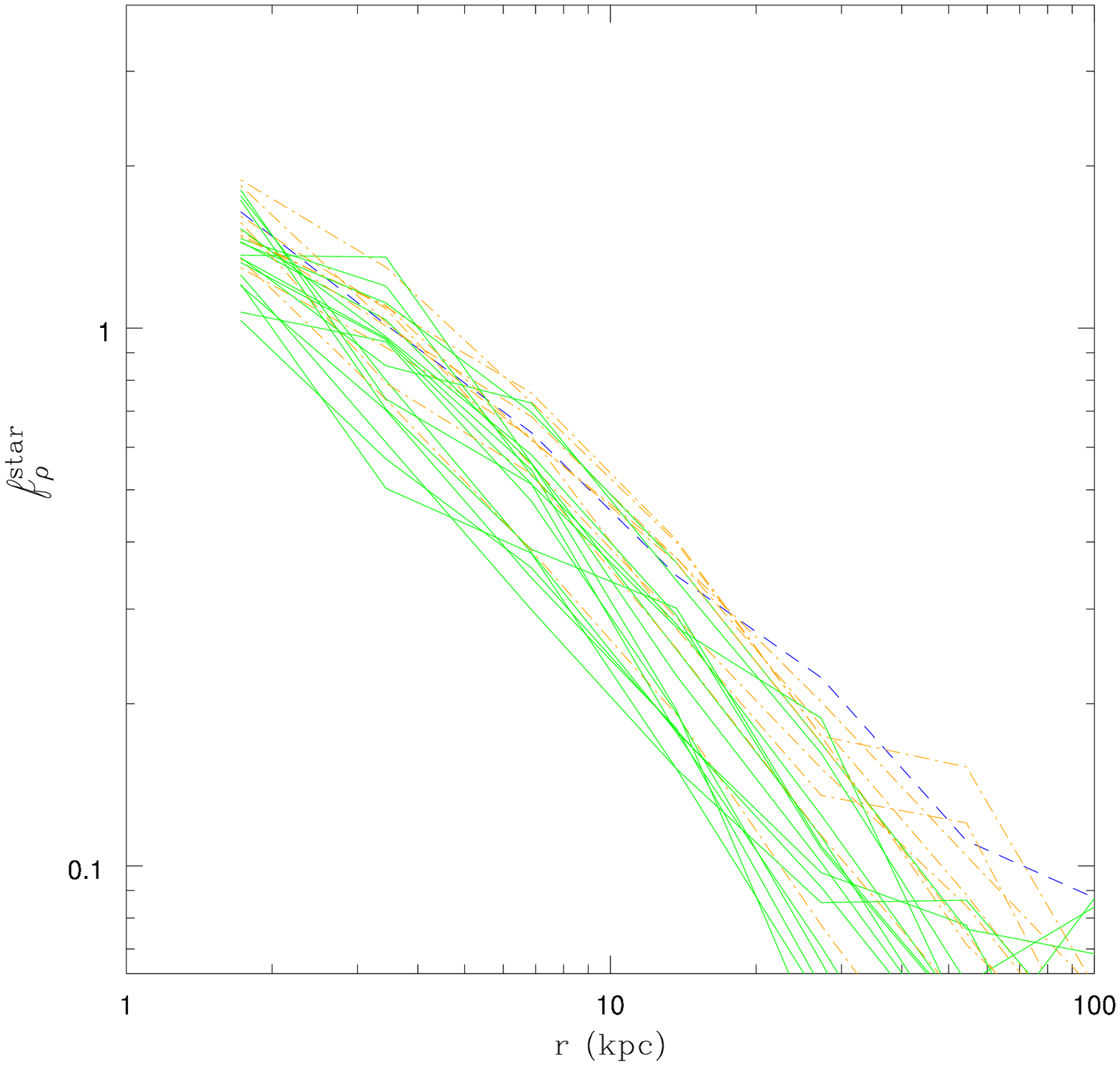}
     \includegraphics[width=0.45\textwidth]{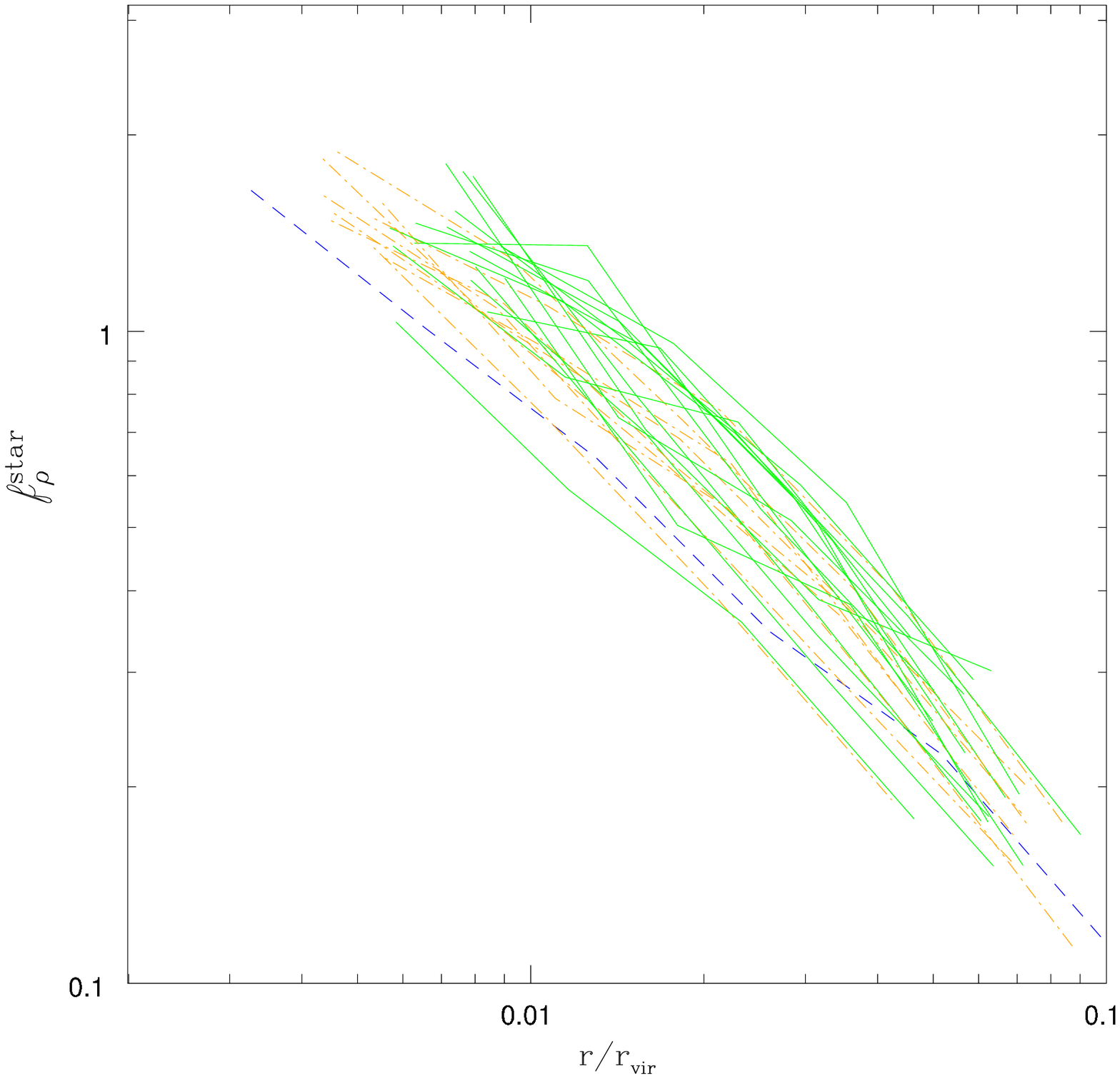}
           \caption{The stellar-to-dark mass profiles for 
	         ELOs in the SF-A sample. In the lower panel, radii are 
		 normalised to the virial radii. Line types and colours
		 are the same as in Figure 9
}
  \label{PerStFrac}
 \end{center}
\end{figure}

To  study
the possibility that the homology in the dark- versus bright-mass
distribution is also broken,  the stellar-to-dark  density ratio  profiles

\begin{equation}
f^{\rm star}_{\rho}(r) = \rho^{\rm star}(r)/ \rho^{\rm dark}(r)
\end{equation}

are plot versus either the radii 
(Figure~\ref{PerStFrac} upper panel) or the radii normalised to 
virial radii (Figure~\ref{PerStFrac} lower panel).
We see that there is, in any case, a clear mass effects at the inner regions,
with the stellar mass
distribution relative to the dark mass one less concentrated with
increasing ELO mass.
For example, in Figure~\ref{PerStFrac} we see that
the fraction of the ELO  virial volume where
$f^{\rm star}_{\rho}(r) > 1$, is smaller as the ELO mass grows; also, at fixed
$r/r_{\rm vir}$,
$f^{\rm star}(r)$ increases with decreasing ELO mass.
So, the homology is broken in the three-dimensional
stellar-to-dark mass distribution, a fact that could be important to explain
the tilt of the observed FP (see O\~norbe et al. 2005, 2006).

\begin{figure}
 \begin{center}
      \includegraphics[width=.45\textwidth]{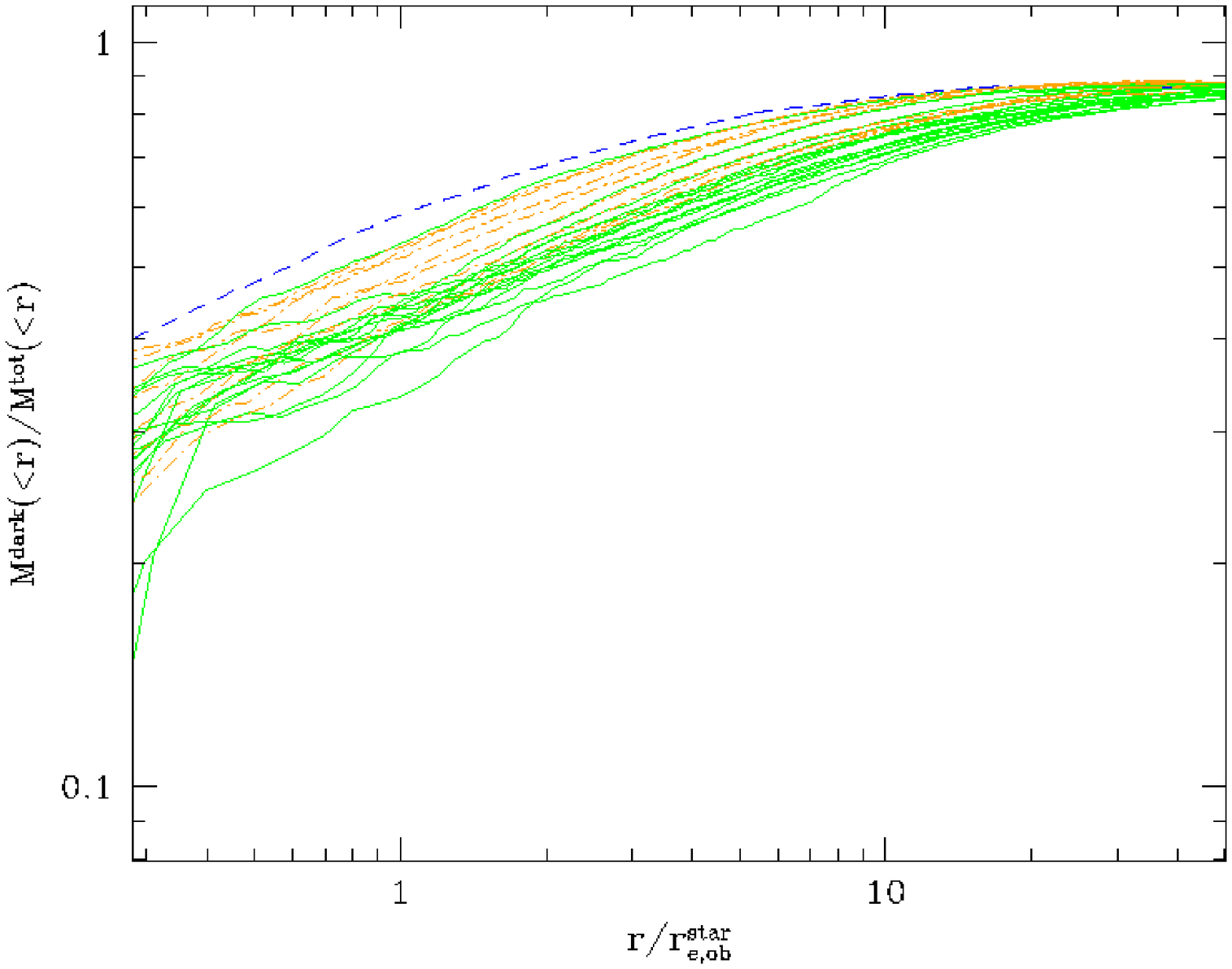}
\caption{The fraction of dark-to-total mass profiles,
           $M^{\rm dark}(< r)/M^{\rm tot}(< r)$ for
ELOs in the SF-A sample; different colours stand for different mass
ranges, as in Figure 9; radii are
normalised to the 3D stellar half-mass radii
}
\label{PerStar}
\end{center}
\end{figure}

\begin{figure}
 \begin{center}
\includegraphics[width=.45\textwidth]{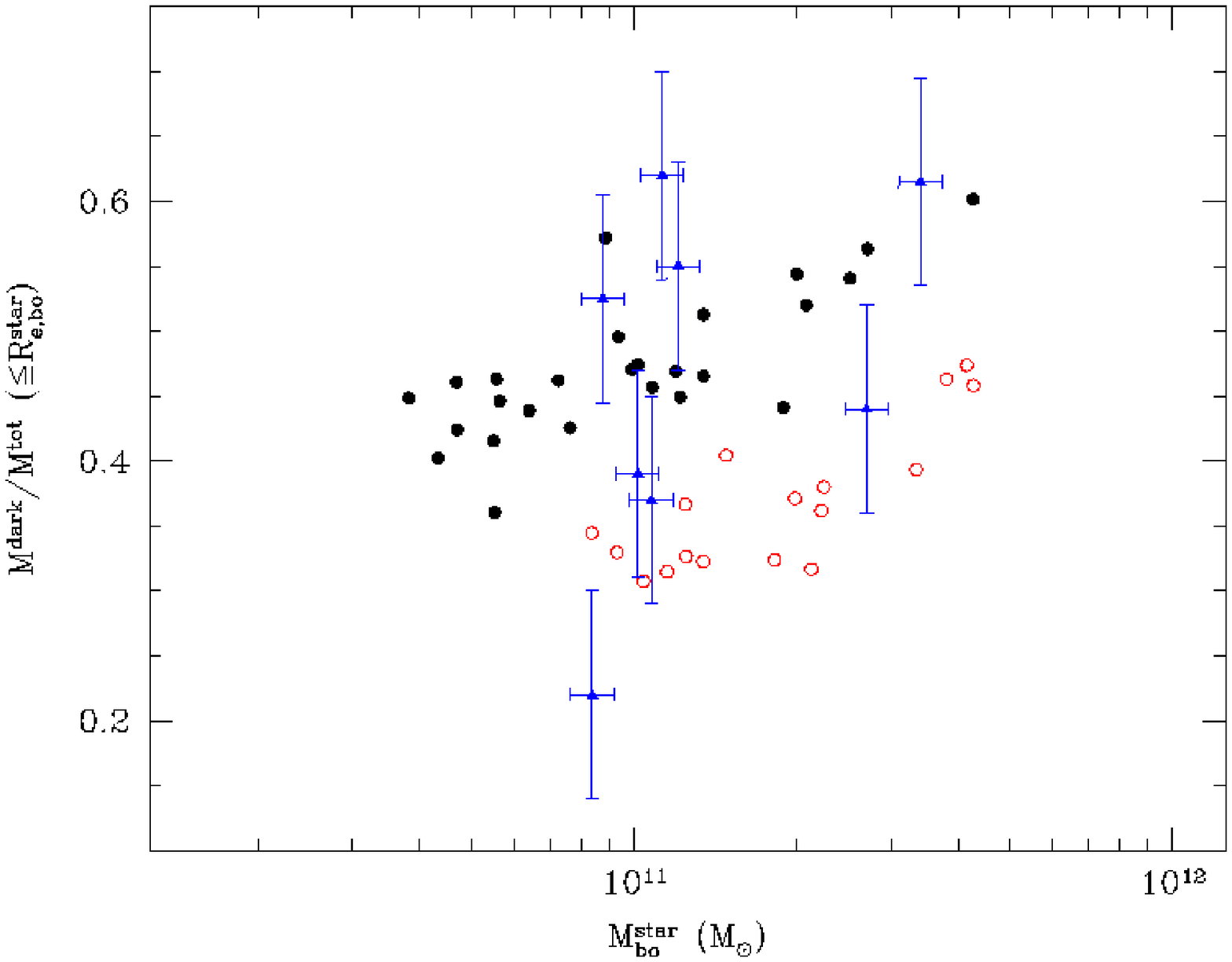}
\includegraphics[width=.45\textwidth]{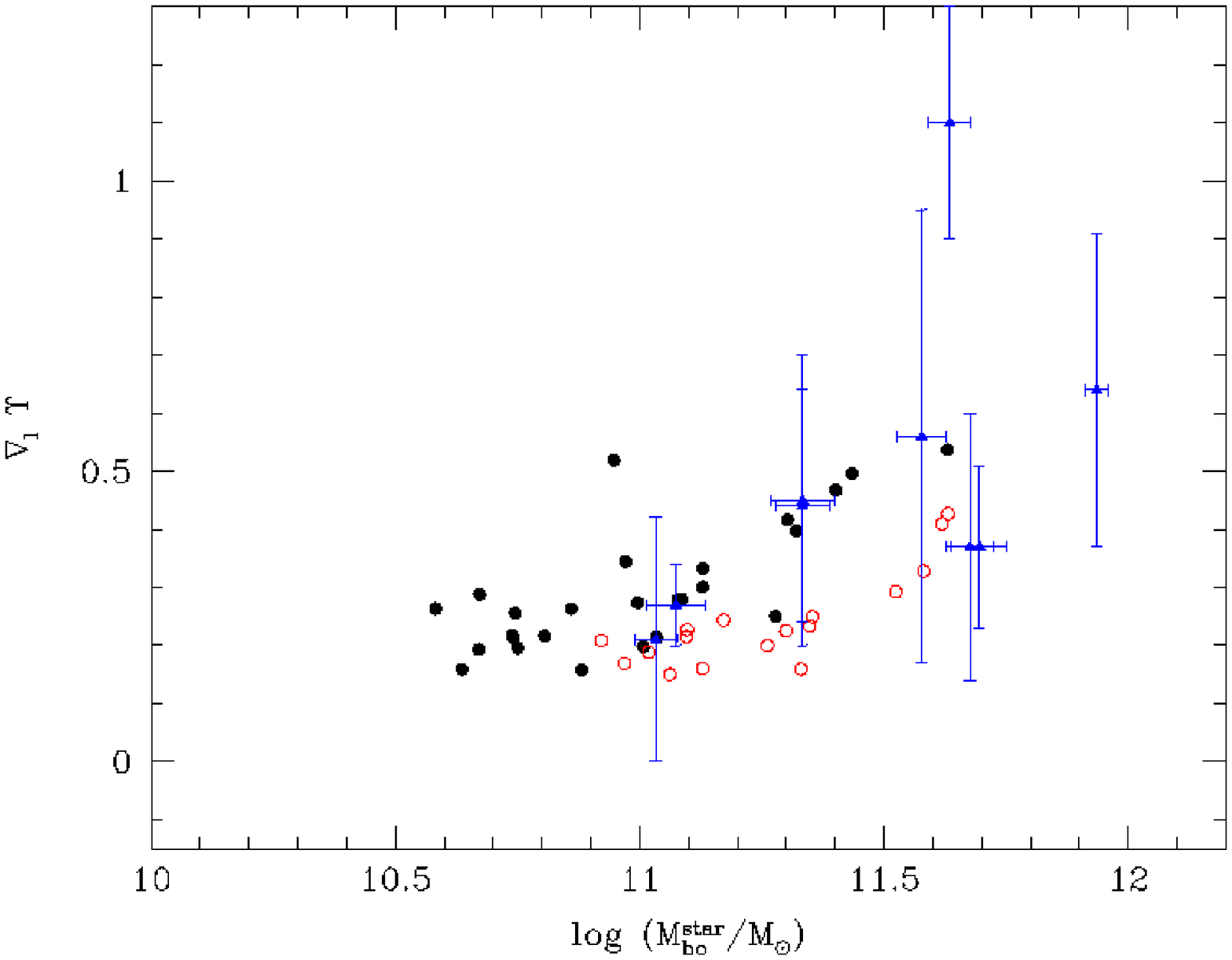}
\caption{
Upper panel: the fraction of dark-to-total mass  at  
$r/R_{\rm e, bo}^{\rm star} = 1$ versus the ELO stellar masses.
Filled (open) symbols: ELOs in SF-A (SF-B) sample.
Points with error bars are the values corresponding to the
SAURON sample of ellipticals.
Lower panel: the gradients of the $M^{\rm dark}(< r)/M^{\rm star}(< r)$
profiles as a function of their stellar masses;
blue triangles with error bars
are the empirical mass-to-light gradients as determined
by Napolitano et al. 2005 for galaxies with the $a_4 \times 100$ shape parameter lower
than 0.1 (that is, for boxy ellipticals)
}
\label{FracRboGrad}
\end{center}
\end{figure}

To further analyse this point  and make the comparison with
observational results easier, the
dark-to-stellar mass ratio profiles,
$M^{\rm dark}(< r)/M^{\rm tot}(< r)$,  are drawn
in Figure~\ref{PerStar} for the same ELOs,
with the radii in units of
the three dimensional stellar 
half-mass radii \footnote[5]{The effective or stellar half-mass radii
$r_{\rm e, bo}^{\rm star}$ are
defined as the radii of the spheres enclosing half the ELO stellar mass.
This is the relevant  three-dimensional
size parameter at ELO scales. See O\~norbe et al. (2006)  for details}.
We see that there is, in any case, a positive gradient,
and again  a  clear mass effect,
with a tendency of the dark matter fraction at fixed values of
$r/r_{\rm e, bo}^{\rm star}$
to be higher as the mass scale increases.
To be more quantitative and  compare with observational data,
we plot in Figure~\ref{FracRboGrad}, upper panel, 
the fraction of dark-to-total masses  at
$r/R_{\rm e, bo}^{\rm star} = 1$ for ELOs in the SF-A 
and SF-B samples, versus their stellar masses. The differences among 
results for both samples come from the smaller $R_{\rm e, bo}^{\rm star}$
values of SF-B sample ELOs as compared with their SF-A counterparts
(see details in O\~norbe et al. 2006).
Blue triangles with error bars are results  from integral field SAURON data and
models by Cappellari et al. 2006. We see that these empirical determinations
of the dark matter fraction at the centre of ellipticals is consistent
with the values found in ELOs of both samples, 
including its growth with the mass scale.

In the lower panel of Figure~\ref{FracRboGrad} we
give the gradients of the $M^{\rm dark}(< r)/M^{\rm star}(< r)$
profiles as a function of their stellar masses.
Blue triangles with error bars
are the empirical mass-to-light gradients as determined
by Napolitano et al. 2005 for EGs with isophotal shape
$a_{4} \times 100 < 0.1$, that is, boxy ellipticals.
We have used as inner and outer radii
$r/r_{\rm e, bo}^{\rm star, in} = 0.5$ and 
$r/r_{\rm e, bo}^{\rm star, out} = 4$, roughly
the average values of the inner and outer radii
these authors give in their Table 1.
We see that there is a mass effect and that our results are
consistent with those found by these authors in
the range of stellar mass values our samples span,
especially when we consider that ELOs in our samples are boxy
(see \S\ref{LosKin}).
A SF effect in the stellar mass distribution
also appears in Figure~\ref{FracRboGrad}, again due to the compactness
of the SF-B sample ELOs relative to their SF-A sample counterparts.

\begin{figure}[H]
 \begin{center}
     \includegraphics[width=.45\textwidth]{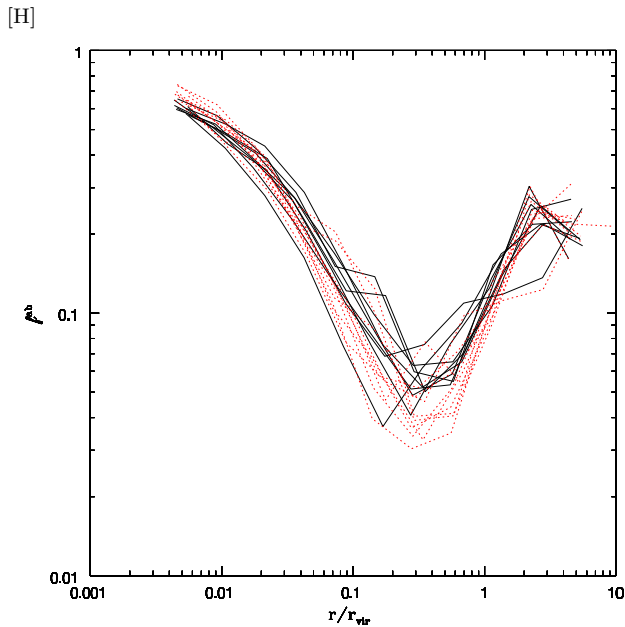}
               \caption{Baryon fraction profiles for ELOs
	       in SF-A sample (black full lines) and SF-B sample (red point lines), 
in the same range of virial mass, $1.5 \times 10^{12}$M$_{\odot} \leq M_{\rm vir} < 5 \times 10^{12}$M$_{\odot}$
	                          }
   \label{Baryonfrac1}
 \end{center}
\end{figure}

\begin{figure}[H]
  \begin{center}
   \includegraphics[width=.45\textwidth]{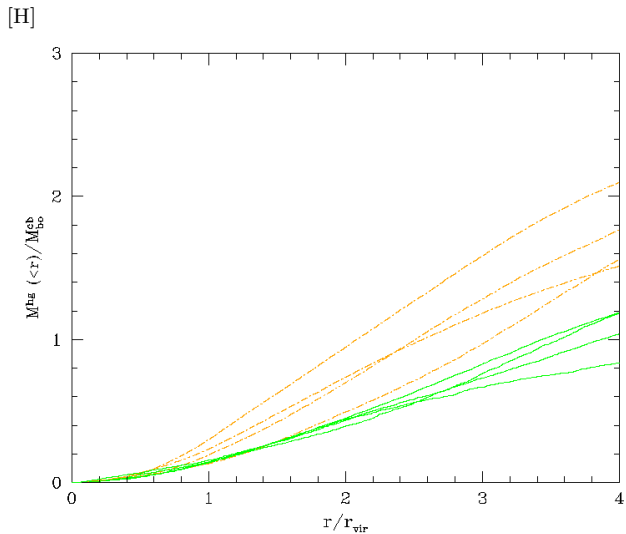}
    \caption{The $M^{\rm hg} (< r)/M_{\rm bo}^{\rm cb} $ profiles for typical
    ELOs. $M^{\rm hg} (< r)$ is the mass of hot gas within a sphere of radius
    $r$. Orange point-dashed lines: ELOs with $1.5 \times 10^{12}  \leq M_{\rm vir} < 5 \times 10^{12}$M$_{\odot}$;
    green full lines: ELOs with $M_{\rm vir} < 1.5 \times 10^{12}$M$_{\odot}$.
    Only isolated ELOs have been considered.
 }
  \label{HGasfrac}
 \end{center}
\end{figure}

We now turn to analyse the baryon space distribution at halo scales.
To have an insight on how baryons of any kind are distributed relative to
the  dark   matter at the halo scale and beyond, 
the baryon fraction profile

\begin{equation}
f^{\rm ab}(r) = \rho^{\rm ab}(r)/ \rho^{\rm tot}(r),
\end{equation}

where "ab" stands for baryons of any kind (i.e., stars, cold gas and hot gas)
and "tot" stands for matter of any kind (i.e., dark plus baryons of any kind),
is drawn in Figure~\ref{Baryonfrac1} for ELOs in the SF-A sample (black full lines) and in the SF-B sample (red point lines) in the same range of virial mass, $1.5 \times 10^{12} $M$_{\odot} \leq M_{\rm vir} < 5 \times 10^{12}$M$_{\odot}$.
 Despite individual characteristics, the $f^{\rm ab}(r)$ curves show a typical
   pattern in which their values are high at the centre, then they
     decrease and have a minimum 
lower than the global value, 
$f_{\rm cosmo}^{\rm ab} \equiv \Omega_b / \Omega_m = 0.171$,
at a radius $r_{\rm min}^{\rm ab}$, then they
       increase again, reach a maximum value and then they decrease and fall 
to the   $f^{\rm ab}_{\rm cosmo}$ value  at a rather large
	   $r$ value, larger than the corresponding virial radii. This result,
i.e., that EGs are not baryonically closed, is also indicated by
recent X-ray observations (Humphrey et al. 2006).
Notice  (Figure~\ref{HGasfrac}) that the increase of $f^{\rm ab}(r)$
at $r > r_{\rm min}^{\rm ab}$ is mainly  contributed by
hot gas, almost absent at $r < r_{\rm min}^{\rm ab}$,
indicating that $r_{\rm min}^{\rm ab}$ separates the (inner) region
where gas cooling has been possible from the (outer) region where gas
has not had time enough  to cool in the ELO lifetime.
Note also in Figure~\ref{HGasfrac}
that an important  amount of hot gas is outside the spheres of radii
$r_{\rm vir}$, that is, it is not bound to the self-gravitating
 configuration defined by the ELO halo. In fact, the mass of hot gas increases 
 monotonically up to $r \simeq 4 r_{\rm vir}$, and maybe also beyond this value,
 but it is difficult at these large radii to properly dilucidate
 whether or not a given hot gas mass element
 belongs to a given ELO or to another
 close one (to alleviate this difficulty, only those ELOs not having massive
 neighbours within  radii of  6$\times r_{\rm vir}$
 have been considered to draw this Figure).
 Another important result is that the hot gas mass fraction,
 relative to the cold mass fraction at the ELO scale, increases
with $M_{\rm vir}$ at given $r/r_{\rm vir}$, and the differences
between massive and less massive ELOs can be as high
as a factor of $\sim 2$ at $r/r_{\rm vir} < 4$.
We see that, in massive ELOs, this excess of baryons in the form of hot gas 
at the outer parts of their
configurations, compensates for the lack of baryons in the form of stars
at the ELO scales.

\subsection{Total three-dimensional mass density profiles}
\label{TotProf}

\begin{figure}
 \begin{center}
   \includegraphics[width=.45\textwidth]{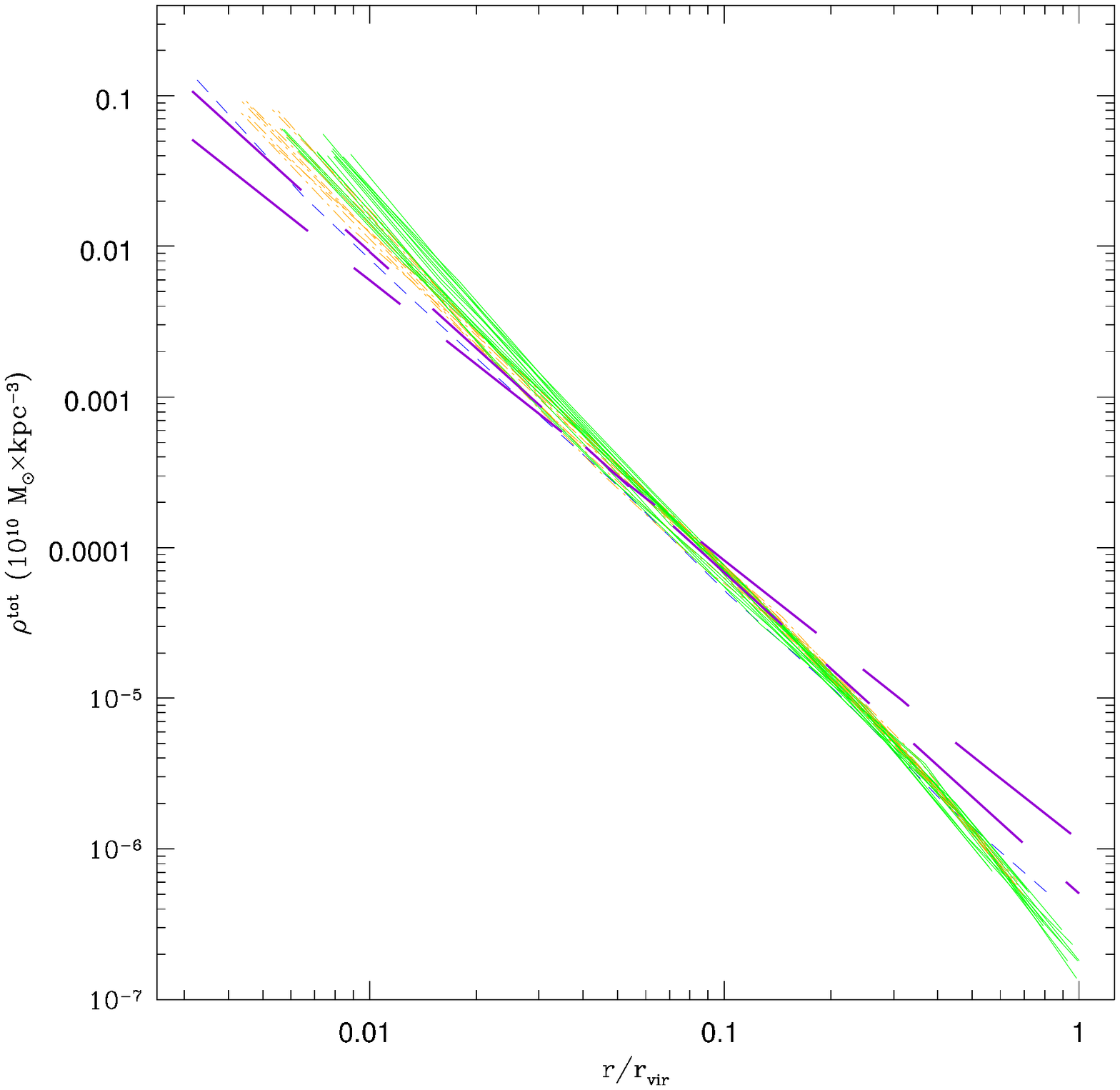}
   \includegraphics[width=.45\textwidth]{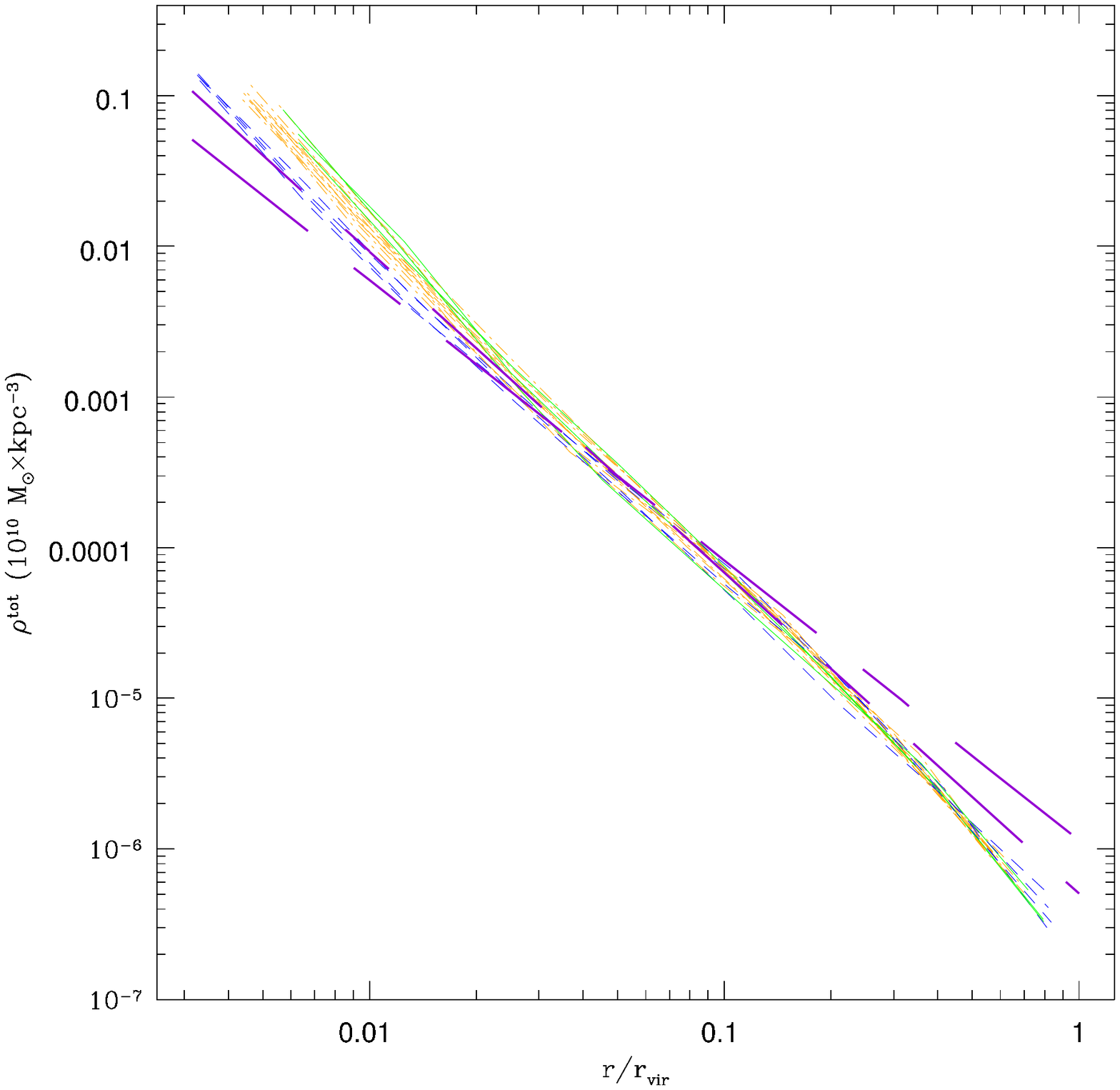}
   \caption{The total mass density profiles for ELOs
     in the SF-A sample (up) and in the SF-B sample (down).
     Green full lines: ELOs with $M_{\rm vir} < 1.5 \times 10^{12}$M$_{\odot} $;
     orange point-dashed lines:
ELOs with $1.5 \times 10^{12}$M$_{\odot} \leq M_{\rm vir} < 5 \times 10^{12}$
M$_{\odot}$; blue dashed lines: ELOs with $5 \times 10^{12}$M$_{\odot} \leq M_{\rm vir} $. The violet long-dashed lines are the one sigma interval for the slope resulting from fits to power-law profiles of lens ellipticals from Koopmans et al. (2006).
     }
  \label{TotPerf}
 \end{center} 
\end{figure}

\begin{figure}
 \begin{center}
\includegraphics[width=.45\textwidth]{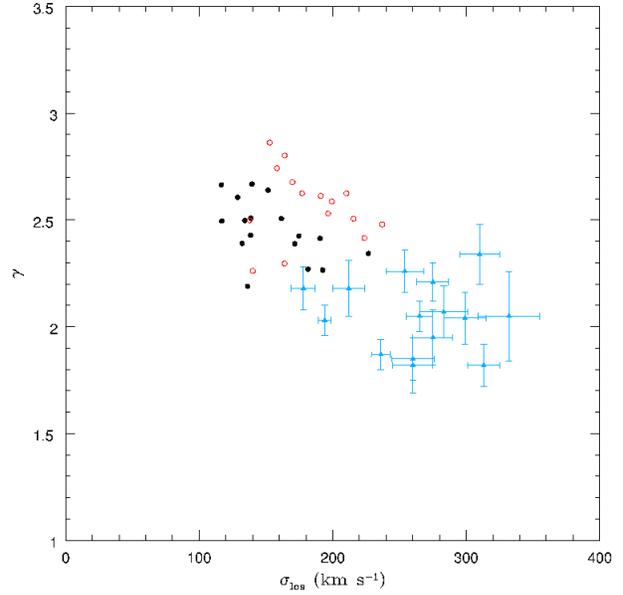}
\caption{The logarithmic  slopes corresponding to the 
total mass density    profiles for  ELOs
in the SF-A (black filled circles) and the SF-B samples (red open circles), versus their central L.O.S. stellar velocity
dispersions. Triangles with error bars correspond to data on
SLACS lens ellipticals, as given in Table 1 of Koopmans et al. (2006)
}
\label{GammaKoopmans}
\end{center}
\end{figure}

We now address the issue of the total mass (i.e., baryonic plus dark)
density profiles. In Figure~\ref{TotPerf} they are drawn for 
ELOs in the SF-A sample (upper panel) as well as for those in the
SF-B sample (lower panel). In both cases, the profiles corresponding
to ELOs in different mass intervals have been drawn with different 
line and colour codes. 
Some important results are that
i), they are well fit by power-law expressions 
$\rho^{\rm tot}(r) \propto r^{-\gamma}$ in a range of 
$r/r_{\rm vir}$ values larger than two decades, 
  ii), the slope
of the power-law increases
with decreasing ELO mass, and,
iii) a slight SF effects appears, but only
at the very inner regions, with SF-B sample ELOs showing 
a worse fit to a power law than their SF-A counterparts.
Koopmans et al. 2006 have also found
that the total mass density profiles of their massive 
($< \sigma_{\rm ap}> = 263$ km s$^{-1}$) lens EGs
 can be fit by power-law expressions within their Einstein radii
($< R_{\rm Einst}> = 4.2 \pm 0.4$ kpc, with 
$<R_{\rm Einst}/R_e^{\rm light}> = 0.52 \pm 0.04$, i.e., the inner region),
whose average  slope is $<\gamma > = 2.01^{ + 0.02}_{-0.03} \pm 0.05$
(68 percent C.L.), with an intrinsic scatter of 0.12.
To elucidate how well the total mass density profiles of ELOs
compare with these results, in Figure~\ref{GammaKoopmans}
we plot the slopes $\gamma$ for ELOs, as well as for SLACS lens ellipticals
(Table 1, Koopmans et al. 2006),  versus  their
central L.O.S. stellar velocity dispersions.
The fitting range for ELOs used to draw this
Figure is $r < R_{\rm e, bo}^{\rm star}$.
We see that results for ELO and SLACS lens galaxy samples  are consistent
in the range of velocity dispersion values where they coincide.

\section{Kinematics}
\label{Kine}

\subsection{Three-dimensional velocity distributions}
\label{3DVelDis}

\begin{figure*}
    \begin{center}
        \includegraphics[width=.45\textwidth]{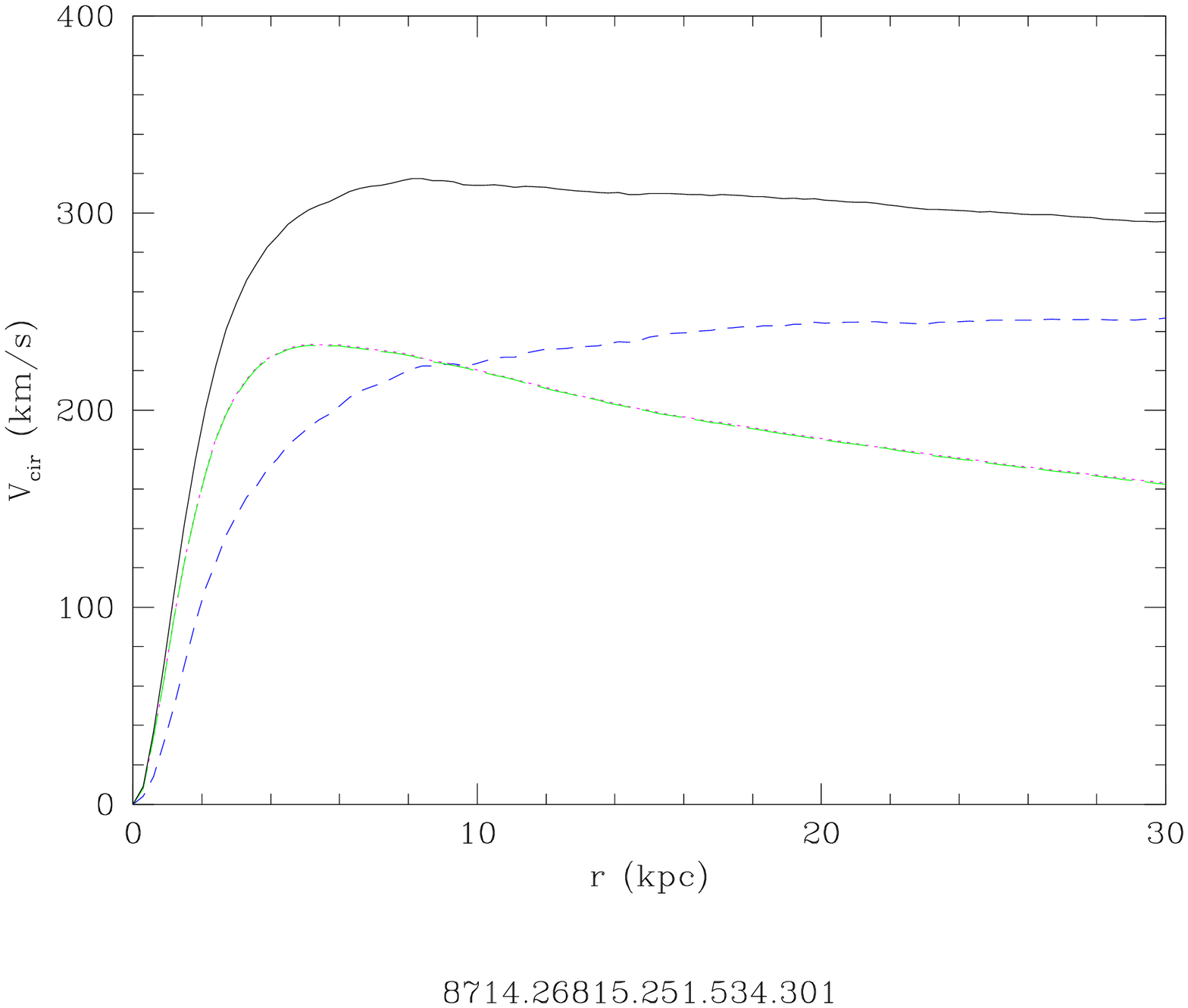}
	\includegraphics[width=.45\textwidth]{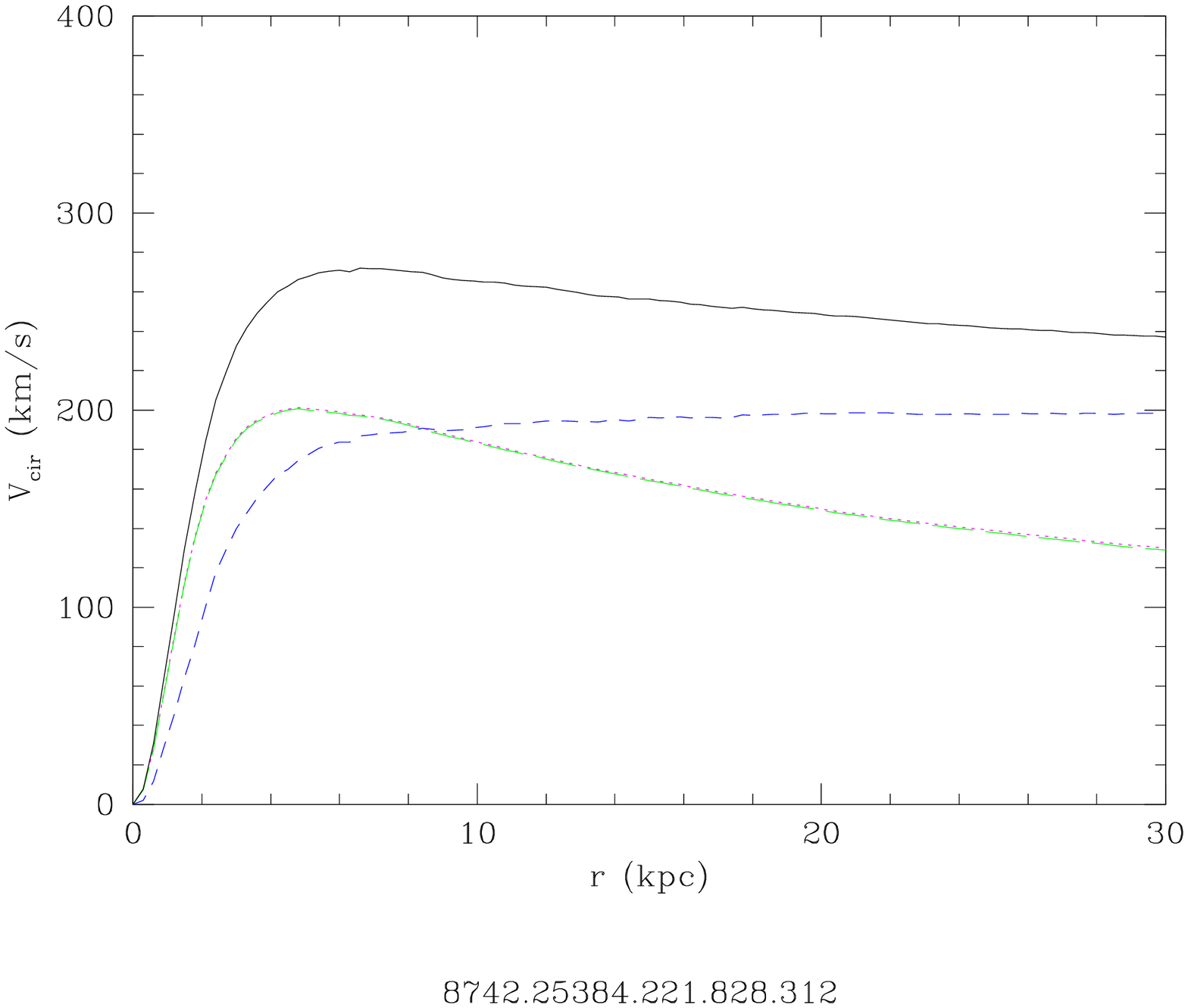}
	\includegraphics[width=.45\textwidth]{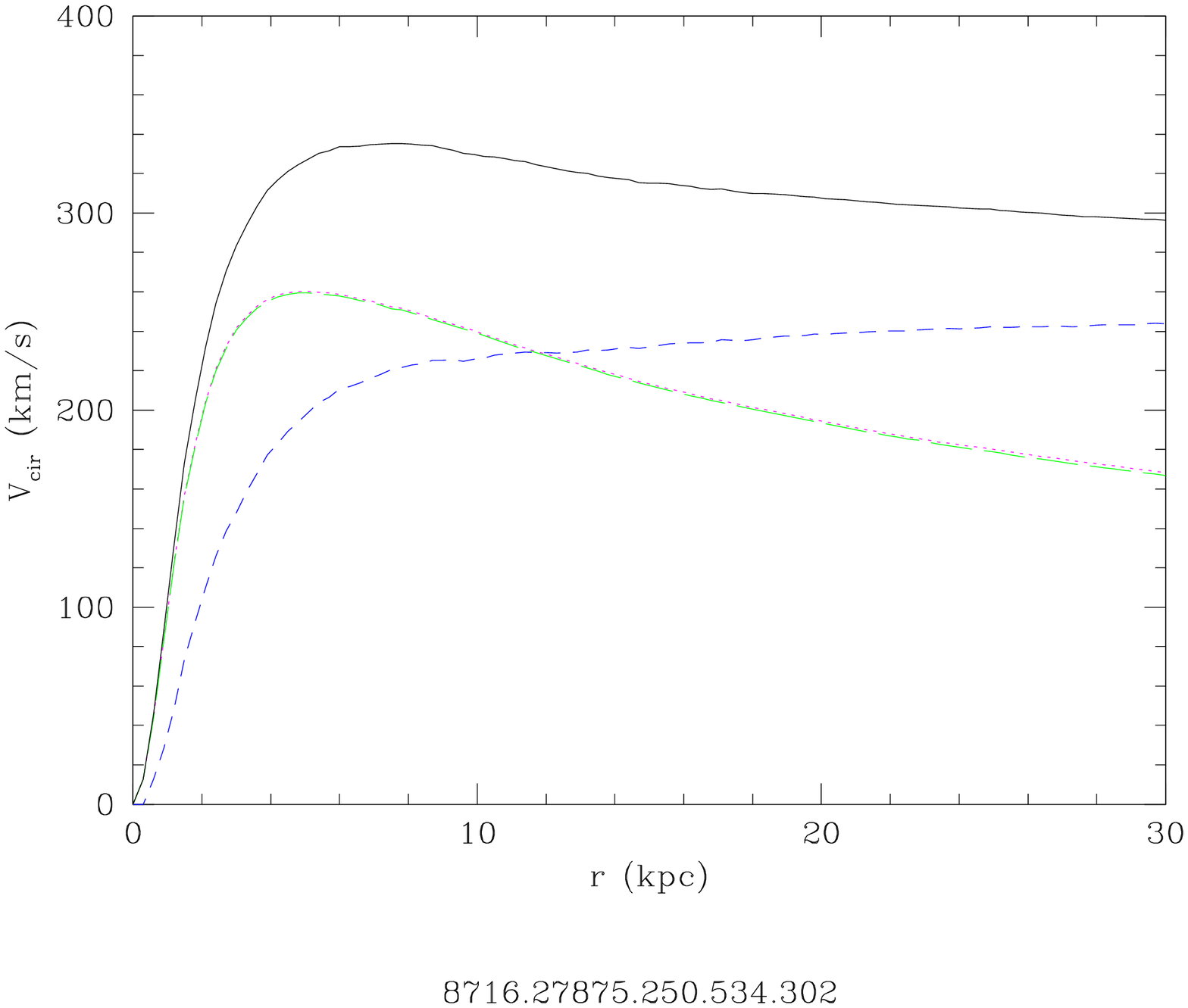}
\includegraphics[width=.45\textwidth]{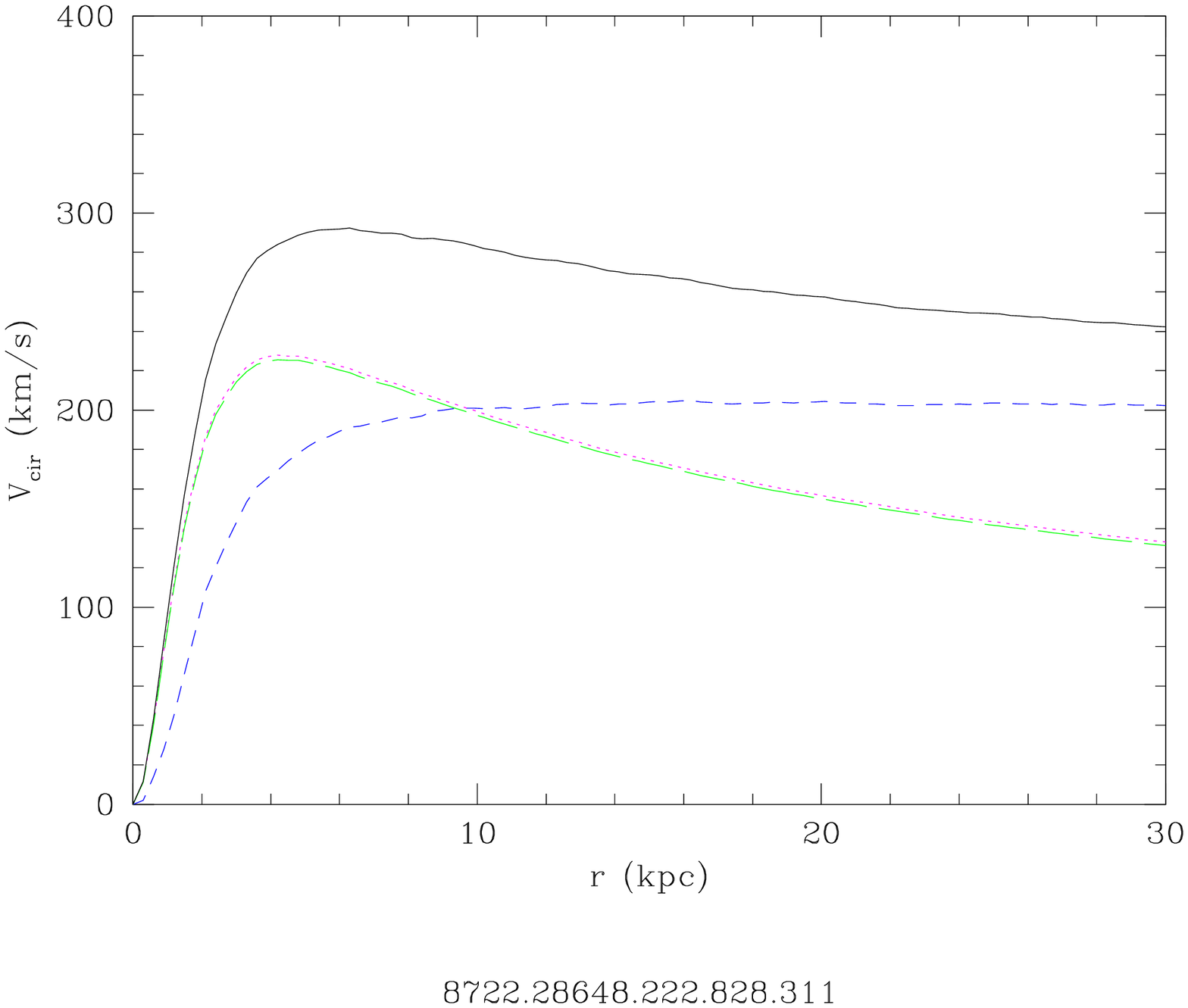}
\caption{
The circular velocities
profiles
of two typical ELOs in the SF-A sample (upper panels)
and their SF-B sample counterparts (lower panels).
Black continuous line: total mass;
blue short-dashed line: dark matter contribution;
green long-dashed line: stellar mass contribution;
red point line: cold baryon contribution
}
   \label{VelCir3DProf_1}
 \end{center} 
\end{figure*}

Shapes and mass density profiles (i.e., positions)
are  related to the 3D velocity distributions
of relaxed  EGs through
 the Jeans equation (see Binney \& Tremaine 1987).
  Observationally,  informations
on such 3D distributions are  not available for external galaxies,
 only the line-of-sight velocity distributions
  (LOSVD) can be inferred from their spectra
 (Binney \& Tremaine 1987; van der Marel \& Franx 1993; de Zeeuw \& Franx 1991).
      The complete six dimensional phase space informations
       for each of the particles sampling the ELOs provided
        by numerical simulations,
	 allow us  to calculate the 3D
	  profiles for the velocity dispersion, $\sigma_{\rm 3D}(r)$,
	   as well as the circular
	      velocity profiles, $V_{\rm cir}(r)$ .
In Figure~\ref{VelCir3DProf_1}
we draw  the  $V_{\rm cir}(r)$ profiles (full line),
as well as their dark matter (short-dashed line)
and baryonic contributions (stars, long-dashed line;
stars plus cold gas, point line).

The $V_{\rm cir}(r)$  profiles
provide another
measure of ELO mass distribution.
We note in Figure~\ref{VelCir3DProf_1}
that the baryon mass distribution is more
concentrated than the dark matter one due to energy losses by
the gaseous component before being transformed into stars.
This is a general property of the circular velocity
profiles of the ELO samples.
Moreover, objects in SF-B  sample are more concentrated than
their SF-A sample counterparts,
because of the SF implementation: the amount of baryons
at their central volumes relative to dark matter is always lower in
SF-A than in SF-B objects;
this is a small scale effect as
$r \sim $ 30 kpc or $r \sim $ 40 kpc radii enclose roughly similar
amounts of baryons or dark matter in any cases.

\begin{figure*}
\begin{center}
\includegraphics[width=.45\textwidth]{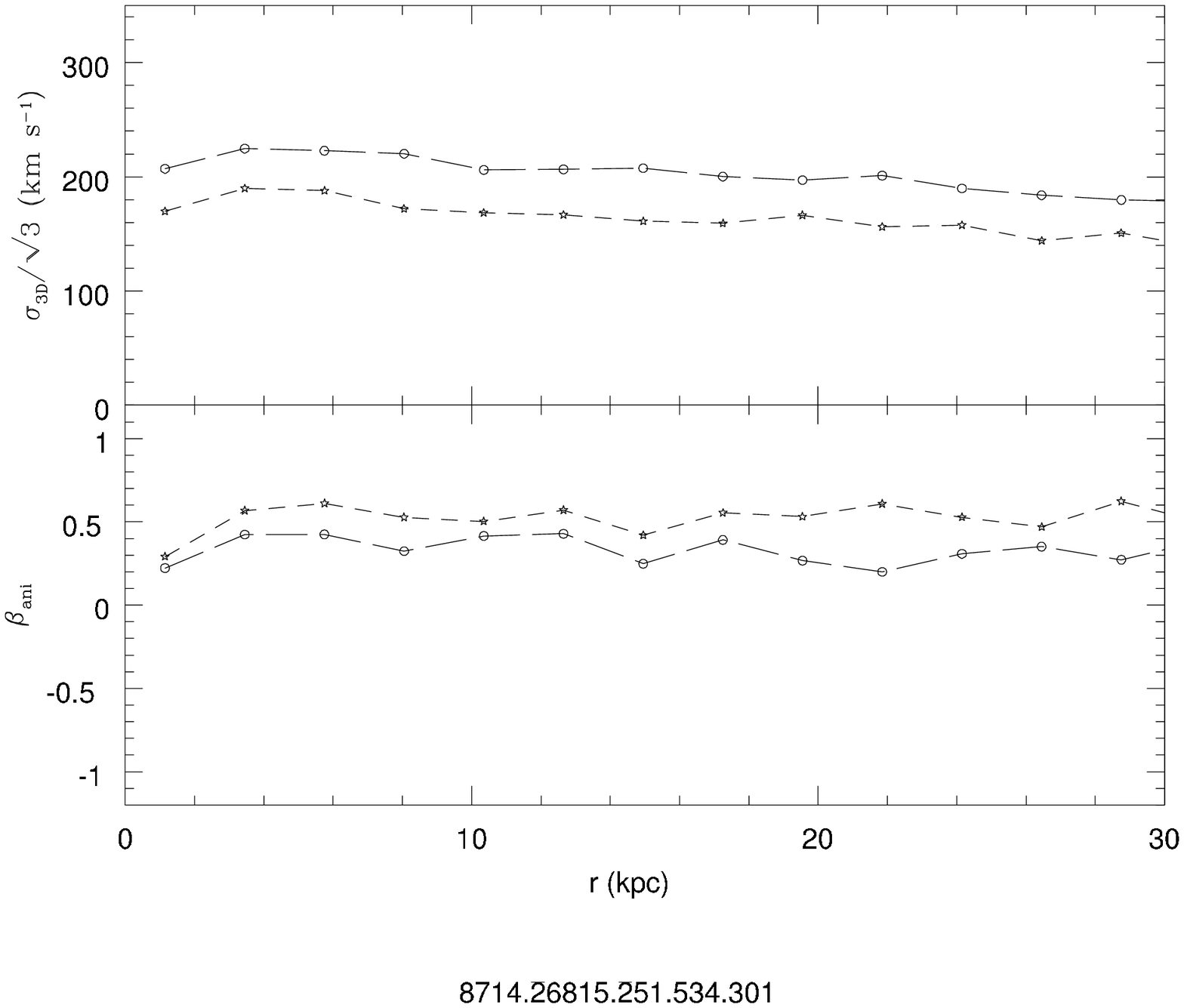}
\includegraphics[width=.45\textwidth]{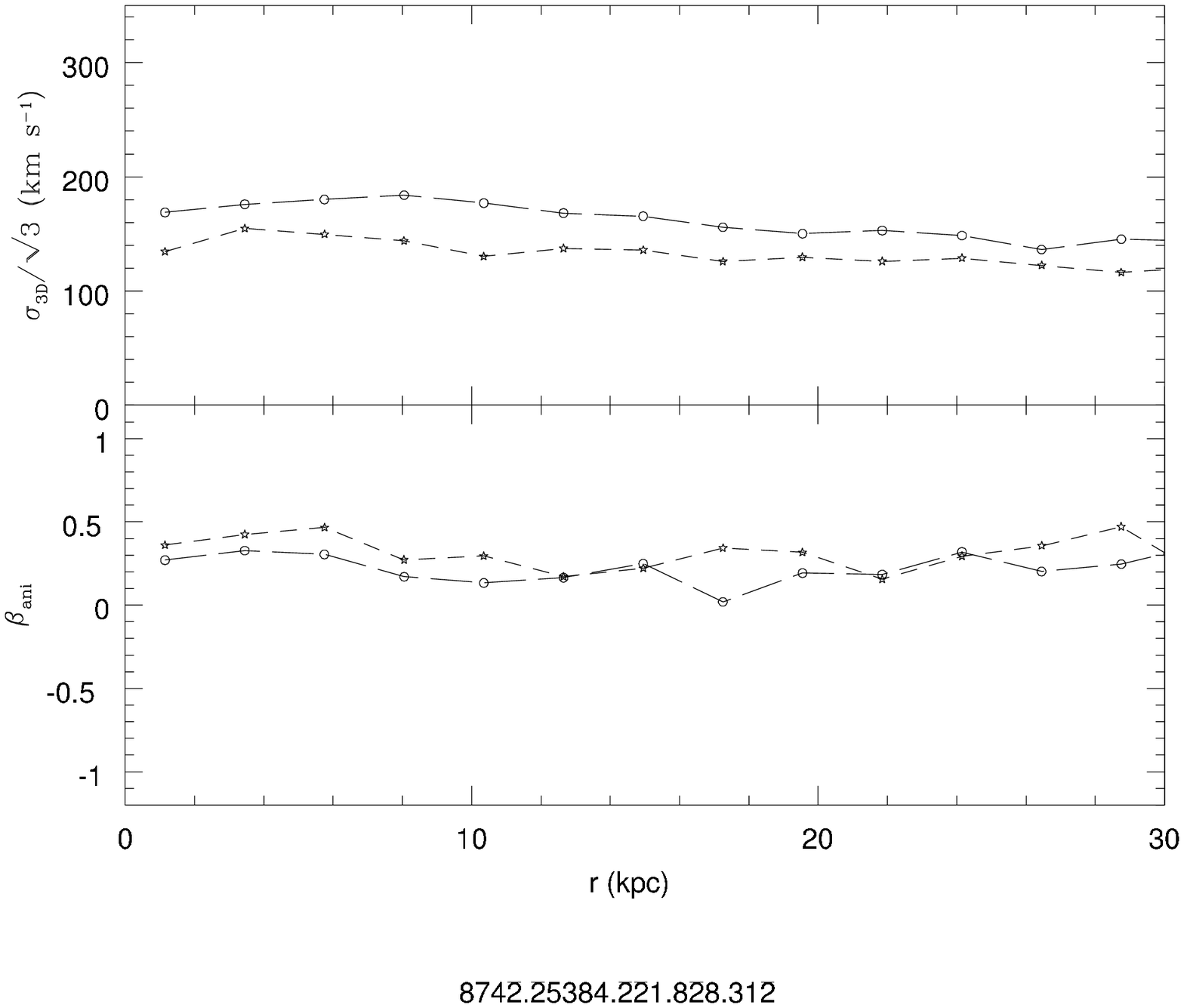}
\includegraphics[width=.45\textwidth]{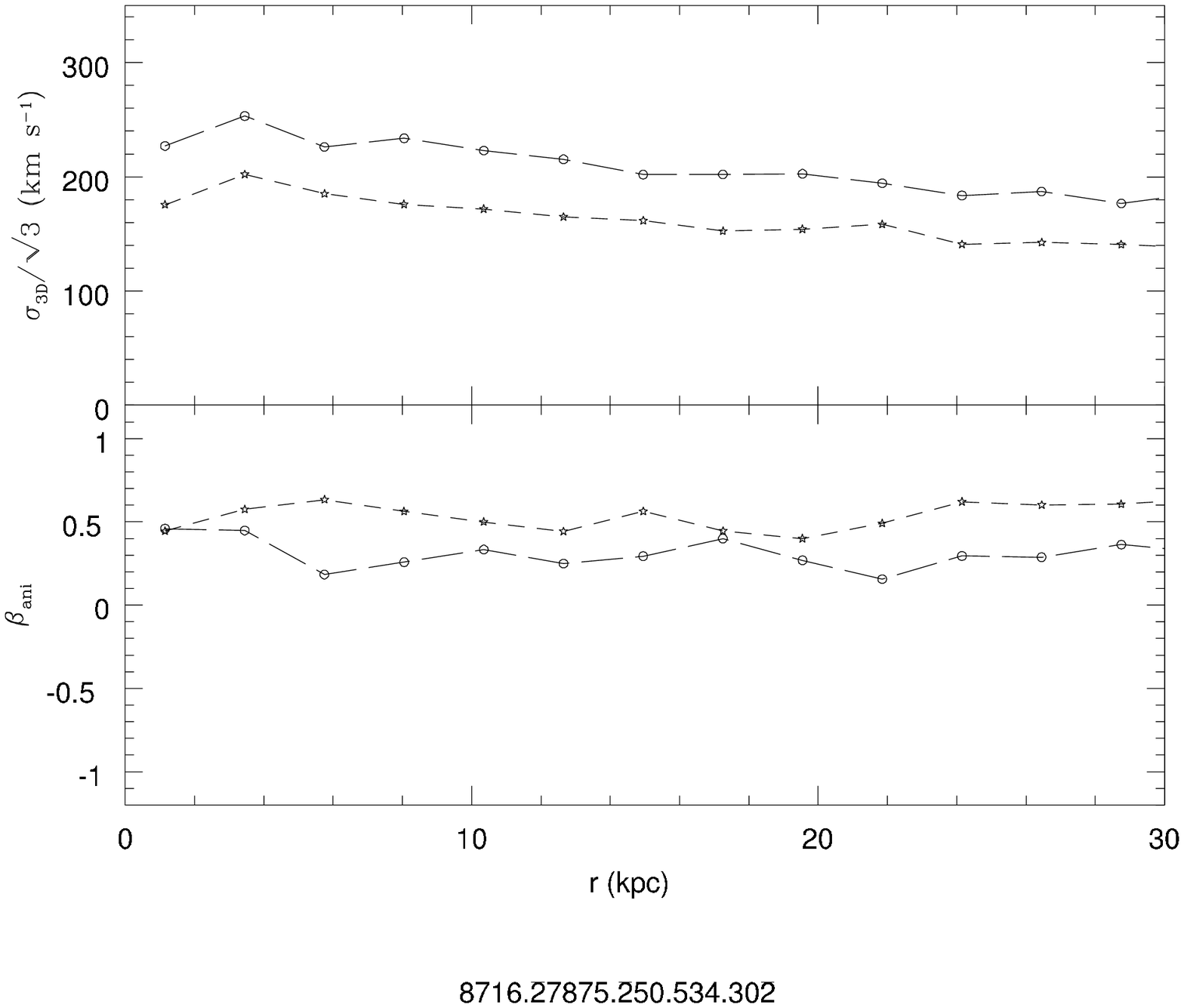}
\includegraphics[width=.45\textwidth]{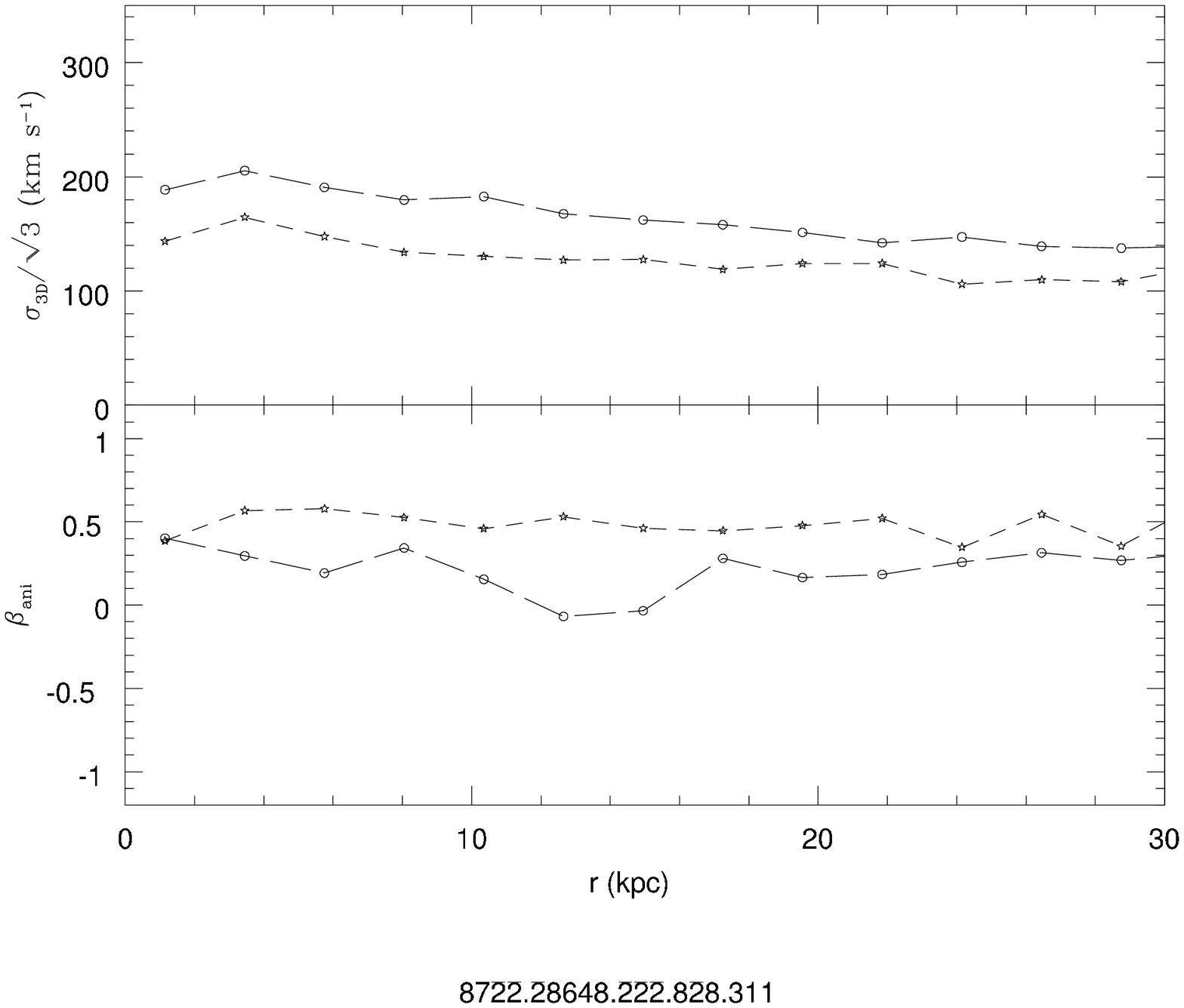}
\caption{
The $\sigma_{\rm 3D}(r)$ profiles
of two typical ELOs in the SF-A sample (upper panels)
and their SF-B sample counterparts (lower panels).
Also shown are the anisotropy profiles $\beta_{\rm ani}(r)$.
Long-dashed lines: dark matter;
short-dashed lines: stars}
\label{Sig3DAni}
\end{center}
\end{figure*}

\begin{figure*}
\begin{center}
\includegraphics[width=.45\textwidth]{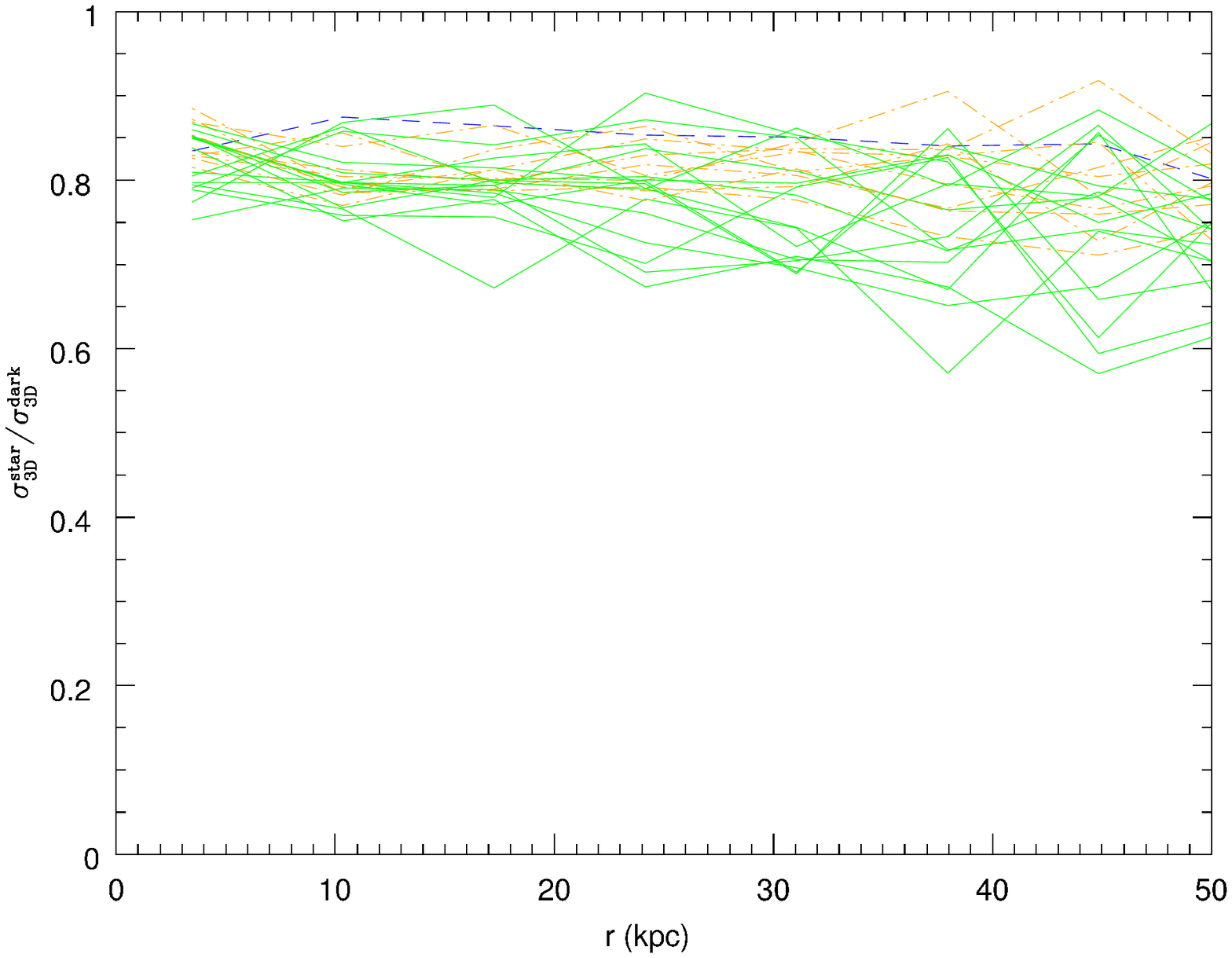}
\includegraphics[width=.45\textwidth]{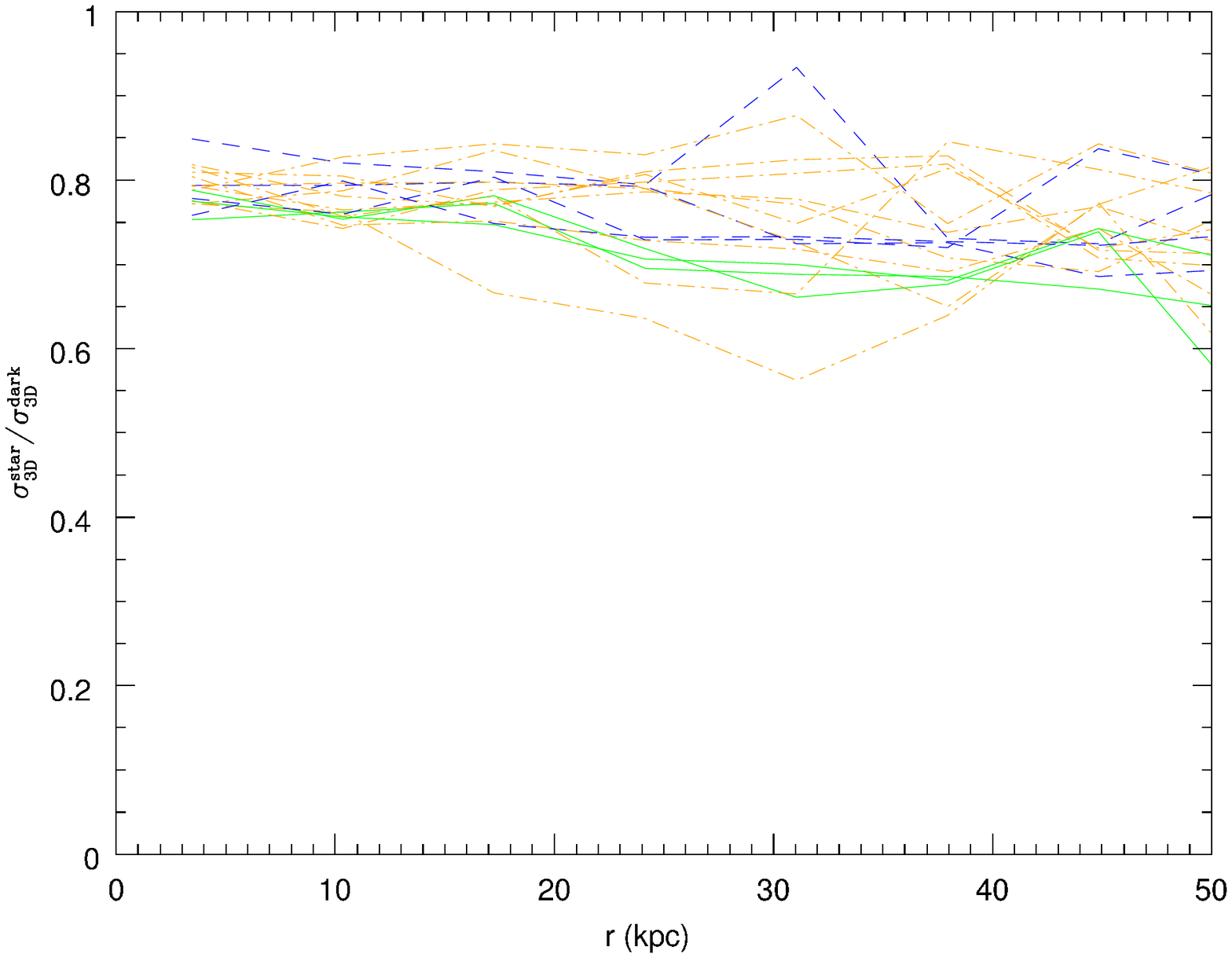}
\caption{The $\sigma^{\rm star}_{3D}(r)/\sigma^{\rm dark}_{3D}(r)$ ratio profiles
for ELOs in SF-A (upper panel) and SF-B (lower panel) samples. Different colour
and line codes stand for ELO mass intervals, as in Figure 9
}
 \label{KinSegre}
\end{center}
\end{figure*}

In Figure~\ref{Sig3DAni}
we draw, for the same ELOs, the $\sigma_{\rm 3D}(r)$ profiles
  as measured by stars, ($\sigma^{\rm star}_{3D}(r)$, starred symbols
      and short-dashed lines),
            and by dark matter,
($\sigma^{\rm dark}_{3D}(r)$, open circles and long-dashed lines) 
as proof particles in the overall potential well.
These profiles are in any case  decreasing outwards,
both for the dark matter and for the stellar components.  
An outstanding result  illustrated by Figure~\ref{Sig3DAni} is that
$\sigma^{\rm d}_{3D}(r)$ is always higher than
$\sigma^{\rm star}_{3D}(r)$ (because stars are made out of cooled gas),
with $\sigma^{\rm star}_{3D}(r)/\sigma^{\rm dark}_{3D}(r) \sim 0.8$,
in consistency with the values found by Loewenstein (2000) on 
theoretical grounds and by Dekel et al. (2005)
from pre-prepared simulations of mergers of disc galaxies.
This is the so-called kinematical segregation (S\'aiz 2003, S\'aiz et al. 2004).
To further analyse this issue, in  Figure~\ref{KinSegre} we plot
the $\sigma^{\rm star}_{3D}(r)/\sigma^{\rm dark}_{3D}(r)$ ratios  for
the ELOS in both the SF-A sample and in the SF-B sample, with different
colour and line codes depending on the ELO mass range.
We see that the kinematical segregation does not show either a clear
mass dependence, or a radial dependence. Moreover, the SF parametrization
effect is only mild.

Another relevant quantity is  the anisotropy of the
3D velocity distributions of the ELO sample,
defined as:

\begin{equation}
\beta_{\rm ani} = 1 - \frac{\sigma^2_{\rm t}}{2\,\sigma^2_{\rm r}},
\label{AniDef}
\end{equation}

where $\sigma_{\rm r}$ and $\sigma_{\rm t}$ are the radial and
tangential velocity dispersions ($\sigma_{\rm t}^2 = \sigma_{\theta}^2 + \sigma_{\phi}^2$), relative to the centre of the
    object.
The anisotropy  profile, $\beta_{\rm ani}(r)$, is
represented in Figure~\ref{Sig3DAni}
for typical ELOs in the sample, for their
dark matter and stellar particle
components. The anisotropy is always positive 
(i.e., an excess of dispersion in  radial motions), 
the profiles are almost constant,
except at the innermost regions,
and the stellar component is always
more anisotropic than the dark matter one,
presumably as a consequence of the mergers involved in
the ELO mass assembly (see \S\ref{ELOForm} and
DTal06).
In fact, the characteristics of the stellar anisotropy
profiles (roughly constant and $\beta_{\rm ani}^{\rm star}(r) \simeq 0.5$
in most cases) are consistent with those found by Dekel et al. (2005),
where they conclude that large radial anisotropy is generic to the
stellar component of merger remnants of any kind.

\subsection{Stellar LOS velocity and velocity dispersion profiles}
\label{LosKin}

\begin{figure}
\begin{center}
\includegraphics[width=.45\textwidth]{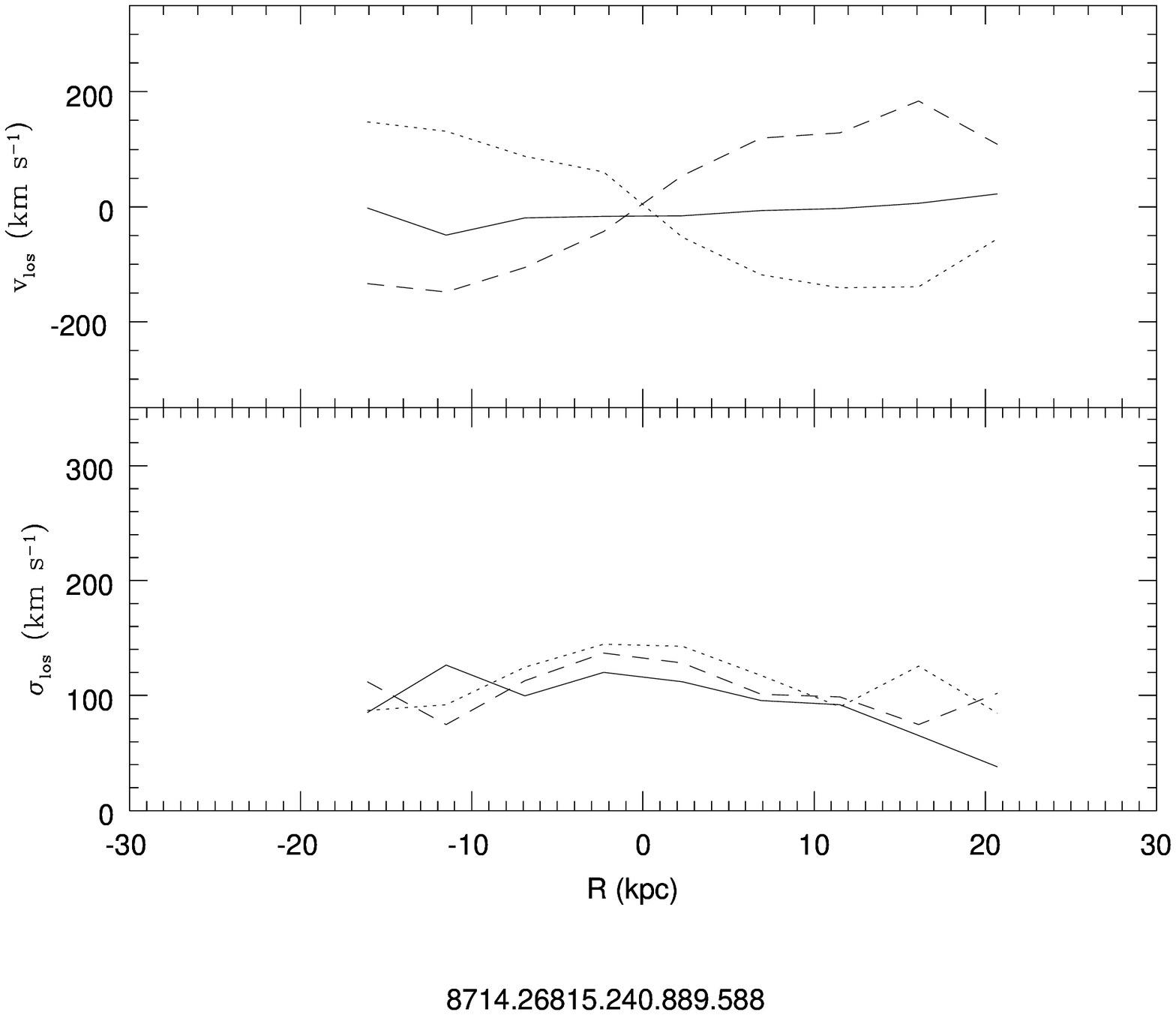}
\caption{Upper panel, full line: the major axis  stellar LOS velocity profile
along the  spin direction for an ELO in SF-A sample. Point and dashed lines:
same as the continuous line taking the LOS direction normal to the ELO spin
vector.
This particular ELO rotates.
Lower panel: same as the upper panel for the major axis LOS 
stellar velocity dispersion profiles
}
\label{RotCurvSI}
\end{center}
\end{figure}

\begin{figure}
\begin{center}
\includegraphics[width=.45\textwidth]{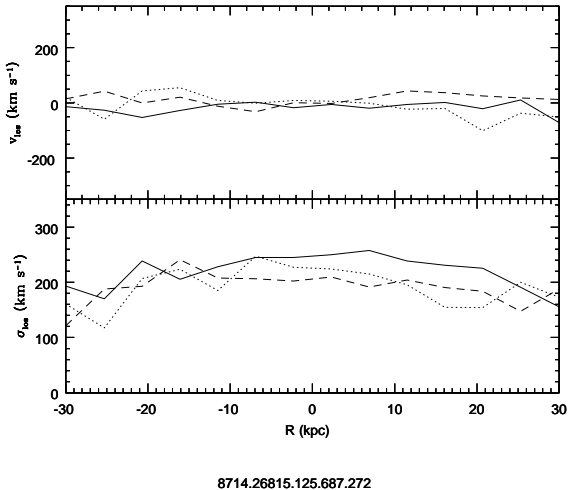}
\caption{Same as the previous Figure for another ELO.
In this case, the rotation is only mild.
}
\label{RotCurvNO}
\end{center}
\end{figure}

The two last Figures provide
an illustration of the general characteristics of the
lower-order moments of the 3D velocity distribution.
The profiles plot in these Figures are not observationally available,
but only the lower-order moments of the LOSVD are.
We have measured  the stellar line-of-sight velocity
and the stellar velocity dispersion profiles,
$V_{\rm los}^{\rm star}(R)$ and $\sigma_{\rm los}^{\rm star}(R)$,
along three orthogonal projections
for all ELOs in the sample.
To mimic observational techniques used in stellar kinematics
of EGs, we have measured these profiles along the
major and minor axes of projected ELOs, where the major
axis is defined as that orthogonal to the ELO spin vector projected
on the plane normal to the LOS, and the minor axis is parallel 
to the spin projection. 
We have found that in some cases ELOs do indeed show a clear rotation curve,
while in most cases the rotation is only modest or even very low, as
illustrated in Figure~\ref{RotCurvSI} and in Figure~\ref{RotCurvNO},
respectively.

\begin{figure}
\begin{center}
\includegraphics[width=.45\textwidth]{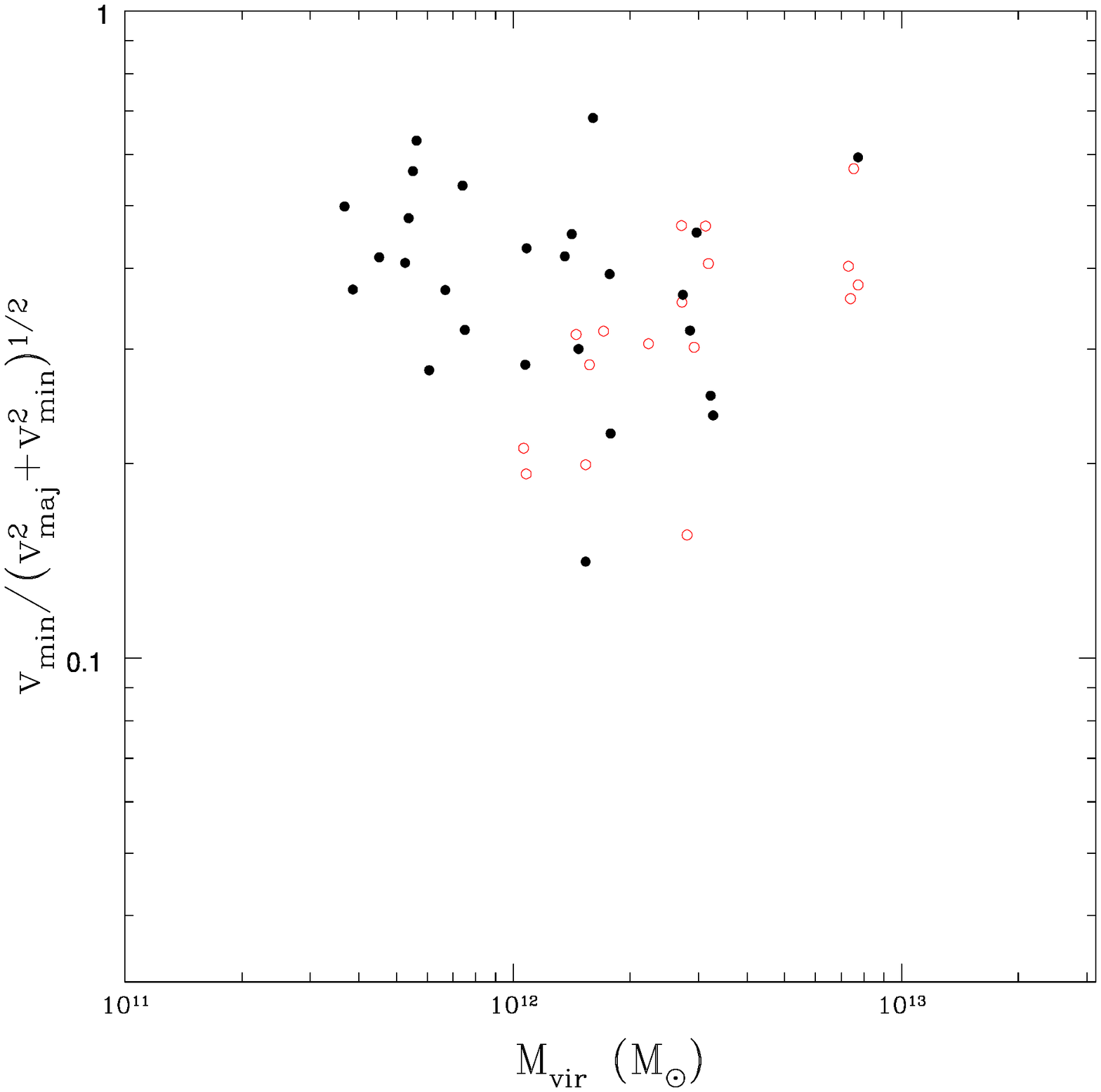}
\caption{The $c_{\rm rot}$ ratios as a function of the virial mass
for ELOs in the samples
}
\label{RotCoeff}
\end{center}
\end{figure}

To quantify the amount of rotation in ELOs and its possible dependence on 
the mass scale, in Figure~\ref{RotCoeff} we plot the ratios
$c_{\rm rot} = 
V_{\rm min}/(V_{\rm maj}^2 + V_{\rm min}^2)^{1/2}$ as a function of the
ELO virial masses, for ELOs in both the SF-A and the SF-B samples
($V_{\rm maj}$ and $V_{\rm min}$ are the maximum values of 
the $V_{\rm los}^{\rm star}(R)$ profile when measured along
the major and the minor axes, respectively).
When the ELO shows a clear rotation curve, $V_{\rm min}$ is much
lower than $V_{\rm maj}$, and the $c_{\rm rot}$ ratio is low; 
by contrast, when the rotation is unimportant, then
$V_{\rm min} \simeq V_{\rm maj}$ and  $c_{\rm rot} \sim 0.7$.
For a given ELO, the $c_{\rm rot}$ value depends on the direction taken as
LOS direction, in such a way that it is maximum
when the ELO spin is taken as LOS direction,
and minimum  when the LOS direction is normal to the ELO spin vector,
that is, when rotation stands out. This is the LOS direction taken
to draw this Figure, where we see that there is not a clear mass dependence,
that most ELOs are in between the two
situations described above and that the
values of the $c_{\rm rot}$ ratio of ELOs are typical
of boxy ellipticals (see, for example, Binney \& Merrifield, figure 4.39).

\begin{figure*}
   \begin{center}
        \includegraphics[width=.45\textwidth]{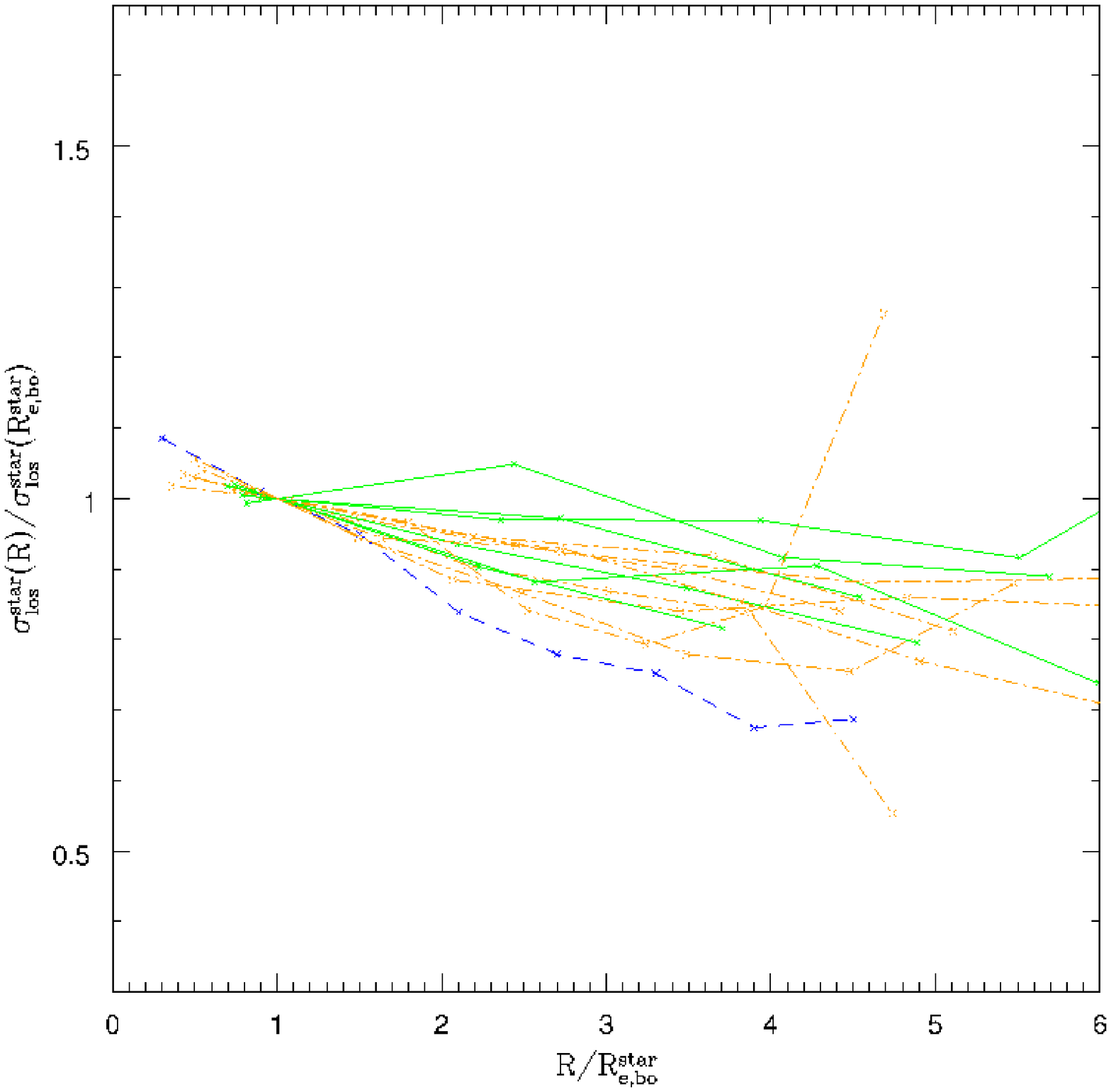}
	\includegraphics[width=.45\textwidth]{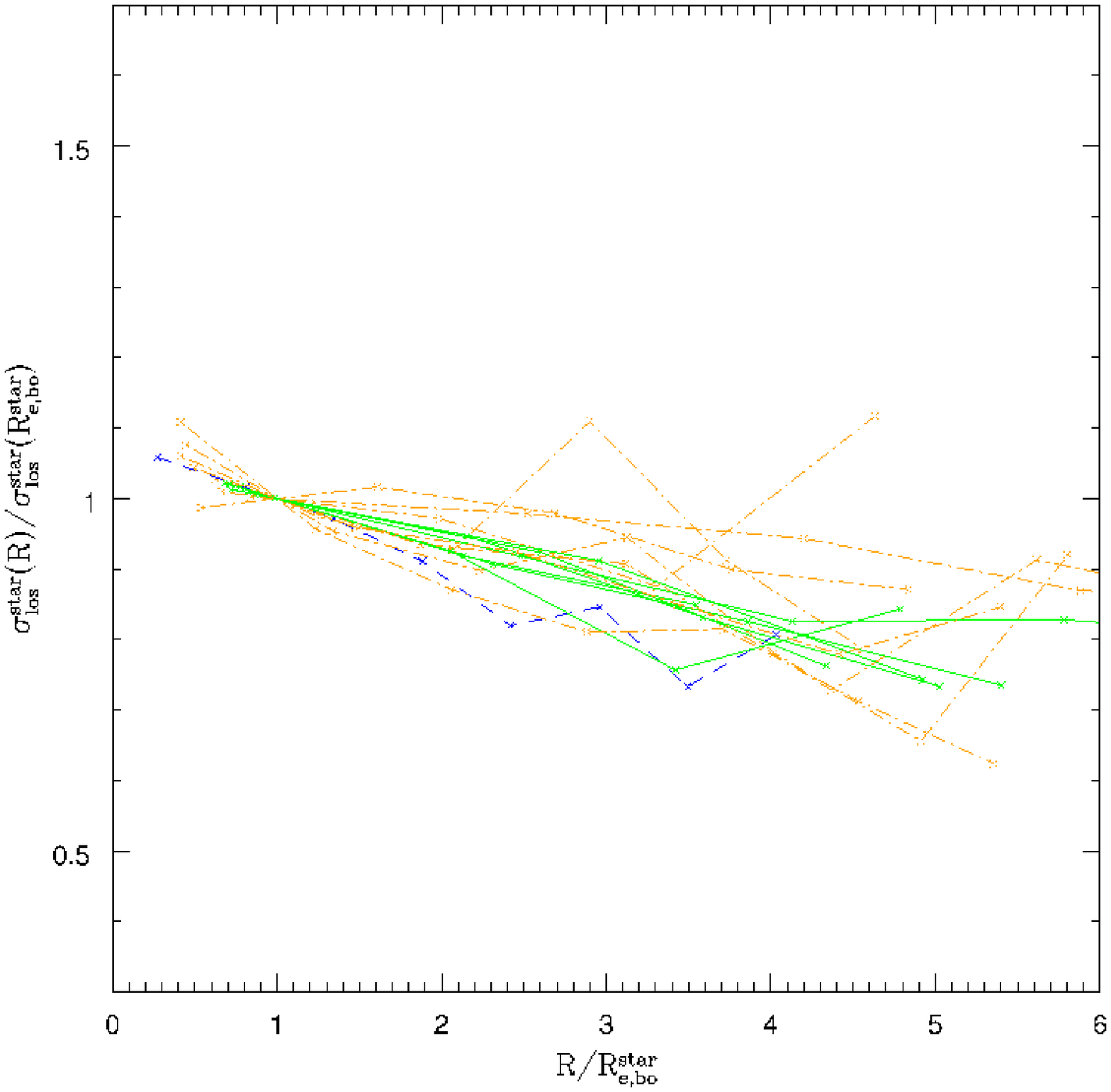}
	\includegraphics[width=.45\textwidth]{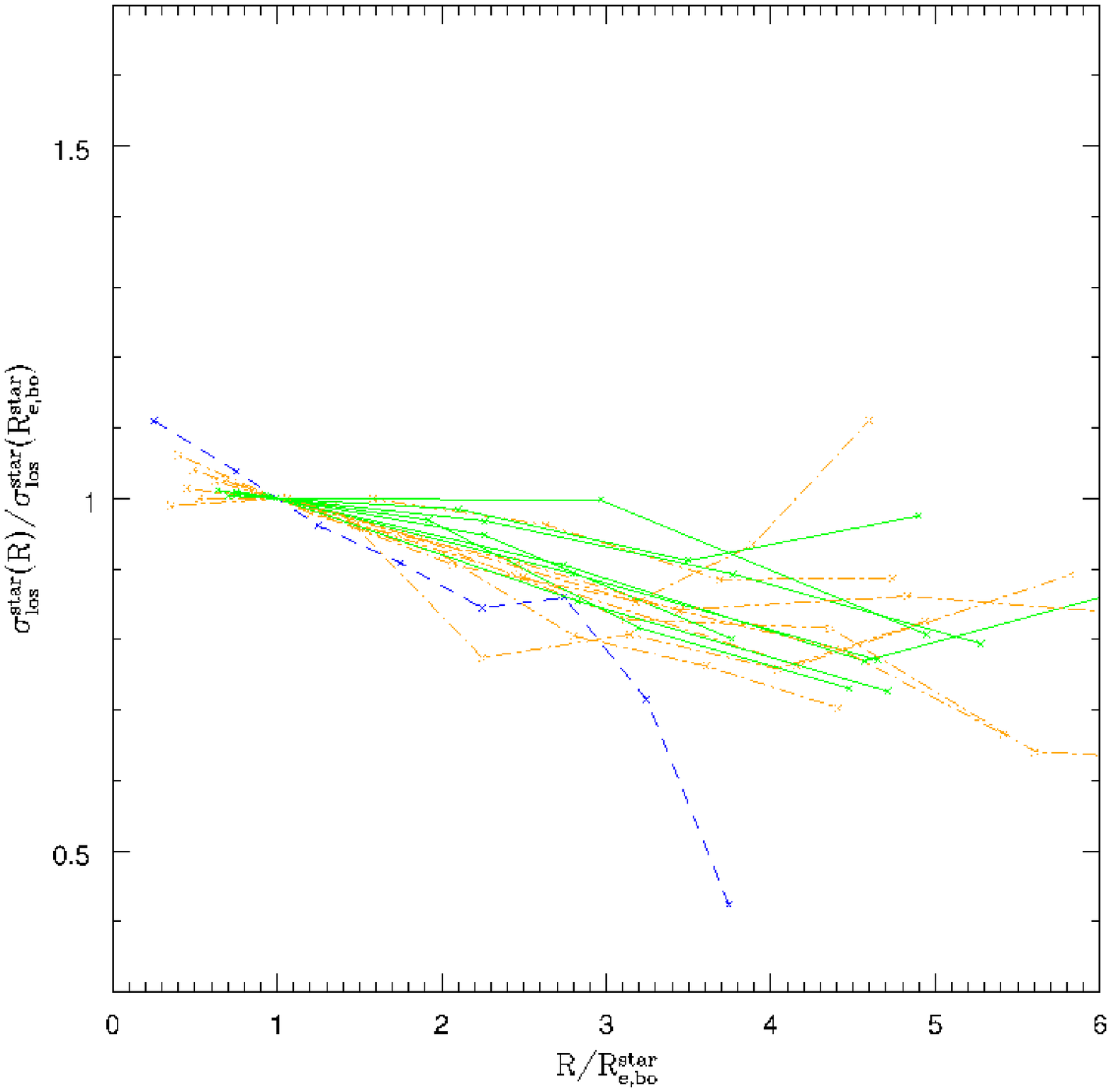}
	\caption{
	LOS velocity dispersion velocity profiles 
	along three different
	orthogonal projections for ELOs in SF-A sample up to 6 effective radii.
	The profiles are normalised to their value at 
	$R_{\rm e, bo}^{\rm star}$ for each ELO. Different colour
and line codes stand for ELO mass intervals, as in Figure 9
	}
\label{VelDispLarge}
\end{center}
\end{figure*}

We now comment on the major axis LOS stellar velocity dispersion profiles
of ELOs (Figure~\ref{RotCurvSI} and in Figure~\ref{RotCurvNO}).
Their most outstanding feature is  
the decrease of the $\sigma_{\rm los}^{\rm star}(R)$ profiles in some
cases and particularly so along some LOS directions at large $R$.
These profiles  are suited to compare with stellar kinematics data.
In other cases, for example to compare with planetary nebulae data,
the LOS
velocity dispersion 
profiles must be  calculated by averaging over  the LOS
velocities of stars placed within  cylindrical shells,
with their axes in the LOS direction. 
Figure~\ref{VelDispLarge} is a plot of such  profiles normalised
to $\sigma_{\rm los}^{\rm star}(R_{\rm e, bo}^{\rm star})$ for
the SF-A sample ELOs;
each panel corresponds to a different orthogonal
projection. 

\begin{figure}
\begin{center}
\includegraphics[width=.45\textwidth]{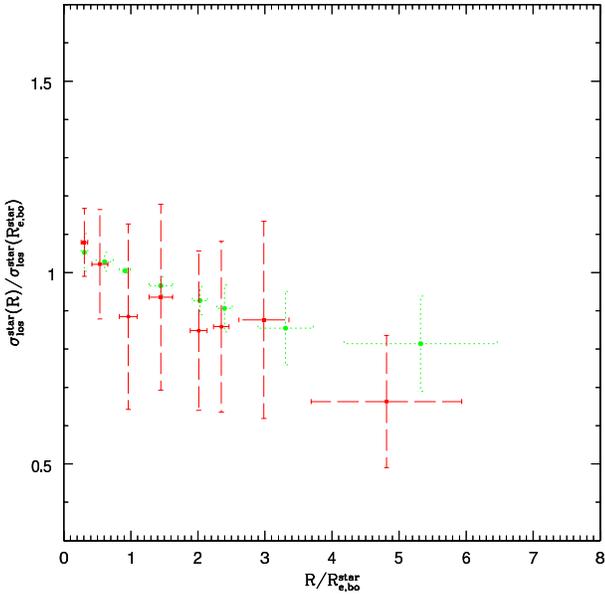}
\caption{The SF-A sample average LOS velocity dispersion profiles
normalised to their values at $R_{\rm e, bo}^{\rm star}$ for each ELO
(green points) along with their 1 $\sigma$ dispersions.
Orange points and error bars: the same for the young stellar particles,
with the same normalisation
}
\label{VelDispData}
\end{center}
\end{figure}

To make clearer  the decline of the $\sigma_{\rm los}^{\rm star}(R)$ profiles,
 in Figure~\ref{VelDispData} we plot, at different
$R$ values, the averages of the stacked profiles shown in 
Figure~\ref{VelDispLarge} with their dispersions (green points and error
bars), as well as the averages of the profiles corresponding to
young stars (age $\le 3$  Gyears, orange squares and error bars),
normalised for each ELO to their corresponding
$\sigma_{\rm los}^{\rm star}(R_{\rm e, bo}^{\rm star})$. The decline
of these  velocity dispersion profiles can be clearly appreciated,
as well as the slightly larger decline of the profiles corresponding
to the younger stellar populations.
These results are consistent, within their dispersions,
with that shown by Dekel et al. (2005) in their figure 2 (lower panel).
They are also marginally consistent with 
the decline shown by PN data 
in the  NGC 821, NGC 3379, NGC 4494 and NGC 4697 galaxies
(Romanowsky et al. 2003;
Romanowsky 2006).
Note, however, that our ELOs are boxy, while the $a_4 \times 100/ a$
shape parameters
for these galaxies are
2.5, 0.2, 0.3 and 1.4, respectively, that is, they are rather discy ellipticals.

\section{Summary, discussion  and conclusions}
\label{summary}

\subsection{Summary: method and main results}
This paper belongs to a series aimed at studying galaxy formation in
a cosmological context through hydrodynamical simulations.
Here we present an analysis of the sample of
elliptical-like-objects (ELOs) formed in ten different
cosmological  simulations,
run within the same
global flat $\Lambda$ cosmological model, roughly consistent
with observations.
The normalisation parameter has been taken slightly high,
$\sigma_8 = 1.18$, as compared with the average fluctuations
of 2dFGRS or SDSS  galaxies, 
to mimic an active region of the Universe.
Newton laws and hydrodynamical
equations have been integrated in this context,
with a standard cooling algorithm and a SF parameterisation
through a Kennicutt-Schmidt-like law,
containing our ignorance about its details at sub-kpc scales,
and where subresolution processes affecting SF are
implicitly taken into account through the values given to these parameters.
No further hypotheses to model  the assembly processes have been made.
Individual galaxy-like objects
naturally appear as an output of the simulations, so that
the physical processes underlying mass assembly can be studied.
Five out of the ten simulations (the SF-A type simulations)
share the SF parameters
and differ in the seed  used to build up the initial conditions.
To test the role of SF parameterisation,
the same initial conditions have been run with different SF parameters
making SF  more difficult, contributing another set of five
simulations  (the SF-B type simulations).
ELOs have been identified in the simulations as those
galaxy-like objects  at $z = 0$ having a prominent,
dynamically relaxed spheroidal   component made out
of stars, with no extended discs and very low gas content.
These stellar component is embedded in  a dark matter halo
that contributes an important fraction
of the mass at distances from the ELO centre larger than
$\sim 10 - 15$ kpc on average, within which some clumps
made out of cold dense gas and stars, associated in some cases
with dark matter substructures, orbit.
No ELOs with stellar masses below 3.8 $\times$ 10$^{10}$ M$_{\odot}$
or virial masses below 3.7 $\times$ 10$^{11}$ M$_{\odot}$ have been
found that met the selection  criteria
(see Kauffmann et al. 2003 for a similar result in SDSS galaxies
and Dekel \& Birnboim 2006, and Cattaneo et al. 2006 for a possible
physical explanation).
ELOs have also an extended halo of hot, diffuse gas.
Stellar and dark matter particles constitute a dynamically hot
component with an important velocity dispersion, and, except
in the very central regions, a positive anisotropy.

The first step in the program of studying the origins of EGs
through self-consistent simulations, is to make sure that they
produce ELO samples  that have counterparts
in the real local Universe.
This objective has been partially fulfilled in previous works.
An analysis of the structural and dynamical ELO parameters that
can be constrained from observations has shown that they
are consistent with those measured in the SDSS DR1 elliptical sample
(S\'aiz et al. 2004), including the FP relation (O\~norbe et al. 2005;
O\~norbe et al. 2006). Also, ELO stellar populations  have age distributions
with the same trends as those inferred from observations, i.e.,
most stars have formed at high $z$ on short timescales, and, moreover
more massive objects have older  means  and narrower spreads
in their stellar age distributions than less massive ones
(DSS04).

In this paper we address  the important issue of the
amount and distribution of dark matter in virtual ellipticals
and, in particular, its amount and distribution
relative to the bright matter distribution.
We also address the kinematics of the dark matter component
and its relationship with the kinematics of the bright matter
component. We want to answer to the question of the mass and
extension of dark matter haloes in EGs.
To answer to this question, we have first compared our virtual
results with new observational
data, obtaining a very satisfactory agreement.
To be specific:
\begin{itemize}
\item
The projected stellar {\it mass} profile, $\Sigma^{\rm star}(R)$,
can be adequately
fitted by a S\'ersic-like law. The shape parameter values $n$ we
have obtained are consistent with observations, including their
correlations with the ELO luminosity,  size and velocity dispersion.
\item
The fraction of dark-to-total mass inside the projected half-mass radii
are consistent with the observational ones obtained by
Cappellari et al. (2006) from SAURON data.
\item
The gradients of the dark-to-stellar 
$M^{\rm dark}(<r)/M^{\rm star}(<r)$ profiles
as a function of their stellar masses, are consistent with
those observationally found by Napolitano et al. (2005)
for boxy ellipticals.
\item
The total mass (i.e., baryonic plus dark)
density profiles can be well fit by a power-law expression
in a large range of $r/r_{\rm vir}$ values, with power-law slopes that
are consistent with, within the dispersion, or slightly  higher than
those observationally found
by Koopmans et al. 2006
for massive lens  ellipticals within their Einstein radii.
\item
{The line-of-sight velocity profiles along the major axis show, in some
cases, a  clear  rotation, even if in most cases
the rotation is modest or low. The values of the rotation ratio
along the major and minor axis (a measure of the rotation in ELOs)
does not depend on the mass scale}
\item
The values spanned by the rotation ratios of ELOs are typical of boxy ellipticals.
\item
The line-of-sight velocity dispersion profiles, $\sigma_{\rm los}(R)$,
decline outwards at large $R$,
and the slope slightly increases when only the younger
stellar populations are considered.
These profiles are only marginally consistent with data on
PNs at large radii;  but these correspond to discy ellipticals
while our virtual ellipticals are rather boxy.
\end{itemize}

These agreements  strongly suggest that
the intrinsic three-dimensional dark and bright matter mass and velocity
distributions we get in our simulations might also adequately describe
real ellipticals. We now summarise our most important findings
on the three-dimensional mass and velocity structure of ELOs:

\begin{itemize}
\item
ELOs are embedded in extended massive dark matter haloes.

\item
The best fits  of  their spherically-averaged dark matter density profiles to
usual analytical formulae (Hern90, NFW, TD, JS, Eina)
are generally provided by the two last
formulae.
The quality of the fits is good, so that ELO haloes
form a two-parameter family where the two parameters are correlated. This
is consistent with those produced in purely N-body simulations. 
The JS inner slope parameter, $\alpha$,
is always higher than the NFW value ($\alpha$ =1).
\item
The slope parameters grow as the
ELO mass scale decreases, indicating that the halo concentration grows when the
mass decreases.
\item
Halos have suffered from adiabatic contraction.
This can be made quantitative by
comparing  the plot of the density at  the Einasto
scale, $\rho_{-2}$, versus the scale $r_{-2} = a_h$,  with the plot provided
by Navarro et al. 2004 (results of purely N-body simulations).

\item
At the ELO scale, most baryons have turned into stars.
The three dimensional stellar-mass density profiles can be fit by
Einasto  or JS profiles, but with small $r_{-2}$ values.
\item
The mass distribution homology is  broken in the stellar mass as well as in the
dark- versus bright-mass distributions, with the stellar mass
distribution relative to dark mass one less concentrated with
increasing ELO mass. That is, massive ELOs miss baryons at short scales
as compared with less massive ones, when we normalise to the dark matter
content.
This result is consistent with the observational ones by
Cappellari et al. (2006) from SAURON data, as well as by
Napolitano et al. (2005) we quoted above.
\item
At the halo scale, the baryon fraction profiles
have been found to show a typical pattern,
where their values are high at the centre, then they
decrease and have a minimum  roughly at
0.3 $< r_{\rm min}^{\rm ab}/r_{\rm vir} <$0.7,
well below the global value, $\Omega_b / \Omega_m = 0.171$,
then they increase again,
reach a maximum value and then they decrease and fall
to the   global $\Omega_b / \Omega_m$ value well beyond the virial radii
$r_{\rm vir}$.
This suggests that the baryons that massive ELOs miss at short
scales (stars) are found at the outskirts of the configuration as diffuse hot gas.
This result could reflect the presence of a stable virial shock that
prevents gas infall more efficiently as mass increases
(Dekel \& Birnboim 2006).
\item
Concerning kinematics,
ELO velocity dispersion profiles in three dimensions
are slightly decreasing for increasing $r$, both for
dark matter and stellar particles, $\sigma^{\rm d}_{3D}(r)$
and $\sigma^{\rm star}_{3D}(r)$.
\item
The dark and bright matter components of ELOs are kinematically
segregated, as we have found that $(\sigma^{\rm d}_{3D}(r))^2 \sim$ (1.4 -- 2)
$(\sigma^{\rm star}_{3D}(r))^2$,
confirming previous results
(S\'aiz 2003; Loewenstein 2000; Dekel et al. 2005).
This is so because stars are formed from gas that had lost
energy by cooling.

\item
This kinematical segregation does
not show any clear mass or radial dependence.
\item
The anisotropy is always positive (i.e., an excess of radial motions) and almost
non-varying with $r$ inside the ELOs.
Recall, however, that ELOs have been identified as
dynamically relaxed objects: there are not recent mergers
in our samples.
\item
The stellar component generally shows
more anisotropy than
the dark component, maybe derived from the radial motion  of the gas particles
that gave rise to the stars.

\end{itemize}

\subsection{Possible resolution effects}

\begin{figure}
  \begin{center}
      \includegraphics[width=.45\textwidth]{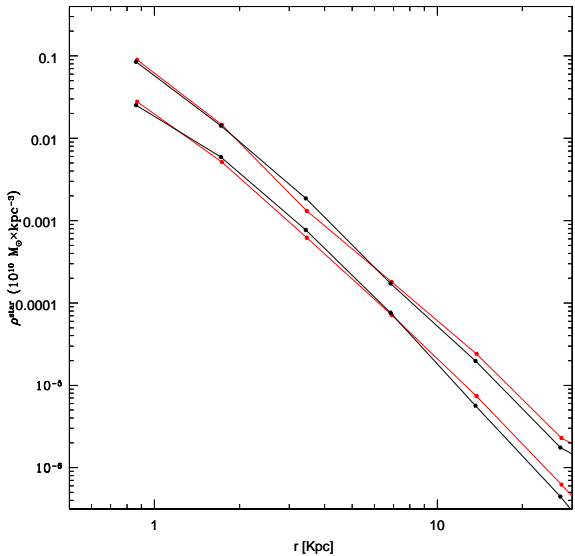}
      \caption
      {Three-dimensional stellar mass density profiles for 2 ELOs identified at
      $z=1$ in S128 (red lines) and their counterparts in S64 (black lines).
      }
         \label{ComPerStar}
	   \end{center}
	   \end{figure}

To make sure that the results we report in this paper are not unstable under
resolution changes, a control simulation with 128$^3$ dark matter and
128$^3$ baryonic particles, a gravitational softening of $\epsilon = 1.15$ kpc
and the other parameters as in SF-A type simulations (the S128 simulation),
has been run. The results of its analysis have been compared with those of a
2 $\times 64^3$ simulation (the S64 simulation),
whose initial conditions have been built up
by randomly choosing 1 out of 8 particles in the S128 initial conditions,
so that every ELO in S128 has a counterpart in the lower resolution
simulation and conversely. Due to the very high CPU time requirements
for S128, the comparison has been made at $z=1$.
The results of this comparison are very
satisfactory, as Figure~\ref{ComPerStar} illustrates.

Otherwise,
Figures 20 and 21  show that two-body relaxation effects
(typically the most  stringent requirement for convergence) have not
been important at least for $r$ larger than $\sim  1$ kpc.
In fact, two-body relaxation effects cause energy equipartition.
But the  the values of the
$\sigma^{\rm star}_{3D}(r)/\sigma^{\rm dark}_{3D}(r) $ ratios we have obtained
($\simeq 0.8$)
exclude energy equipartition among dark matter and stellar particles in ELOs,
because such equipartition would demand
$\sigma^{\rm star}_{3D}(r)/\sigma^{\rm dark}_{3D}(r) = \left [m^{\rm dark}/m^{\rm star} \right]^{0.5} = 2.194$,
where $m^{\rm dark} =  1.29 \times 10^8 $M$_{\odot}$
and $m^{\rm star} = 2.67 \times 10^7 $M$_{\odot}$ are the mass
of dark and stellar particles, respectively.
This result is important because it shows that two-body relaxation effects
have played no important role in the gravitational interaction.

\subsection{The physical processes underlying mass homology breaking
and their observational implications}

One of the most important findings of this paper is the
homology breaking ELO samples show both in the relative content and
in the relative distribution of the  baryonic and the dark mass components.
As explained in O\~norbe et al. (2005, 2006), this has as a consequence
the observed tilt of the Fundamental Plane relation relative
to the virial one. Which are the physical processes underlying
this breaking of homology? According with our simulations, they lie in
he systematic decrease,
with increasing ELO mass,
of the relative amount of
dissipation experienced by the baryonic mass component along ELO
stellar mass assembly (DTal06, O\~norbe et al. 2006).
This possibility, already  suggested by Bender, Burstein \& Faber (1992),
Guzm\'an, Lucey \& Bower (1993) and Ciotti, Lanzoni \& Renzini (1996), was first
addressed through numerical methods by
Bekki (1998).  He studied  elliptical formation through pre-prepared simulations
of dissipative mergers of disc galaxies,
where the rapidity of the SF in mergers is controlled by
a free efficiency parameter $C_{\rm SF}$. He shows that the SF
rate history of galaxies determines the differences in dissipative
dynamics, so that to explain the lack of homology in EGs
he {\it needs to assume}
that more luminous galaxies are formed by galaxy mergers with a shorter
timescale for gas transformation into stars.
Recently, Kobayashi (2005) and Robertson et al. (2006) have confirmed
the importance of dissipation and the timescale for
SF to explain mass homology breaking
in ellipticals.

In this paper we go an step further and study mass and velocity distributions
of two samples of virtual
ellipticals formed in  self-consistent cosmological simulations.
As explained in Section 3, the simulations provide us with clues on
the physical processes involved in elliptical formation.
They  indicate
that most of the dissipation involved in the mass assembly of
a given ELO occurs  in the   violent early phase at high $z$ and on very short
timescales (and earlier on and on shorter timescales as the ELO mass grows,
see details in DSS04 and DTal06),   as a consequence
of ELO assembly out of gaseous material and its transformation into stars.
Moreover, the dissipation rate history is reflected by the star
formation rate history.
During the later  slower phase of mass assembly,
ELO stellar mass growth  essentially  occurs without
any dissipation and the SF rate substantially
decreases (see more details and Figure 1 in DTal06).
So, the mass homology breaking appears in the early, violent phase of mass
assembly and it is essentially preserved during the later, slower phase.
A consequence is that the dynamical plane
appears in the violent phase and is roughly preserved
along the slower phase, see discussion in DTal06 and O\~norbe et al. 2006.
We see that our results on the role of dissipative
dynamics essentially include previous ones,
but they also add important new informations.
First, our results on the role of dissipative dynamics to break mass homology
agree with the previous ones, but it is important to note that, moreover,
ELO stellar populations show age effects, that is,
more massive objects produced in the
simulations {\it do have} older  means  and narrower spreads
in their stellar age distributions than less massive ones
(see details DSS04); this  is equivalent to downsizing (Cowie et al. 1996)
and naturally appears in the
simulations, so that it need not be considered as an additional
assumption.
Second, the preservation of the dynamical plane
in the slow phase of mass aggregation
in our simulations also
agrees with previous work based on dissipationless simulations
of pre-prepared mergers (Capelato et al. 1995; Dantas et al. 2003;
Gonz\'alez-Garc{\'\i}a \& van Albada 2003;  Nipoti et al. 2003;
Boylan-Kochin et al. 2005).
But, again, it is important to note that the important decrease of
the dissipation rate in the slow phase of evolution naturally appears in the
simulations and we do not have to assume this decrease.
Also, the decrease of the merger
rate in the later phase of mass assembly results from the global
behaviour of the merger rate history in the particular cosmological context
we have considered.
Third, it turns out that the physical processes involved in ELO
formation unveiled by our simulations,  not only explain mass homology breaking
(and its implications in the formation and preservation of the dynamical
plane), and stellar age effects or downsizing in ellipticals,
but they might also explain other elliptical properties
recently inferred from observations
 (for example, the appearance of
   blue cores,  Menanteau et al. 2004; the increase of the stellar mass
      contributed by the elliptical population since higher $z$,
          Bell et al. 2004; Conselice, Blackburne, {\&} Papovich 2005;
	       Faber et al. 2005;
	             see more details in DTal06).

\subsection{Conclusions}

We conclude that the
simulations provide an unified scenario where most current
observations on ellipticals can be interrelated.
In particular, this scenario  proofs the importance of dark matter haloes
in relaxed virtual ellipticals, and suggests that real ellipticals
must  also have extended, massive dark matter haloes.
Also, this scenario explains the homology breaking
in the relative dark- to bright-mass content and distribution of ellipticals,
that could have important implications to explain the physical
origin of the Fundamental Plane relation and its preservation.
It is worth mentioning that this scenario
shares some characteristics of previously proposed scenarios,
but it has also significant differences, mainly
that most stars  in EGs
form out of cold gas that had never been shock heated
at the halo virial temperature and then formed a disc, 
as the conventional recipe for galaxy formation propounds
(see discussion in Keres et al. 2005 and references therein).
The scenario for elliptical formation emerging from our simulations
has the advantage that it results from simple
physical laws acting on initial conditions that are
realizations of power spectra consistent
with observations of CMB anisotropies.

We thank H. Artal for computing assistance.
This work was partially supported by the MCyT (Spain) through grants
AYA-0973, AYA-07468-C03-02 and AYA-07468-C03-03
from the PNAyA,  and also by  the
regional government of Madrid through the ASTROCAM Astrophysics network (S--0505/ESP--0237).
We  thank the Centro de Computaci\'on
Cient\'{\i}fica (UAM, Spain) for computing facilities.
AS  thanks  FEDER financial support from UE.

\bibliographystyle{mn}

\begin{thebibliography}{87}
\expandafter\ifx\csname natexlab\endcsname\relax\def\natexlab#1{#1}\fi

\bibitem[]{Balles:06} Ballesteros-Paredes J., Klessen R.~S., McLow M.~-M.,
V\'azquez-Semadeni E., 2006, astro-ph/0603357 preprint

\bibitem[]{Bekki:98} Bekki K, 1998, \apj, 496, 713

\bibitem[]{Bell:04} Bell E.~F., et al. 2004, ApJ, 608, 752

\bibitem[]{Benderetal92}
Bender R., Burstein D., Faber S.~M., 1992, \apj, 399, 462


\bibitem[]{Ber:03} Bergond G., Zepf S.~E., Romanowsky A.~J., Sharples R.~M.,
 Rhode K.~L., 2006, A\&A, 448, 155

\bibitem[]{Bernardi:03} Bernardi M., et al., 2003, AJ, 125, 1866  

\bibitem[]{Bert:02} Bertin G., Ciotti L., Del Principe M., 2002, A\&A, 386, 149

\bibitem[]{BT:87} Binney J., Tremaine S., 1987, Galactic Dynamics, Princeton 
University Press (Princeton, New Jersey)

\bibitem[]{BM:98} Binney J., Merrifield M., 1998, Galactic Astronomy, Princeton
 University Press (Princeton, New Jersey)

\bibitem[]{BFFP86} {Blumenthal} G.~R., {Faber} S.~M., {Flores} R., {Primack} J.~R., 1986, \apj,
    301, 27

\bibitem[] {Bond:84}
Bond J.~R., Centrella J., Szalay A.~S., Wilson J.~R.\ 1984,
\mnras, 210, 515

\bibitem[] {Bor:03} Borriello A., Salucci P., Danese L., 2003, \mnras, 341, 1109 

\bibitem[]{Boy:05} Boylan-Kolchin M., Ma C.~-P., Quataert E., 2005, \mnras, 362, 184


\bibitem[] {Bry:98} Bryan G.~L., Norman M.~L., 1998, \apj, 495, 80

\bibitem]{Bullock+01}{Bullock} J.~S., {Kolatt} T.~S., {Sigad} Y., {Somerville} R.~S., {Kravtsov}
  A.~V., {Klypin} A.~A., {Primack} J.~R., {Dekel} A., 2001, \mnras, 321, 559


\bibitem[]{Busarello} Busarello G., Capaccioli M., Capozziello S., Longo G., Puddu E., 1997, A\&A, 320, 415

\bibitem[]{CCDO93}{Caon} N., {Capaccioli} M., {D'Onofrio} M., 1993, \mnras, 265, 1013

\bibitem[]{Cap:95} Capelato H.~V., de Carvalho R.~R, Carlberg R.~G., 1995, \apj, 451, 525 

\bibitem[]{Cappellari}Cappellari M. et al., 2006, \mnras, 366, 1126

\bibitem[]{Cattaneo:06} Cattaneo A., Dekel A., Devriendt J., Guiderdoni B., Blaizot J., 2006, \mnras, 370, 1651 

\bibitem[]{Cio:96}Ciotti L., Lanzoni B.,  Renzini A., 1996, \mnras, 282, 1

\bibitem[]{Cons:05} Conselice C.~J., Blackburne J.~A., Papovich C., 2005, ApJ, 620, 564

\bibitem[]{Couchman} Couchman H.~M.~P., 1991, \apj, 368, L23

\bibitem[]{Cowie:96} Cowie L.~L., Songaila A., Hu E.~M., Cohen J.~G., 1996, AJ, 112, 839

\bibitem[]{Dalcanton} Dalcanton J., Spergel D.~N., Summers F.~J., 1997, \apj, 482, 659 

\bibitem[]{Dan:03} Dantas C.C., Capelato H. V., Ribeiro A. L. B.,
 de Carvalho R. R., 2003, \mnras, 340, 398

\bibitem[]{Dekel:04} Dekel A.,  Birnboim Y., 2006, \mnras, 368, 2

\bibitem[]{Dekel+05}
  {Dekel} A., {Stoehr} F., {Mamon} G.~A., {Cox} T.~J., {Primack} J.~R., 2005,
    \nat, 437, 707
    
\bibitem[]{deLu:05} de Lucia G., Springel V., White S.~D.~M., Croton D., Kauffmann G., 2006, \mnras, 366, 499
 
\bibitem[]{deVaucouleurs48}
de Vaucouleurs G.,  1948, Annales d'Astrophysique, 11, 247

\bibitem[]{dZ} de Zeeuw T., Franx M., 1991, A.R.A.A. 29, 239

\bibitem[]{SAUR} de Zeeuw P.~T. et al., 2002, \mnras, 329, 513


\bibitem[]{DjorgovskiD87}
{Djorgovski} S., {Davis} M., 1987, \apj, 313, 59

\bibitem[]{DT:04} Dom{\'{\i}}nguez-Tenreiro R., S\'aiz A., Serna A., 2004,
\apj, 611, L5 (DSS04)

\bibitem[]{DT:06} Dom\'{\i}nguez-Tenreiro R., O\~norbe J., S\'aiz A., Artal H., Serna A.,  2006, \apj, 636, L77 (DTal06)

\bibitem[]{DOnofrio01} {D'Onofrio} M., 2001, \mnras, 326, 1517

\bibitem[]{dou02} Douglas N. et al., 2002, PASP, 114, 1234

\bibitem[]{Dressleretal87}
{Dressler} A., {Lynden-Bell} D., {Burstein} D., {Davies} R.~L., {Faber} S.~M., {Terlevich} R., {Wegner} G., 1987, \apj, 313, 42 

\bibitem[]{Eina:65} Einasto J., 1965, Trudy Inst. Astrofiz. Alma-Alta, 5, 87

\bibitem[]{Eina:68} Einasto J., 1968, Tartu Astr. Obs. Publ. Vol. 36, Nr 5-6, 414

\bibitem[]{Eina:69} Einasto J., 1969, Astrofizika, 5, 137

\bibitem[]{Eina:89} Einasto J., Haud U., 1989, A\&A, 223, 89

\bibitem[]{Elmegreen} Elmegreen B., 2002, \apj, 577, 206

\bibitem[]{Evrard} Evrard A., Silk J., Szalay A.~S., 1990, \apj, 365, 13

\bibitem[]{Faberetal87}
{Faber} S.~M., {Dressler} A., {Davies} R.~L., {Burstein} D.,
  {Lynden-Bell} D., 1987, in Nearly Normal Galaxies. From the Planck Time to  the Present, 
ed. S.~M. {Faber}, New York, Springer-Verlag, p. 175

\bibitem[]{Fab:05} Faber S.~M. et al., 2005, astro-ph/0506044 preprint

\bibitem[]{Ferr:05} Ferreras I., Saha P., Williams L.L.R., 2005, ApJ, 623, L5

\bibitem[]{Gerhard} Gerhard O., Kronawitter A., Saglia R.~P., Bender R., 2001, AJ, 121, 1936

\bibitem[]{Dib:06} Gibson, B.K., S\'anchez-Bl\'azquez, P., Courty, S., Kawata, D., 2006, astro-ph/0611086 preprint

\bibitem[]{GKKN04}
{Gnedin} O.~Y., {Kravtsov} A.~V., {Klypin} A.~A., {Nagai} D., 2004, \apj, 616,
    16

\bibitem[]{GG:03} Gonz\'alez-Garc\'ia A.~C., van Albada T.~S., 2003, \mnras, 342, 36

\bibitem[]{GG:06} Gonz\'alez-Garc\'ia A.~C., Balcells M., Olshevsky V.~S., 2006, \mnras, 372, 78

\bibitem[]{Graham98}
{Graham} A.~W., 1998, \mnras, 295, 933

\bibitem[]{Graham} Graham A., Colless M., 1997, \mnras, 287, 221

\bibitem[]{Graham06} Graham A., Merritt, D., Moore, B., Diemond, J., Terzi\'c, B., 2006, AJ, 132, 2711

\bibitem[]{Gustaf06} Gustafsson M., Fairbairn M., Sommer-Larsen J., 2006, Phys. Rev., D74, 123522 

\bibitem[]{guse02} Guzik J., Seljak U., 2002, \mnras, 335, 311

\bibitem[]{Guz:93} Guzm\'an, R., Lucey, J., Bower, R.~G. 1993, \mnras, 265, 731

\bibitem[]{Hernquist90}
{Hernquist} L., 1990, \apj, 356, 359 

\bibitem[]{Hoekstra} Hoekstra H., Yee H.~K., Gladders M.~D., 2004, \apj, 606, 67

\bibitem[]{Humphrey:06}{{Humphrey} P.~J., {Buote} D.~A., {Gastaldello} F., 
	{Zappacosta} L., {Bullock} J.~S., {Brighenti} F., 
	{Mathews} W.~G.}, 2006, \apj, 646, 899


\bibitem[]{JS00}
{Jing} Y.~P., {Suto} Y., 2000, \apj, 529, L69

\bibitem[]{Katz} Katz N., 1992, \apj, 391, 502

\bibitem[]{Kauffmann:03} Kauffmann G. et al., 2003, \mnras, 341, 54

\bibitem[]{Kaw:03} Kawata D., Gibson B.K., 2003, \mnras, 346, 135

\bibitem[]{Kaw:05} Kawata D., Gibson B.K., 2005, \mnras, 358, 16

\bibitem[]{Ken:98} Kennicutt R., 1998, \apj, 498, 541 

\bibitem[]{Ker:05} Keres D., Katz N., Weinberg D.~H., Dav\'e, R., 2005, \mnras, 363, 2
\bibitem[]{Kho:05} Khochfar S., Burkert A., 2005, \mnras, 359, 1379

\bibitem[]{Kob:05}Kobayashi C., 2005, \mnras, 361, 1216

\bibitem[]{LSD} Koopmans L.~V.~E., Treu T., 2003, \apj, 583, 606

\bibitem[]{Koopmans}  Koopmans L.~V.~E., Treu T., Bolton A.~S., Burles S., Moustakas L.~A., 2006, \apj, 649, 599

\bibitem[]{K00} Kronawitter A., Saglia R.~P., Gerhard O., Bender R., 2000, A\&As, 144, 53 

\bibitem[]{LimaNetoetal99}
{Lima Neto} G.~B., {Gerbal} D., {M{\' a}rquez} I. 1999, \mnras, 309, 481
 
\bibitem[]{Loewenstein00}{Loewenstein} M., 2000, \apj, 532, 17

\bibitem[]{MB01} Magorrian J., Ballantyne D., 2001, \mnras, 322, 702

\bibitem[]{ML04a} Mamon G.~A., {\L}okas E.~L., 2005a, \mnras, 362, 95

\bibitem[]{ML04b} Mamon G.~A., {\L}okas E.~L., 2005b, \mnras, 363, 705


\bibitem[]{Manrique} Manrique A., Raig A., Salvador-Sol\'e E., Sanchis T.,
Solanes J.~M., 2003, \apj, 593, 26

\bibitem[]{Marquez+00}
  {M{\' a}rquez} I., {Lima Neto} G.~B., {Capelato} H., {Durret} F., {Gerbal} D.,
    2000, \aap, 353, 873

\bibitem[]{Menan:04} Menanteau F., et al., 2004, \apj, 612, 202


\bibitem[]{Merr:05} Merritt, D., Navarro, J.F., Ludlow, A., Jenkins, A., 2005, ApJL, 624, L85

\bibitem[]{Merr:06} Merritt, D., Graham, A., Moore, B., Diemond, J., Terzi\'c, B., 2006, AJ, 132, 2685


\bibitem[]{Meza:2003} Meza A., Navarro J. F., Steinmetz M., Eke V. R., 2003, \apj, 590, 619

\bibitem[]{Moore} Moore B., Quinn T., Governato F., Stadel J., Lake G.,
1999, \mnras, 310, 1147

\bibitem[]{CLAS} Myers S.~T., et al., 1995, \apj, 447, L5

\bibitem[]{Naab:03} Naab T., Burkert A., 2003, \apj, 597, 893

\bibitem[]{Naab:05} Naab T., Johansson P.~H., Efstathiou G., \& Ostriker J.~P., 2005,
astroph/0512235 preprint

\bibitem[]{Naab:06a} Naab T., Khochfar S., Burkert A., 2006, \apj, 636L, 81

\bibitem[]{Naab:06b} Naab T., Trujillo I., 2006, \mnras, 369, 625

\bibitem[]{Napolitano} Napolitano N.~R., et al., 2005, \mnras, 357, 691

\bibitem[]{Navarro} Navarro J.~F., Frenk C.~S., White S.~D.~M., 1995, \mnras, 275, 720 
 
\bibitem[]{Navarroetal96}
{Navarro} J.~F., {Frenk} C.~S., {White} S.~D.~M. 1996, \apj, 462, 563

\bibitem[]{Navarro+04}
  {Navarro} J.~F., {Hayashi} E., {Power} C., {Jenkins} A.~R., {Frenk} C.~S.,
    {White} S.~D.~M., {Springel} V., {Stadel} J., {Quinn} T.~R., 2004, \mnras,
      349, 1039

\bibitem[]{Nipo:03} Nipoti C., Londrillo P., Ciotti L., 2003, \mnras, 342, 501

\bibitem[]{Nipo:06} Nipoti C., Londrillo P., Ciotti L., 2006, \mnras, 370, 681

\bibitem[]{Ono:05} O\~norbe J., Dom\'{\i}nguez-Tenreiro R., S\'aiz A., Serna A., Artal H., 2005, \apj, 632, L57

\bibitem[]{Ono:06} O\~norbe J., Dom\'{\i}nguez-Tenreiro R., S\'aiz A., Artal H., Serna A., 2006, \mnras, 373, 5030

\bibitem[]{OsPon04a} O'Sullivan E., Ponman T.~J., 2004a, \mnras, 349, 535

\bibitem[]{OsPon04b} O'Sullivan E., Ponman T.~J., 2004b, \mnras, 354, 935

\bibitem[]{Padmanabhan} Padmanabhan N. et al., 2004, New Astronomy, 9, 329   

\bibitem[]{Pahre} Pahre M.~A., de Carvalho R.~R., Djorgovski S.~G., 1998, AJ, 116, 1606

\bibitem[]{PS97}
{Prugniel} P., {Simien} F., 1997, \aap, 321, 111

\bibitem[]{Rob:06} Robertson B., Cox T. J., Hernquist L., Franx M., Hopkins P. F.,
 Martini P., Springel V., 2006, \apj, 641, 21

\bibitem[]{Romanowsky} Romanowsky A.~J., Kochanek C.~S., 2001, \apj, 553, 722

\bibitem[]{ral03} Romanowsky A.~J., Douglas N. G., Arnaboldi M., Kuijken K., Merrifield M. R., Napolitano N. R., Capaccioli M., Freeman K. C., 2003, Science, 301, 1696 

\bibitem[]{romano06} Romanowsky A.~J., 2006, EAS Publication Series, 20, 119

\bibitem[]{Romeo:05} Romeo A.~D., Portinari L., Sommer-Larsen J., 2005, \mnras, 361, 983

\bibitem[]{637} S\'aiz A., Dom{\'{\i}}nguez-Tenreiro R., Tissera P.~B., Courteau S. 2001, \mnras, 325, 119

\bibitem[]{SaiT:03} S\'aiz A., 2003, PhD thesis, Universidad Aut\'onoma de Madrid, Spain

\bibitem[]{Sai:03} S\'aiz A., Dom{\'{\i}}nguez-Tenreiro R., Serna A. 2003, Ap\&SS, 284, 411

\bibitem[]{Sai:04} S\'aiz A., Dom{\'{\i}}nguez-Tenreiro R., Serna A. 2004, \apj, 601, L131 

\bibitem[]{SS:05} Salvador-Sol\'e E.,  Manrique A., Solanes J.~M. 2005,
\mnras, 358, 901    
 
\bibitem[]{Sarson} Sarson G.~R., Shukurov A., Nordlund A., Gudiksen B., Brandenburg A. 2004, Ap\&SS, 292, 267


\bibitem[]{Serna} Serna A., Dom{\'{\i}}nguez-Tenreiro R., S\'aiz A., 2003, \apj, 597, 878
 
\bibitem[]{Sersic68}
{S{\' e}rsic} J.~L., 1968, {Atlas de galaxias australes} (C{\' o}rdoba,
  Argentina: Observatorio Astron{\' o}mico, 1968)

\bibitem[]{Sierra} Sierra-Glez. de Buitrago M.~M.,  Dom{\'{\i}}nguez-Tenreiro R., Serna A., 2003, in Gallego J., Zamorano J., Cardiel N., Highlights of Spanish Astronomy III. Kluwer Academic Press, p. 171

\bibitem[]{Spergel:06} Spergel D.~N., et al., 2006, astro-ph/0603449 preprint

\bibitem[]{Sommer:02} Sommer-Larsen J., Gotz M.,  Portinari L., 2002, Ap\&SS, 281, 519  

\bibitem[]{Thomas} Thomas D.,  Greggio L.,  Bender R.,  1999, \mnras, 302, 537

\bibitem[]{TisseraDT98}
{Tissera} P.~B., {Dom{\'{\i}}nguez-Tenreiro} R., 1998, \mnras, 297, 177
 
\bibitem[]{Tisseraetal97}
{Tissera} P.~B., {Lambas} D.~G., {Abadi} M.~G. 1997, \mnras, 286, 384

\bibitem[]{Treu} Treu T., Koopmans L.~V.~E., 2004, \apj, 611, 739 

\bibitem[]{Trujillo} Trujillo I., Graham A.~W., Caon N., 2001, \mnras, 326, 869

\bibitem[]{Tucke:75}
Tucker W.~H., 1975, Radiation Precesses in Astrophysics (New York,
Wiley)

\bibitem[]{vdBal04} van den Bosch F.~C., Norberg P., Mo H.~J., Yang X., 2004, \mnras, 352, 1302

\bibitem[]{Vazde:04} Vazdekis A., Trujillo I., Yamada Y., 2004, \apj, 601, L36 

\bibitem[]{Vazquez:03a} V\'azquez-Semadeni E., 2004a, IAU Symposium, 221, 51

\bibitem[]{Vazquez:03b} V\'azquez-Semadeni E., 2004b, Ap\&SS, 292, 187

\bibitem[]{Vergasola} Vergassola M., Dubrulle B., Frisch U., Noullez A., 1994, A\&A, 289, 325

\bibitem[]{vanderMarelF93}
{van der Marel} R.~P., {Franx} M. 1993, \apj, 407, 525

\bibitem[]{Wechsler}
Wechsler R.~H., Bullock J.~S., Primack J.~R., Kravtsov A.~V., Dekel A., 2002, \apj, 568, 52

\bibitem[]{Zhao:03}
Zhao D.~H., Mo H.~J., Jing Y.~P., Borner G., 2003, \mnras, 339, 12

 
\end{thebibliography}

\end{document}